\documentclass{article}
\usepackage{arxiv}

\usepackage[utf8]{inputenc} 
\usepackage[T1]{fontenc}    
\usepackage{hyperref}       
\usepackage{url}            
\usepackage{booktabs}       
\usepackage{amsfonts}       
\usepackage{amsmath}
\usepackage{nicefrac}       
\usepackage{microtype}      
\usepackage{cleveref}       
\usepackage{graphicx}
\usepackage{doi}
\usepackage{subcaption}
\usepackage{multirow}
\usepackage[labelfont=bf]{caption}
\usepackage[export]{adjustbox}
\usepackage{array}
\usepackage{graphics}
\usepackage{tabularx}
\usepackage{placeins}
\usepackage[square,numbers,sort&compress]{natbib}
\usepackage{alphalph}
\usepackage{longtable}
\usepackage{pdflscape}

\setcitestyle{square}

\newcommand{\beginsupplement}{%
        \setcounter{table}{0}
        \renewcommand{\thetable}{S\arabic{table}}%
        \setcounter{figure}{0}
        \renewcommand{\thefigure}{S\arabic{figure}}%
     }

\title{Prompt Engineering for Transformer-Based Chemical Similarity Search Identifies Structurally Distinct Functional Analogues}

\author{\href{https://orcid.org/0000-0002-6420-8615}{\includegraphics[scale=0.06]{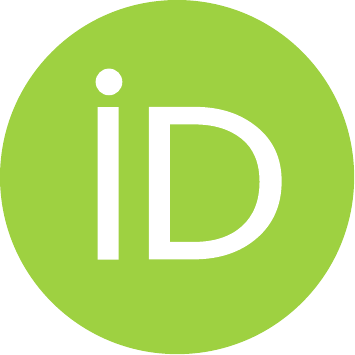}\hspace{1mm}Clayton W. Kosonocky}\\
	Department of Molecular Biosciences\\
	The University of Texas at Austin\\
	\texttt{clayton.kosonocky@utexas.edu} \\
	\And
	\href{https://orcid.org/0000-0002-4476-1026}{\includegraphics[scale=0.06]{orcid.pdf}\hspace{1mm}Aaron L. Feller} \\
	Department of Molecular Biosciences\\
	The University of Texas at Austin\\
	\texttt{aaron.feller@utexas.edu} \\
	\And
	\href{https://orcid.org/0000-0002-7470-9261}{\includegraphics[scale=0.06]{orcid.pdf}\hspace{1mm}Claus O. Wilke\thanks{Corresponding Author}} \\
	Department of Integrative Biology\\
	The University of Texas at Austin\\
	\texttt{wilke@austin.utexas.edu} \\
	\And
	\href{https://orcid.org/0000-0001-6246-5338}{\includegraphics[scale=0.06]{orcid.pdf}\hspace{1mm}Andrew D. Ellington}\\
	Department of Molecular Biosciences\\
	The University of Texas at Austin\\
	\texttt{ellingtonlab@gmail.com} \\
}

\hypersetup{
pdftitle={Prompt Engineering for Transformer-Based Chemical Similarity Search Identifies Structurally Distinct Functional Analogues},
pdfsubject={cs.LG},
pdfauthor={Clayton W.~Kosonocky, Aaron L.~Feller, Claus O.~Wilke, Andrew D.~Ellington},
pdfkeywords={drug discovery, machine learning, chemical similarity search, prompt engineering, SMILES},
}

\begin{document}
\maketitle

\begin{abstract}
    Chemical similarity searches are widely used in-silico methods for identifying new drug-like molecules. These methods have historically relied on structure-based comparisons to compute molecular similarity. Here, we use a chemical language model to create a vector-based chemical search. We extend implementations by creating a prompt engineering strategy that utilizes two different chemical string representation algorithms: one for the query and the other for the database. We explore this method by reviewing the search results from five drug-like query molecules (penicillin G, nirmatrelvir, zidovudine, lysergic acid diethylamide, and fentanyl) and three dye-like query molecules (acid blue 25, avobenzone, and 2-diphenylaminocarbazole). We find that this novel method identifies molecules that are functionally similar to the query, indicated by the associated patent literature, and that many of these molecules are structurally distinct from the query, making them unlikely to be found with traditional chemical similarity search methods. This method may aid in the discovery of novel structural classes of molecules that achieve target functionality.
\end{abstract}


\section{Introduction}

Applications for small molecules in modern society are various and widespread, including treatment of heritable disease, pathogen inhibition, and the generation of functional materials for use in electronics and consumer goods. Molecular function emerges from structure, but it is not always obvious how this emerges from first principles due to the dependence of function on the target molecule \cite{li2020mechanisms}. Traditionally, exploration of natural products has led to the identification of vital pharmaceuticals and specialty chemicals \cite{cragg2005biodiversity, fleming1941penicillin, jiao2006chaetominine, wani2014nature}. These first generation molecules act as starting points, upon which new molecules are engineered for furthering desired functionality \cite{hughes2011principles}. Structural neighbors often share similar functionality, as the relevant chemistry may be unchanged or improved \cite{martin2002structurally}. However, molecules with low structural similarity can act on the same target despite the highly different structure, as is the case with morphine and fentanyl on the $\mu$-opioid receptor \cite{pathan2012basic}.

There are numerous contemporary approaches to applying machine learning to chemistry \cite{pu2019etoxpred, yang2022neural, lee2022drug, sellner2023efficient, wei2022deeplpi, moret2022perplexity, moret2023leveraging, flam2022language}. However, the application of language models to this space has led to surprising success in predicting biochemical features such as drug-likeness and protein-ligand interactions \cite{lee2022drug, wei2022deeplpi}. These methods require string representations of molecules, commonly using the Simplified Molecular-Input Line-Entry System (SMILES) \cite{weininger1988smiles}. Language models are often trained in an unsupervised manner with the reward function tied to sequence reconstruction, i.e. feeding the model a masked or partial input with the goal of reconstructing the original sequence. It was recently demonstrated that a chemical language model, though trained only on SMILES strings, correctly predicted complex biophysical and quantum-chemical properties \cite{ross2022large}. This points to the possibility that these models develop a chemical latent space that allows for the emergence of higher-order biochemical comprehension.

Recently, computationally generated chemical libraries have grown to surpass 37 billion commercially available compounds \cite{tingle2023zinc}. This marked growth has generated a new field of computational pre-screening of chemicals in order to support resource efficient discovery in the laboratory \cite{batool2019structure}. One primary class of computational pre-screening methods are chemical similarity searches. These methods have historically used structure-based comparisons, a notable example being the fingerprint Tanimoto search which computes a hierarchical list of molecules ranked by molecular substructure similarity to a given query \cite{szilagyi2021rapid, stumpfe2011similarity}.

Chemical Language Models (CLM) have been applied to drug discovery, in particular de novo molecule generation and chemical similarity searches. De novo methods generate novel molecules with the decoder portion of a language model after fine-tuning toward a specific molecule or target \cite{moret2022perplexity, moret2023leveraging, flam2022language, ross2022large}. Building off the recent success of generative models such as GPT, de novo molecule generation has shown promise but is limited due to a lack of generalizability and guaranteed synthesizability \cite{brown2020language, gao2020synthesizability}. In contrast, a chemical similarity search based on a CLM has the advantage of computational speed, generalizability, and high database control to ensure synthesizability. Sellner et al. recently created a novel transformer-based chemical similarity search with an optimized loss function to approximate previous structure-based methods \cite{sellner2023efficient}. However, this method tends to identify molecules with high structural similarity to the query molecule, when instead we would often like to find structurally distinct functional analogues. Such a CLM-based search does not currently exist.

Here, we describe a chemical similarity search based on a CLM that identifies molecules with similar function to a given query molecule. This method works by calculating the CLM-computed feature vector similarity between a query SMILES string and a chemical database. Keeping the SMILES canonicalization algorithm constant between the query and database resulted in a chemical language search that approximated recent transformer-based chemical search methods. However, we found that when the query SMILES string was canonicalized with a different algorithm than was used for the database, the reliance on structural similarity diminished while functional similarity was retained. This behavior seemed reasonable given the literature describing how models learn to better represent chemical space when SMILES randomization occurs during training and from the reported emergent understanding of underrepresented languages in predominantly English-trained models \cite{arus2019randomized, mokaya2023testing, wei2022emergent, shi2022language}. We utilize alternatively canonicalized queries as a novel prompt engineering strategy to identify structurally distinct functional analogues of small molecules. Our method fundamentally differs from existing literature in that SMILES augmentation was used for the query of a chemical similarity search rather than for model training and fine-tuning to a specific task. We tested our method across three canonicalizations and eight query molecules and found that with increasingly divergent canonicalizations we were able to identify structurally distinct functional analogues.


\section{Methods}
\paragraph{CheSS Overview.}
The Chemical Semantic Search (CheSS) is a molecular search framework that uses language model-encoded feature vectors to compute similarity scores across molecular space. A database of molecules is encoded as strings using SMILES format \cite{weininger1988smiles}. A chemical language model is then used to generate a feature vector for each molecule in the database as well as the query molecule. The cosine similarity between the query vector and each database vector is computed, resulting in a vector of feature cosine similarities.

\paragraph{Language Model.}
ChemBERTa was used as the language model to generate embeddings \cite{ChemBERTa}. This was a Bidirectional Encoder Representations from Transformers (BERT) model with 12 hidden layers of size 768, and was trained on 10M random non-redundant achiral SMILES strings selected from PubChem \cite{BERT}. ChemBERTa was chosen over newer, higher-parameter BERT models due to the ease of implementation and publicly available dataset. ChemBERTa does not support isomeric SMILES (chirality), and all SMILES were canonicalized before input.

\paragraph{Database.}
The CheSS molecular feature database was built from the $\sim$10M random achiral molecules used to train ChemBERTa \cite{ChemBERTa}. For each molecule, the SMILES string was canonicalized using RDKit \cite{landrum2013rdkit}. We reduced this dataset to exclude all SMILES strings that tokenized to more than 512 tokens, the maximum supported by ChemBERTa. This resulted in a database of 9,999,809 molecules. The database SMILES strings were encoded into feature vectors with ChemBERTa. The [CLS] token vector representations of the final layer were chosen to be the feature vectors, as described in the original BERT paper \cite{BERT}. These feature vectors were then L2 normalized and stored in chunks of 100k SMILES string-feature vector pairs for future cosine similarity calculations.

\paragraph{Canonicalization Query Types.}
ChemBERTa was trained on SMILES strings canonicalized using RDKit \cite{landrum2013rdkit, ChemBERTa}. Different canonicalization algorithms result in different, but equally valid, standardized strings representing the same molecule, which we utilize to create three highly different queries for the same molecule. The first query type used RDKit with its default Python implementation settings. This algorithm was used to canonicalize the database \& train the model. When converting molecules to SMILES, RDKit allows specification of which atom number to root the SMILES string to. The default is Atom 0, and each atom results in a different representation. The feature cosine similarity was calculated between the default RDKit SMILES and the ``Atom n'' RDKit SMILES for each atom in the query molecule, as demonstrated in Figure \ref{fig:rdkit_atom_n_fsim}. From these, we took the most dissimilar ``Atom n'' SMILES strings to be the second query type for each molecule. To obtain a third dissimilar SMILES representation, OEChem 2.3.0, a markedly different canonicalization algorithm than RDKit, was used \cite{oechemversion}. These SMILES strings were obtained from the PubChem website.

\paragraph{Similarity Metrics.}
Various similarity metrics were used throughout, which include feature cosine similarity, Gestalt pattern matching similarity, fingerprint Tanimoto similarity, token vector length similarity, and token similarity \cite{ratcliff1988pattern, bajusz2015tanimoto}. Feature cosine similarity is a distance metric that calculates the angle between two vectors $\textbf{A}$ and $\textbf{B}$:
\begin{equation}
    \text{Cosine\ similarity} = \frac{\mathbf{A} \cdot \mathbf{B}}{||\mathbf{A}||\ ||\mathbf{B}||}.
    \label{eq:cos_sim}
\end{equation}
A cosine similarity of 1 indicates the normalized vectors are the same, 0 means they are orthogonal to one another, and $-1$ means they are opposite of one another.

Gestalt pattern matching was chosen to calculate string similarity. This metric is calculated by dividing twice the number of matching characters ($K_m$) by the total number of characters in both strings ($S_1$, $S_2$):
\begin{equation}
    \text{Gestalt\ similarity} = \frac{2K_m}{|S_1|+|S_2|}.
    \label{eq:gestalt_sim}
\end{equation}
Matching characters are identified first from the longest common substring, with recursive counts in non-matching regions on both sides of the substring. The metric ranges from a perfect match of 1 to a completely dissimilar string of 0. We used the difflib Python implementation of the Gestalt pattern matching algorithm to calculate Gestalt similarity.

Fingerprint Tanimoto similarity was used to calculate the structural similarity between pairs of molecules. This method encodes substructures into a binary vector, and then calculates the Tanimoto similarity between these encoded vectors. The Tanimoto / Jaccard similarity is the number of shared elements (intersection) between two sets $A$ and $B$ over the total number of unique elements in both sets (union) (Eq. \ref{eq:tanimoto_sim}):
\begin{equation}
    \text{Tanimoto\ similarity} = \frac{A \cap B}{A \cup B}.
    \label{eq:tanimoto_sim}
\end{equation}
This metric ranges from 1 (all elements shared) to 0 (no elements shared). The RDKit default implementation of fingerprint Tanimoto similarity was used herein.

All SMILES were encoded into token vectors before being passed into the model. These tokenized vectors were used for additional comparisons to better understand search behavior. The first metric used from these was the ratio of token lengths between two vectors. The second metric was the token Tanimoto / Jaccard similarity (Eq. \ref{eq:tanimoto_sim}) between the two molecules' token vectors, and was used to determine the ratio of shared tokens between the two vectors. This metric ranges from 1 (all tokens shared) to 0 (no tokens shared).

\paragraph{Patent \& Literature Search.}
In order to determine known functionality of the molecules examined herein, a patent \& literature search was performed. The patents and literature articles for each molecule, if available, was obtained from PubChem \cite{kim2016pubchem}. A comprehensive list of all molecules considered herein, \& their associated patents, is provided in Table \ref{tab:patents_list}.


\section{Results and Discussion}

In-silico drug discovery methods have long relied on chemical similarity searches as a computational tool in pharmaceutical pipelines \cite{stumpfe2011similarity}. Recently, language models have been successful in biochemical prediction tasks for chemical properties such as drug-likeness, protein-ligand interactions, and other metrics \cite{lee2022drug, wei2022deeplpi}. Because a multitude of characteristics can be predicted from chemical language models (CLMs), it is plausible that embeddings from these models contain a summary of molecular properties for a given molecule. So-called Simplified Molecular Line Entry System (SMILES) representations can be used to represent chemical structures as strings and can be used as inputs to CLMs, but due to the nature of chemical connectivity, there are often many valid representations for the same molecule \cite{weininger1988smiles}. This multiplicity of input formats causes the CLM to generate different embeddings for the same molecule, which can either be mitigated through string standardization (canonicalization) or utilized as a data augmentation technique that allows CLMs to better span molecular space \cite{arus2019randomized, mokaya2023testing}. The latter reduces overfitting to string artifacts and improves structural comprehension; an analogy in natural language would be that training a model on all languages improves the understanding of language as a whole compared to a model trained only on English \cite{arus2019randomized, mokaya2023testing}.

However, the combination of alternative canonicalizations and data augmentation may actually prove to be advantageous for a chemical similarity search. Recent reports on emergence from the natural language processing literature demonstrated that language models understand Swahili and other languages incredibly well despite them being <0.01\% of training data \cite{wei2022emergent, shi2022language}. SMILES canonicalization formats often share characters and substructures with one another, the primary difference being the specific grammatical rules used to assemble a given molecule. For CLMs trained on one canonicalization, inputs using unseen canonicalizations would be akin to underrepresented languages in natural language training sets, as the underlying string structure of the unseen canonicalization is not entirely unknown to the model, but is novel in the context of the molecule at hand. It seems likely that a vector comparison between embeddings from two different canonicalizations could ignore the differences in string and structural data and instead use whole-molecule properties approximating molecular function, serving as a novel prompt engineering strategy for the discovery of structurally distinct functional analogues. This would be equivalent to querying a predominantly English-trained model with a French phrase to obtain an embedding, and then searching amongst English phrases to find the entry with the closest semantic meaning. To our knowledge this potential behavior is largely unexplored.

\begin{figure}[ht!]
    \centering
    \includegraphics[width=12cm]{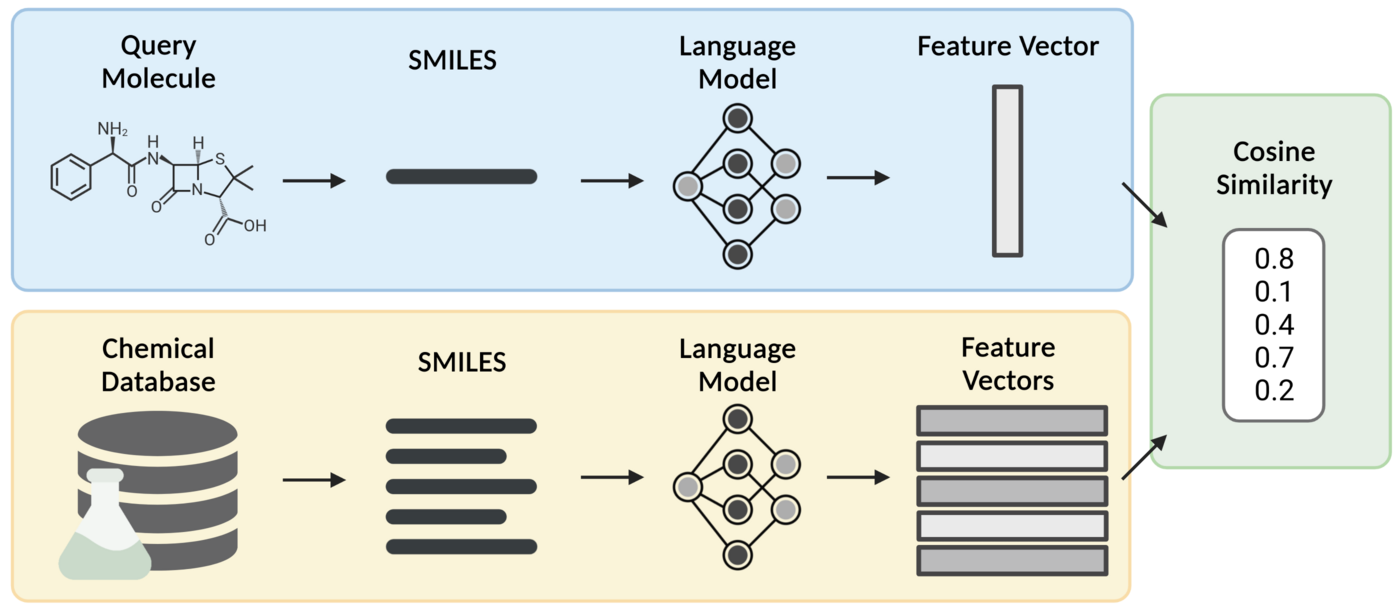}
    \captionsetup{singlelinecheck=false}
    \caption{\textbf{Chemical Semantic Search (CheSS).} The query molecule and chemical database are converted into SMILES strings, which are then inputted into a language model to obtain feature vectors. The cosine similarity between the query feature vector and database feature vectors is computed, resulting in a vector of feature cosine similarities.}
    \label{fig:workflow}
\end{figure}

\begin{figure}[h]
\centering
\begin{subfigure}{.5\textwidth}
    \centering
    \includegraphics[width=5cm]{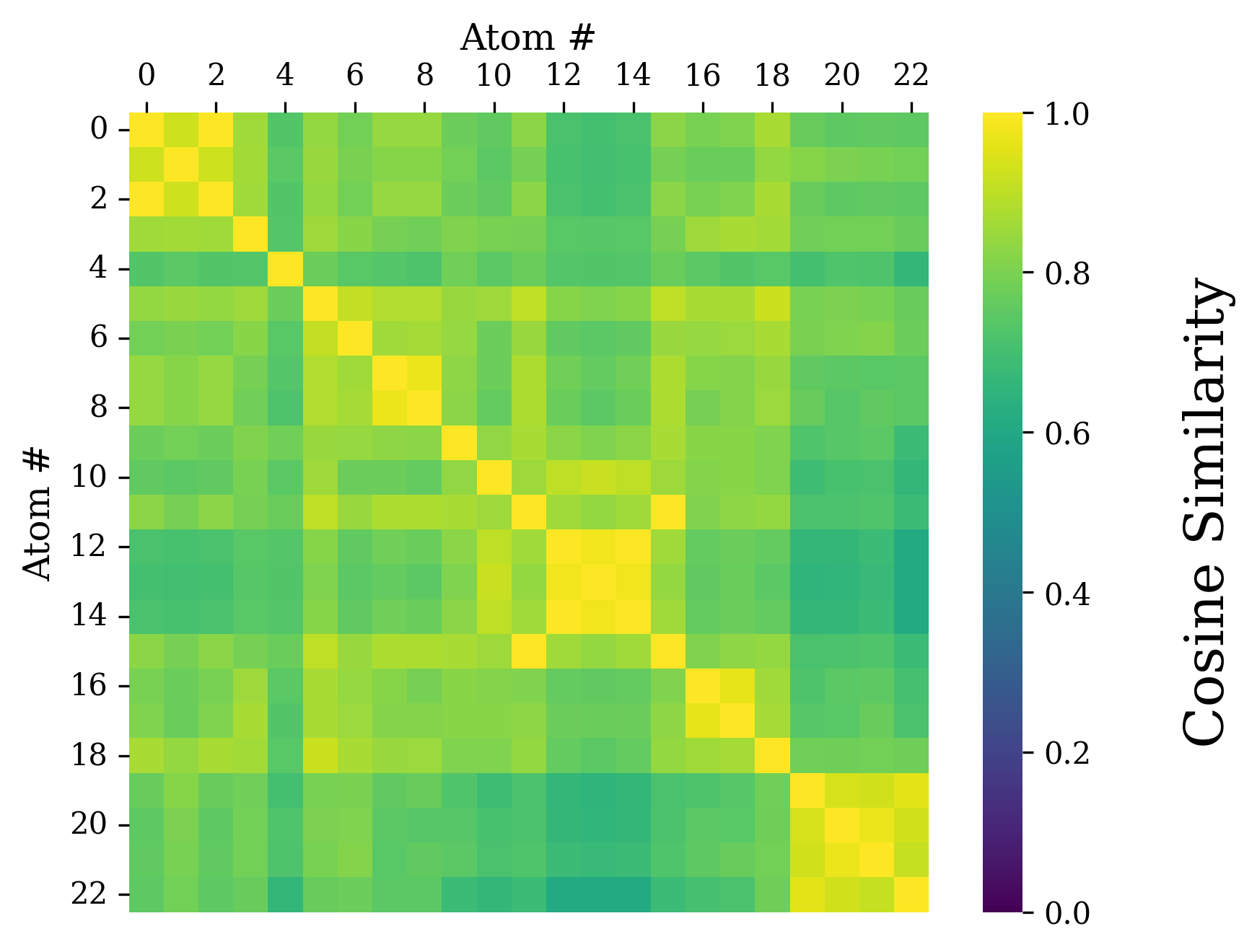}
    \captionsetup{width=5cm}
    \caption{ }
    \label{fig:rdkit_atom_n_fsim_penicillin}
\end{subfigure}%
\begin{subfigure}{.5\textwidth}
    \centering
    \includegraphics[width=5cm]{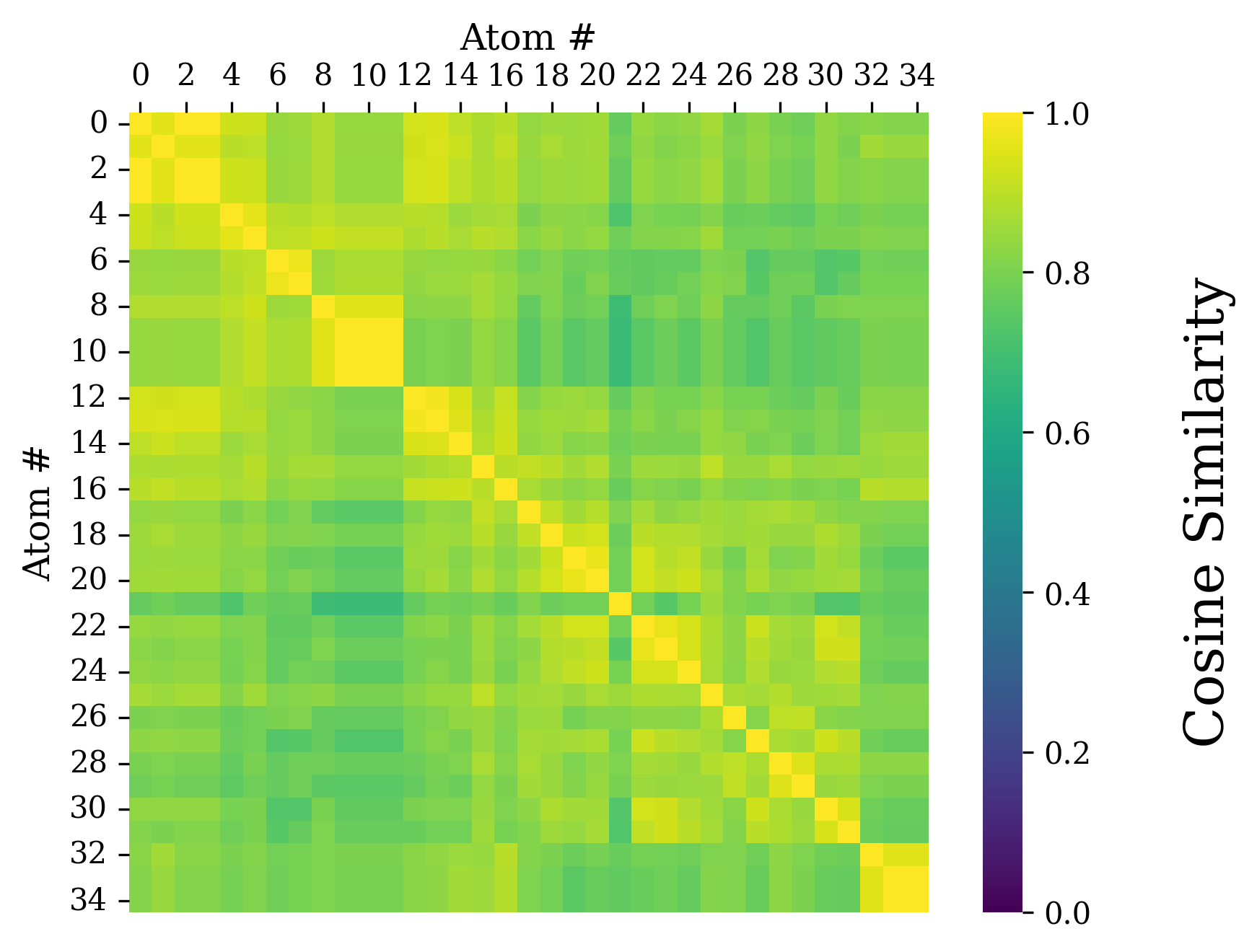}
    \captionsetup{width=5cm}
    \caption{ }
    \label{fig:rdkit_atom_n_fsim_nirmatrelvir}
\end{subfigure}\\
\begin{subfigure}{.5\textwidth}
    \centering
    \includegraphics[width=5cm]{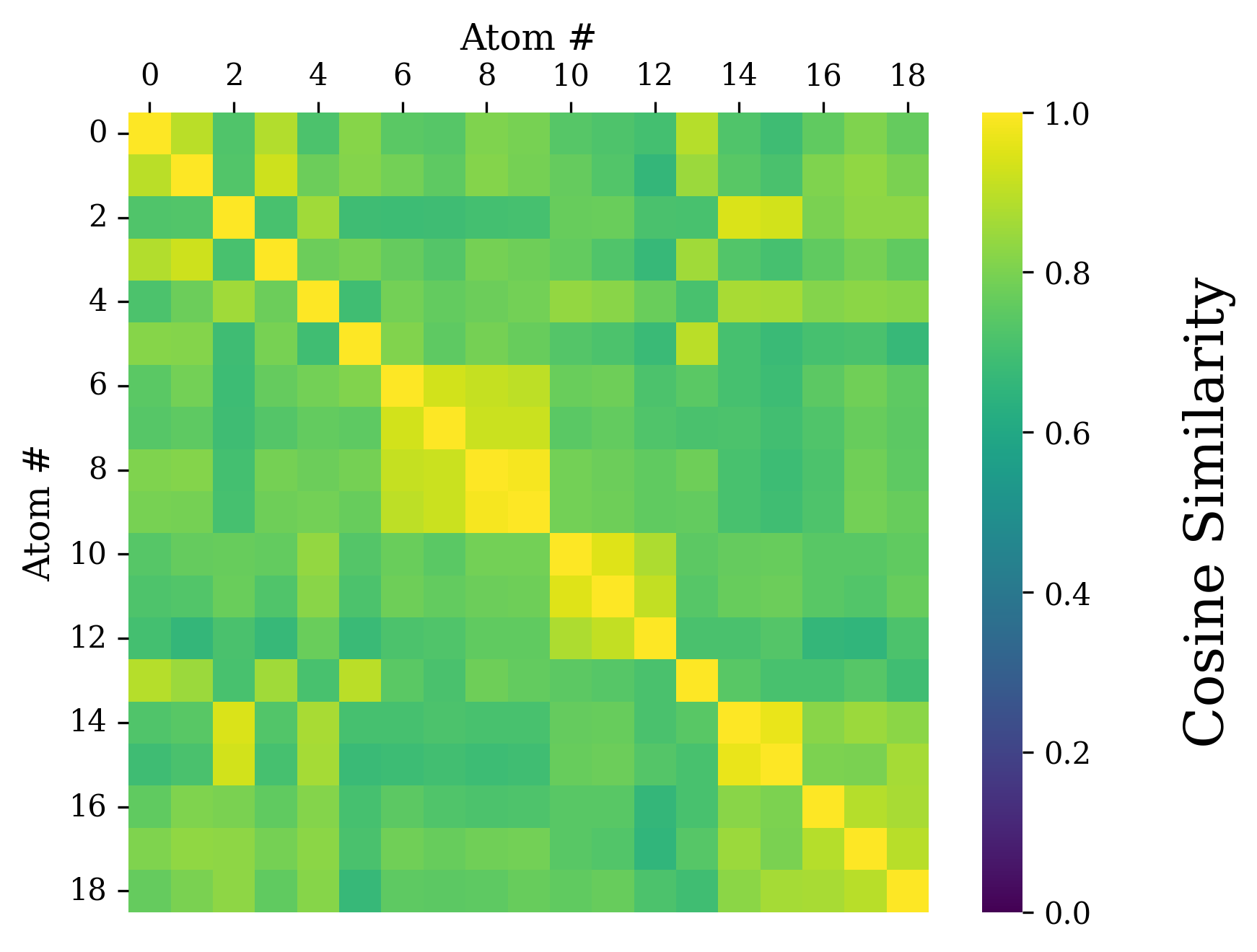}
    \captionsetup{width=5cm}
    \caption{ }
    \label{fig:rdkit_atom_n_fsim_azt}
\end{subfigure}%
\begin{subfigure}{.5\textwidth}
    \centering
    \includegraphics[width=5cm]{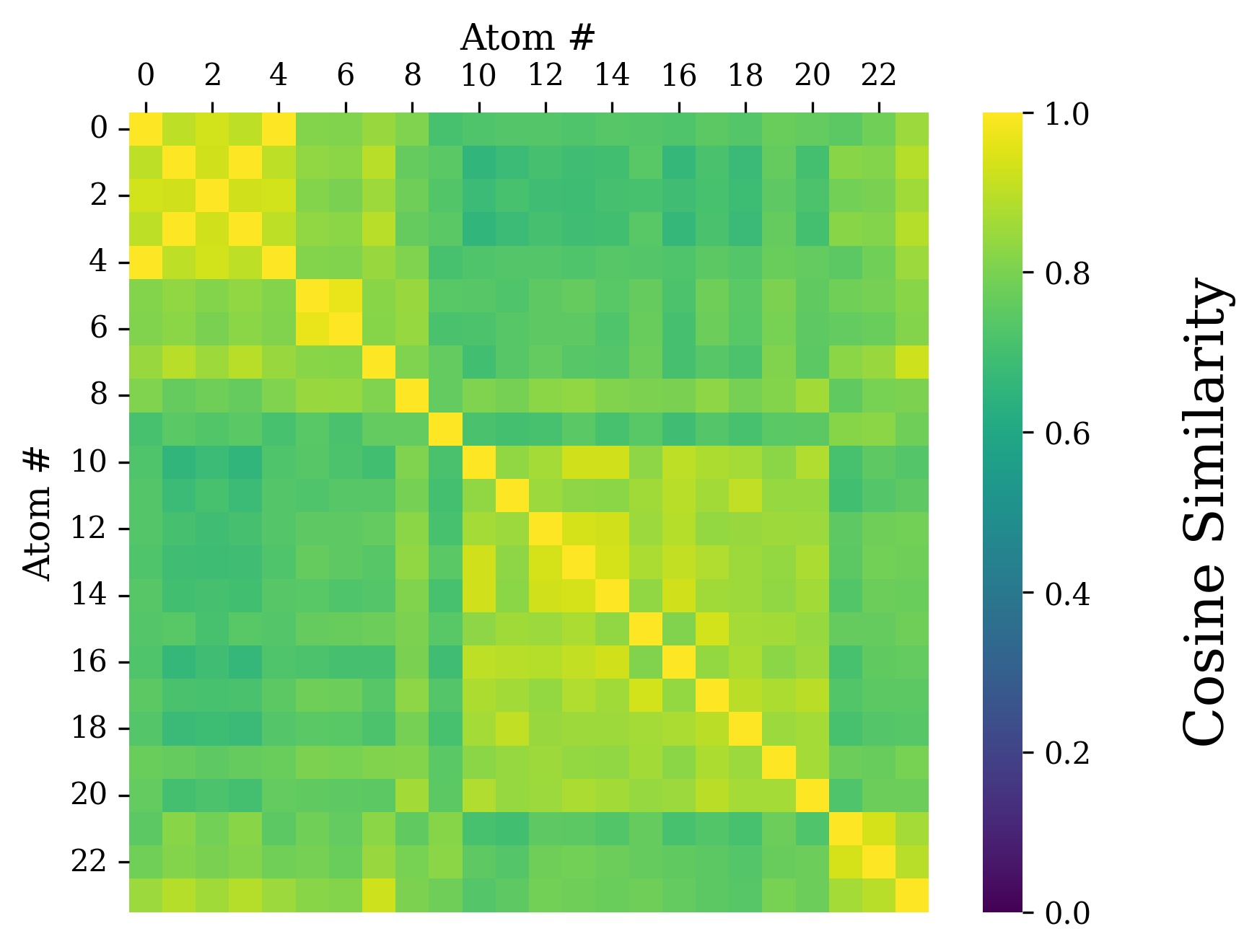}
    \captionsetup{width=5cm}
    \caption{ }
    \label{fig:rdkit_atom_n_fsim_lsd}
\end{subfigure}\\
\begin{subfigure}{.5\textwidth}
    \centering
    \includegraphics[width=5cm]{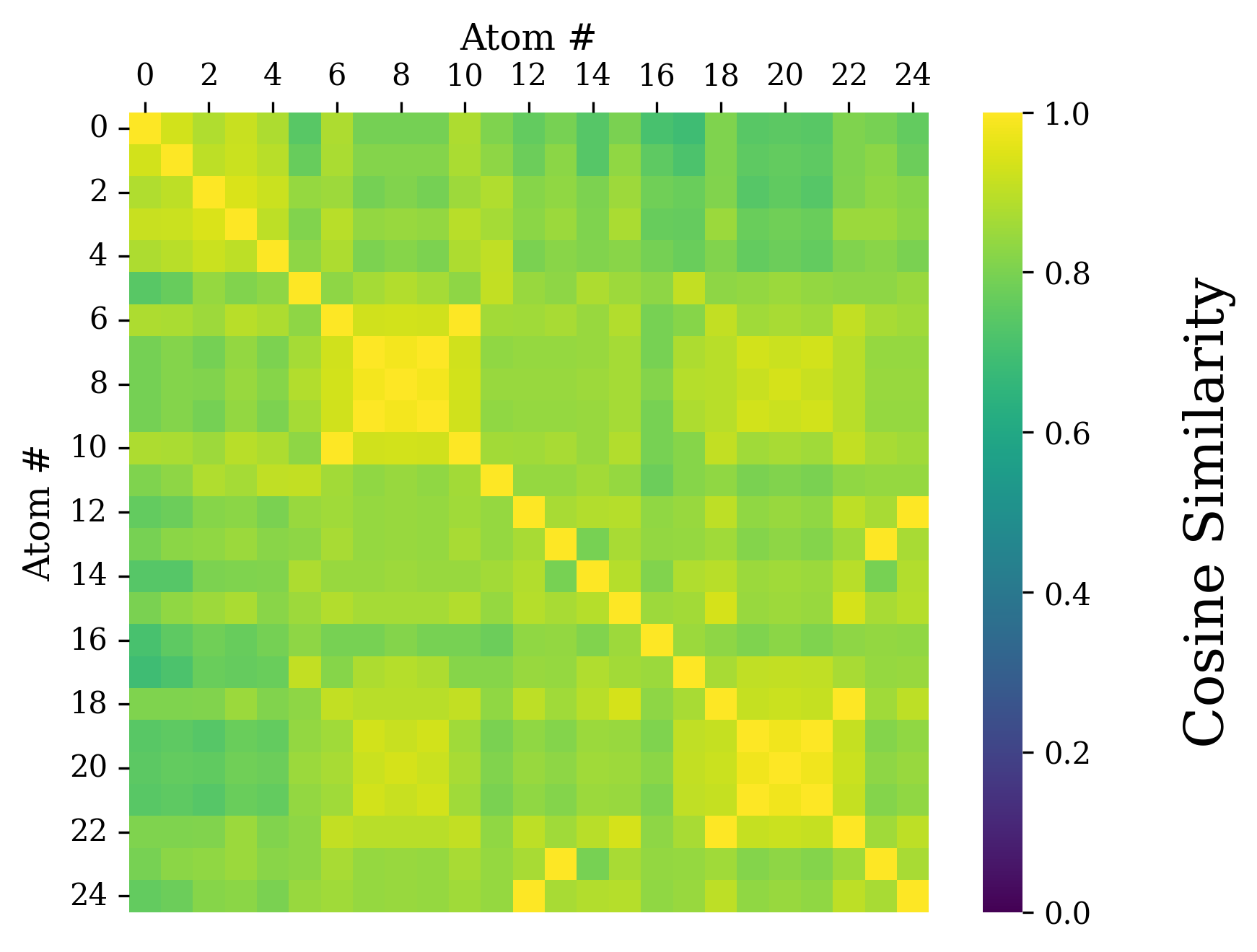}
    \captionsetup{width=5cm}
    \caption{ }
    \label{fig:rdkit_atom_n_fsim_fentanyl}
\end{subfigure}%
\begin{subfigure}{.5\textwidth}
    \centering
    \includegraphics[width=5cm]{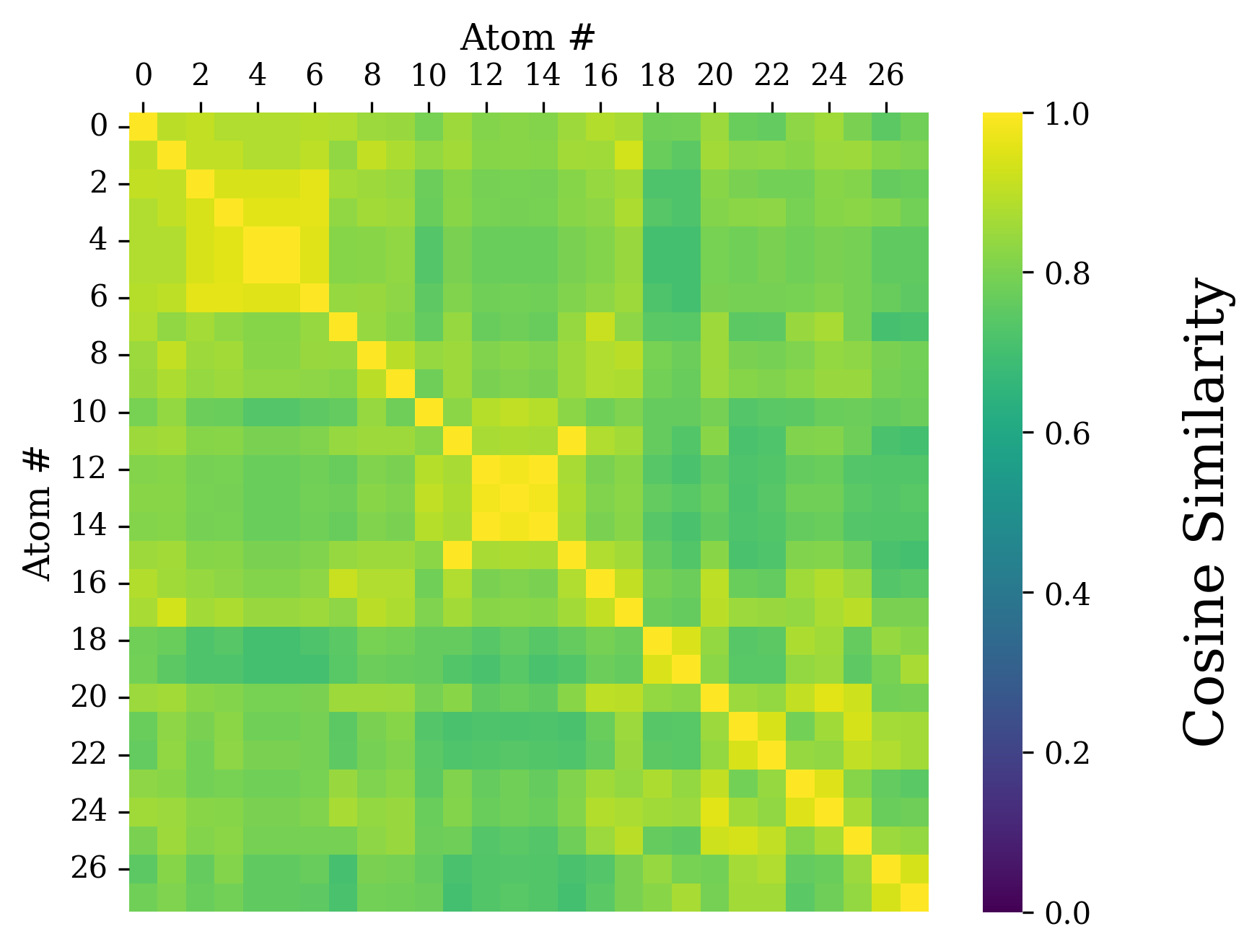}
    \captionsetup{width=5cm}
    \caption{ }
    \label{fig:rdkit_atom_n_fsim_acid_blue_25_free_acid}
\end{subfigure}\\
\begin{subfigure}{.5\textwidth}
    \centering
    \includegraphics[width=5cm]{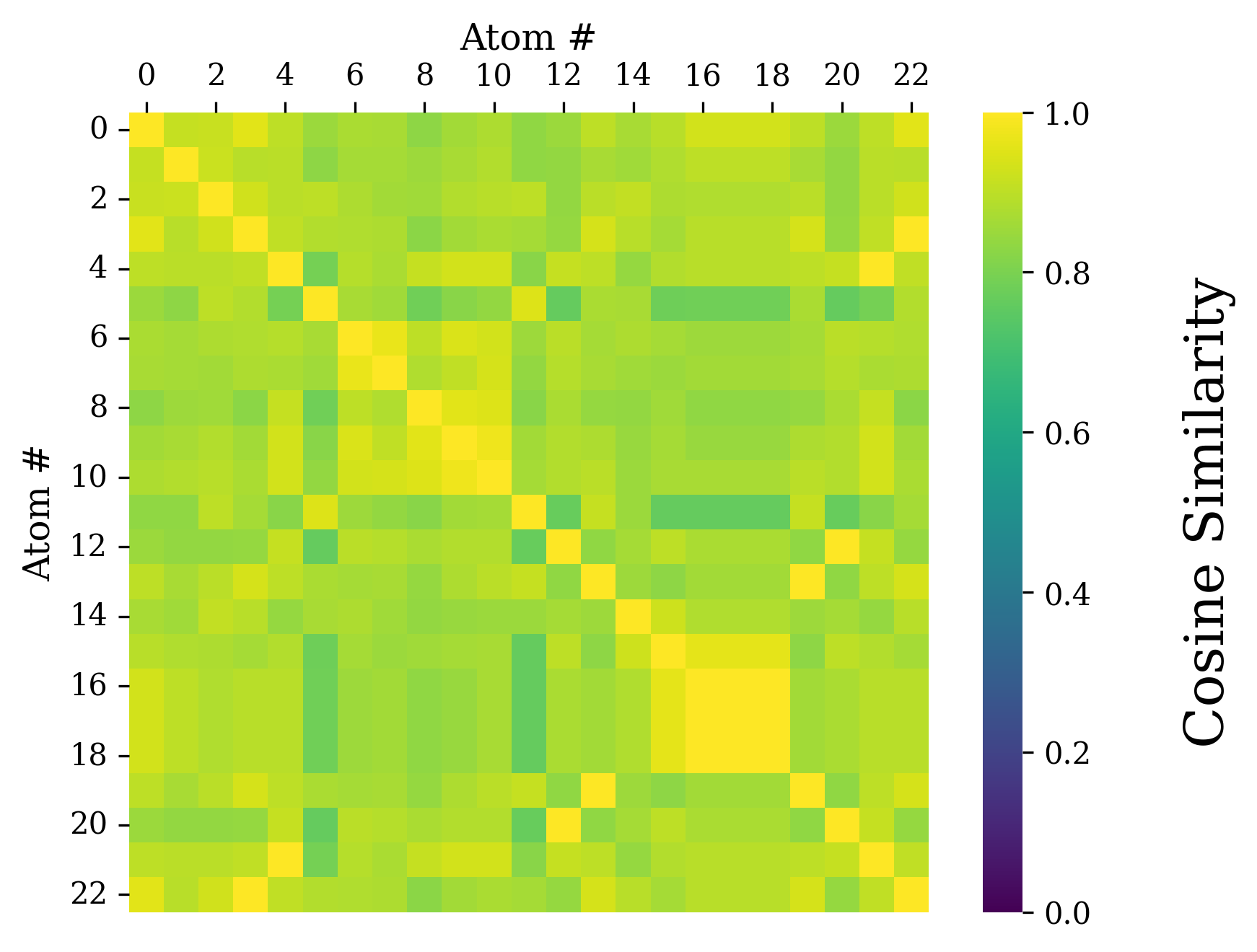}
    \captionsetup{width=5cm}
    \caption{ }
    \label{fig:rdkit_atom_n_fsim_avobenzone}
\end{subfigure}%
\begin{subfigure}{.5\textwidth}
    \centering
    \includegraphics[width=5cm]{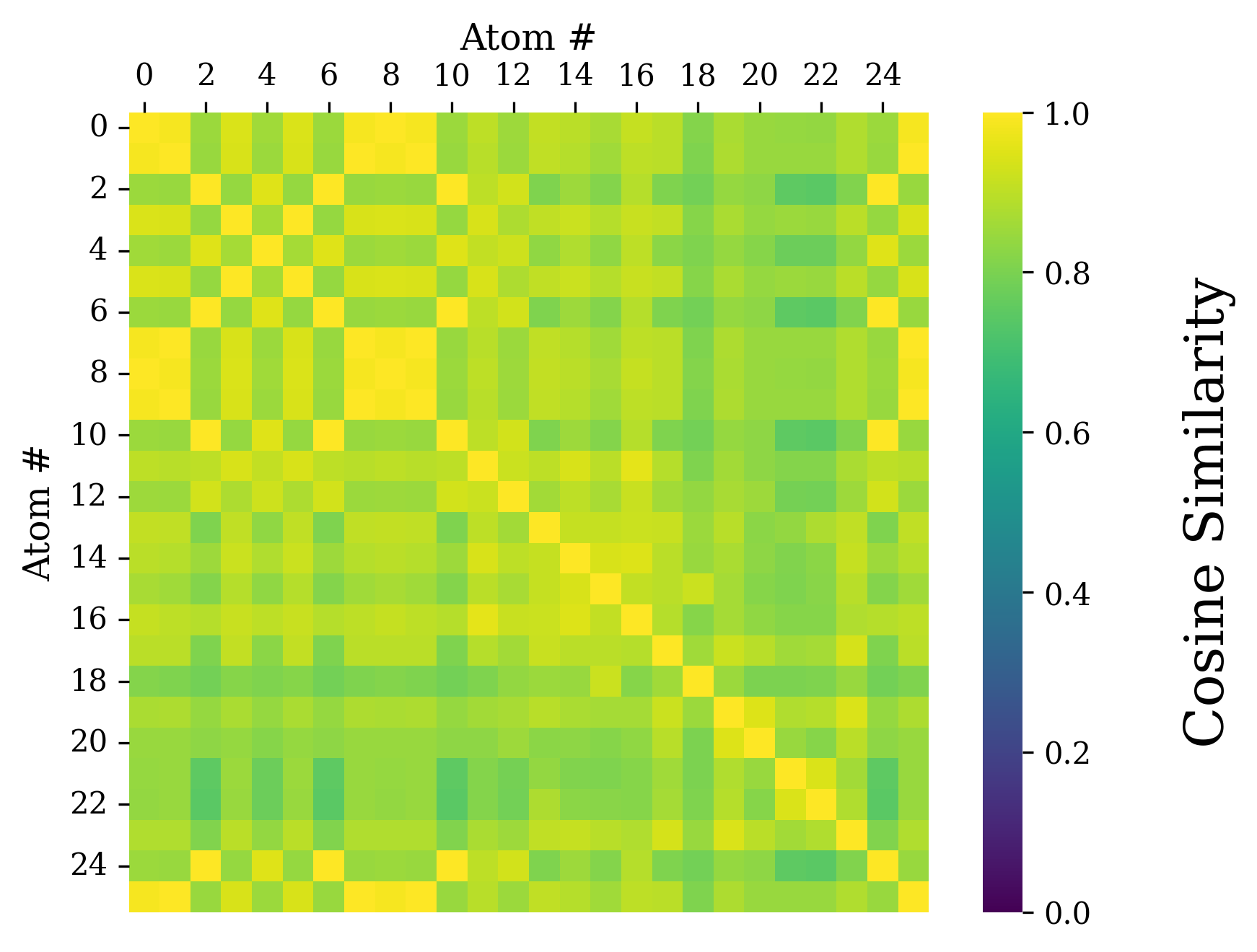}
    \captionsetup{width=5cm}
    \caption{ }
    \label{fig:diphenylaminocarbazole}
\end{subfigure}\\
\caption{\textbf{Feature cosine similarity of each RDKit canonicalized query depending on the chosen root atom number.} \textbf{(a-h).} In order: penicillin G, nirmatrelvir, zidovudine, LSD, fentanyl, acid blue 25 FA, avobenzone, 2-dPAC. The canonicalized variant with the lowest feature similarity to the Atom 0 representation was chosen as the ``RDKit Atom n query''. The root atoms providing most dissimilar feature vectors to the Atom 0 representations were 13 for penicillin G, 21 for nirmatrelvir, 15 for zidovudine, 9 for LSD, 17 for fentanyl, 26 for acid blue 25 FA, 8 for avobenzone, and 18 for 2-dPAC.}
\label{fig:rdkit_atom_n_fsim}
\end{figure}

To test this hypothesis, we built a CLM-embedding-based chemical similarity search utilizing one canonicalization for the CLM and database, and another canonicalization for the query SMILES strings. A pipeline was developed to perform a chemical similarity search using cosine similarity on CLM embeddings obtained from ChemBERTa, an unsupervised transformer encoder-based model \cite{ChemBERTa}. This pipeline was named the Chemical Semantic Search (CheSS), outlined in Figure \ref{fig:workflow}. ChemBERTa was trained on SMILES strings canonicalized using the default RDKit implementation, herein referred to as “RDKit Atom 0” \cite{landrum2013rdkit, ChemBERTa}. We converted ChemBERTa’s training set into a molecular database of $\sim$10M RDKit Atom 0 SMILES strings, upon which CheSS searches were performed. We then explored how the CheSS search results were impacted by three different SMILES canonicalizations. The first was RDKit’s default canonicalization “RDKit Atom 0”. The second canonicalization was created by maximizing the distance of the created string embedding to RDKit Atom 0 by varying the root atom number, herein referred to as “RDKit Atom n” (Fig. \ref{fig:rdkit_atom_n_fsim}). The third used OEChem 2.3.0, a markedly different algorithm, to canonicalize the query and is referred to as “OEChem”. All CheSS searches utilizing these three canonicalizations differed only in the representation of the query molecule, with the database and model remaining constant. 
 
\subsection{Determining Whether Canonicalization Impacts Search Behavior}

\begin{figure}[h]
\centering
\begin{subfigure}{.125\textwidth}
    \centering
    \includegraphics[width=\textwidth]{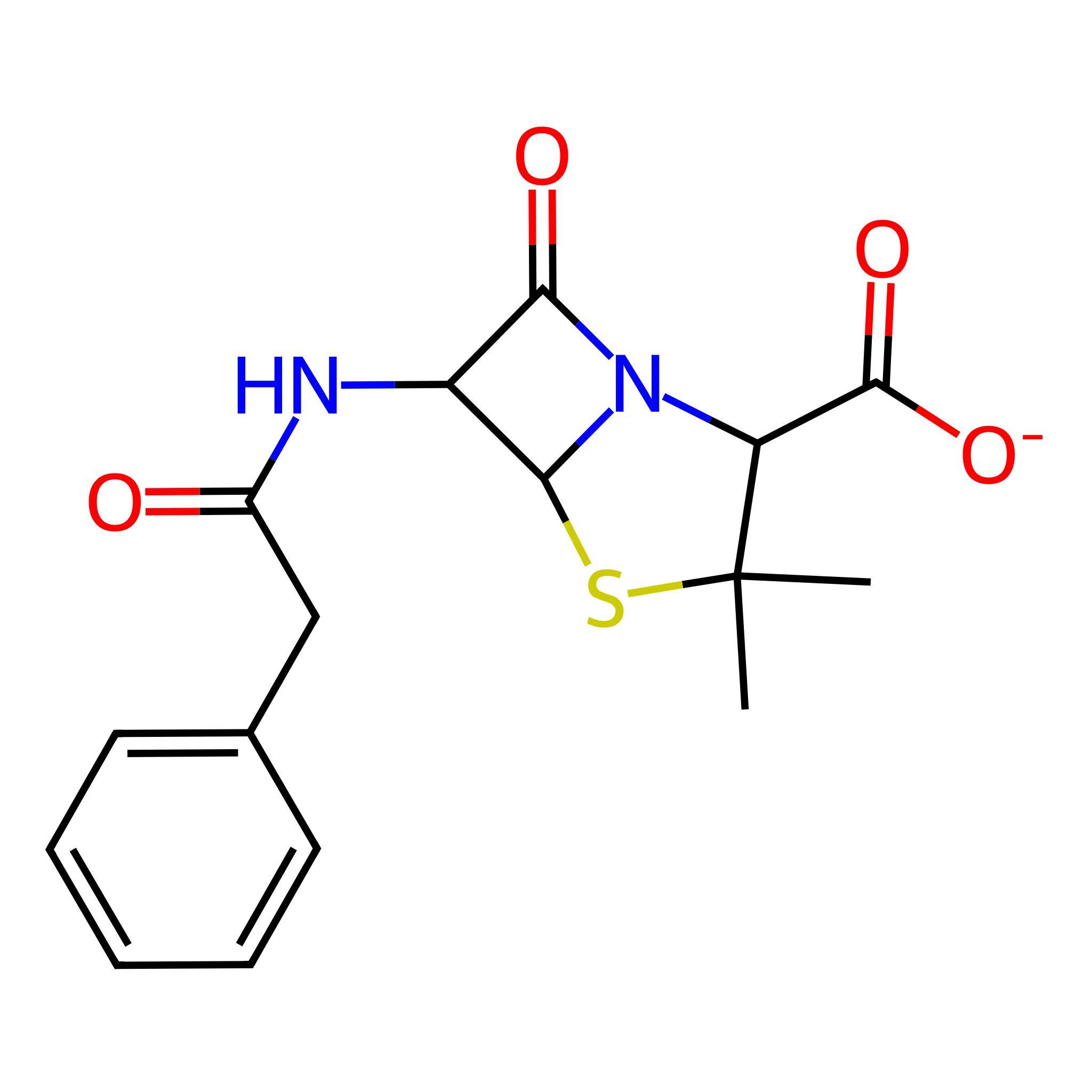}
    \caption{ }
    \label{fig:query_1}
\end{subfigure}%
\begin{subfigure}{.125\textwidth}
    \centering
    \includegraphics[width=\textwidth]{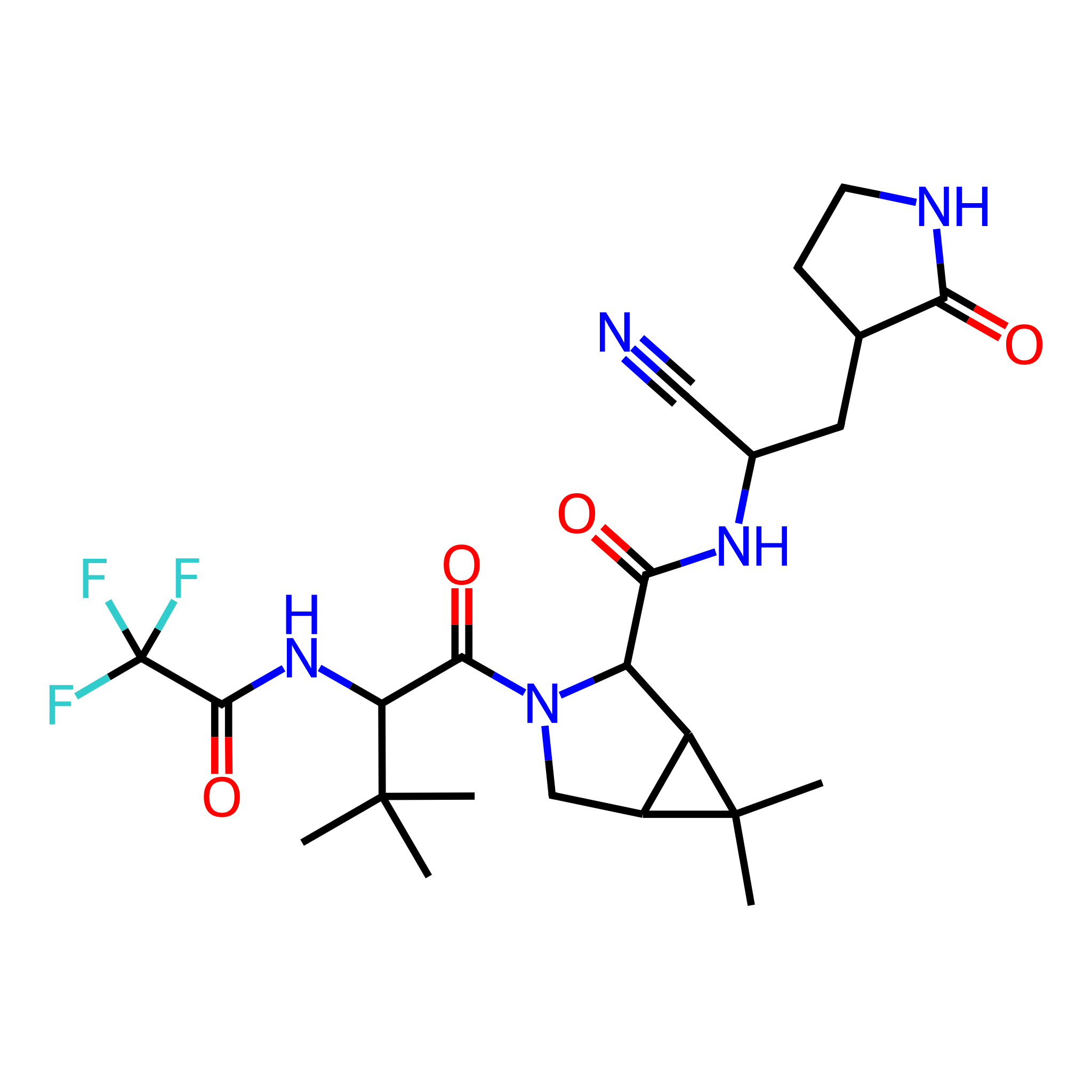}
    \caption{ }
    \label{fig:query_2}
\end{subfigure}%
\begin{subfigure}{.125\textwidth}
    \centering
    \includegraphics[width=\textwidth]{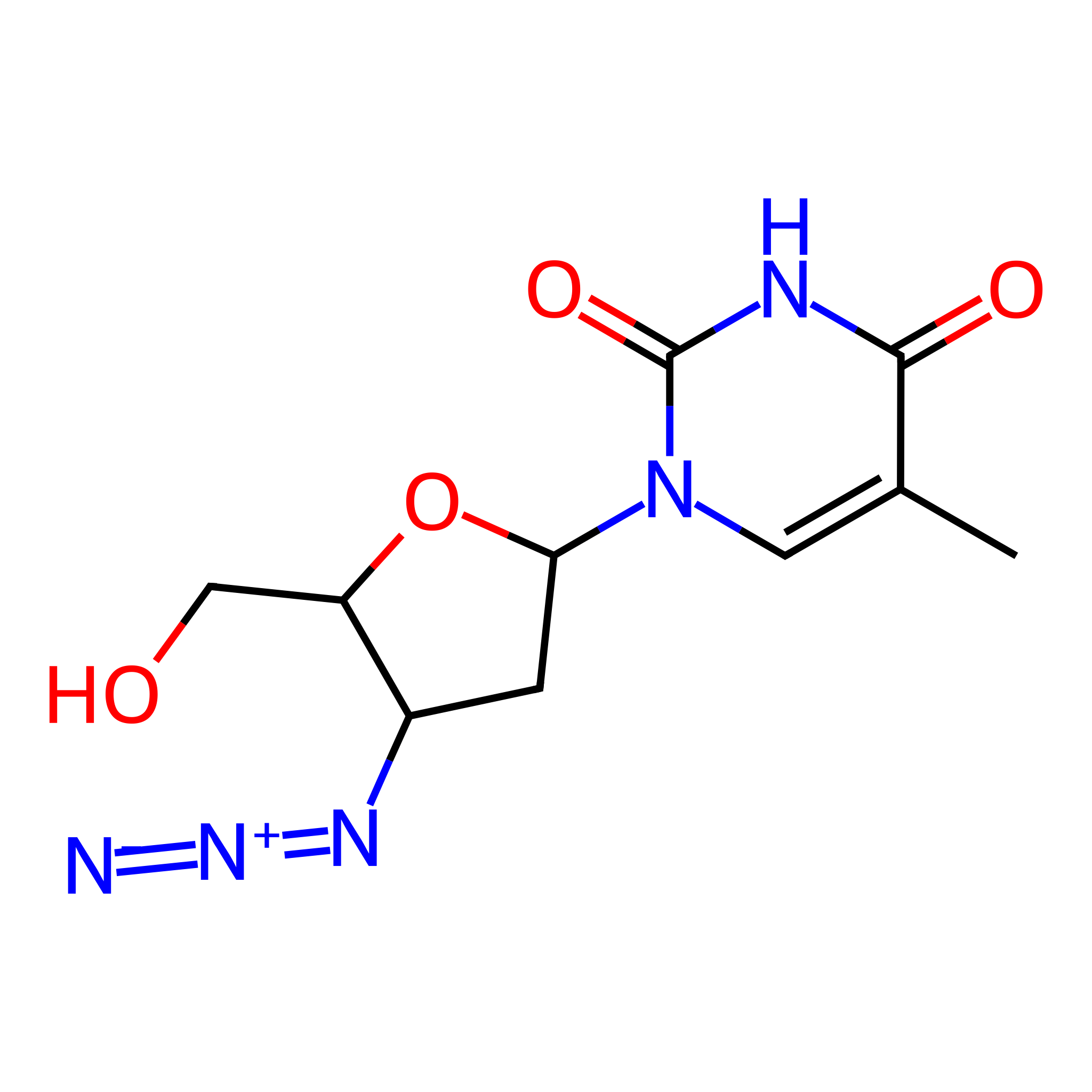}
    \caption{ }
    \label{fig:query_3}
\end{subfigure}%
\begin{subfigure}{.125\textwidth}
    \centering
    \includegraphics[width=\textwidth]{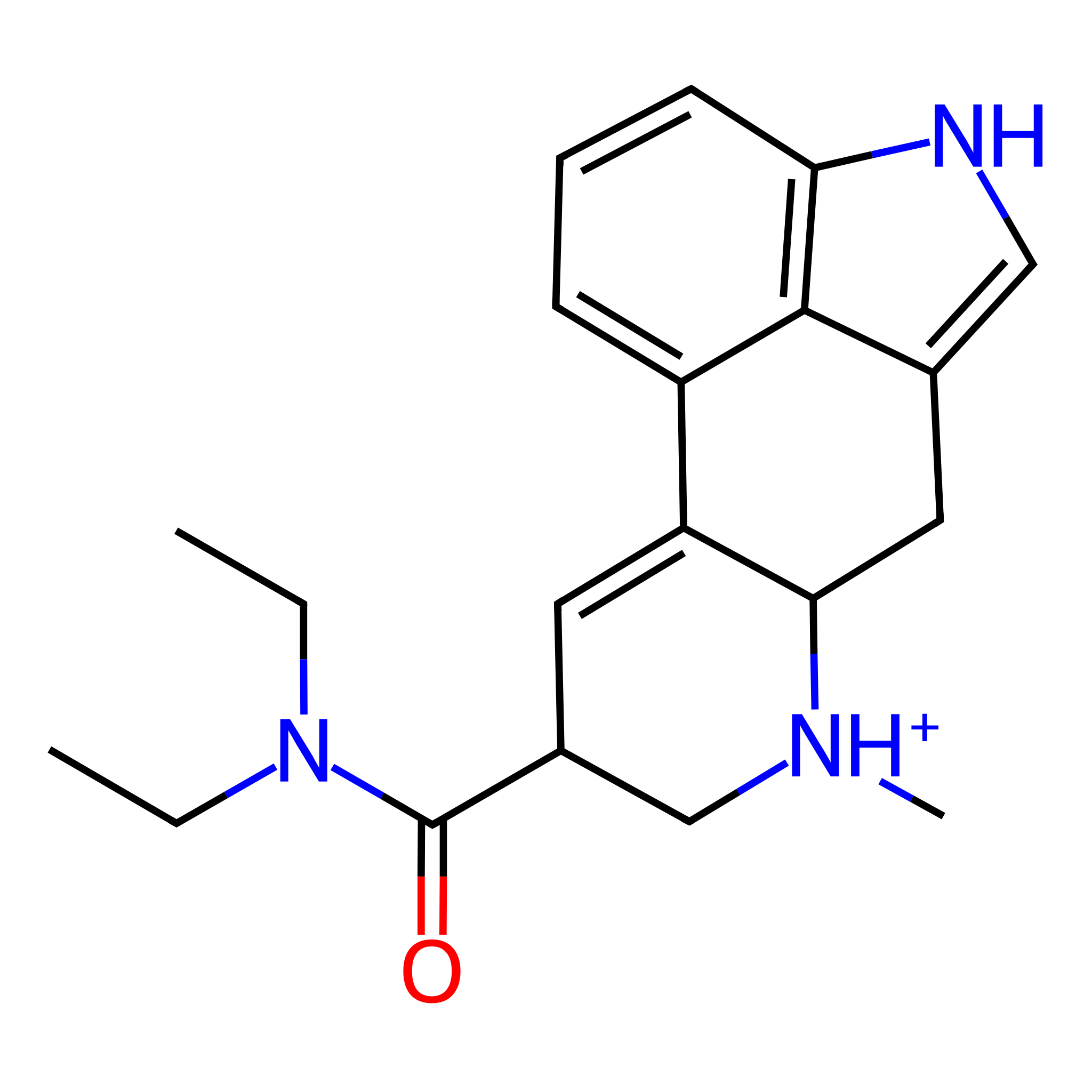}
    \caption{ }
    \label{fig:query_4}
\end{subfigure}%
\begin{subfigure}{.125\textwidth}
    \centering
    \includegraphics[width=\textwidth]{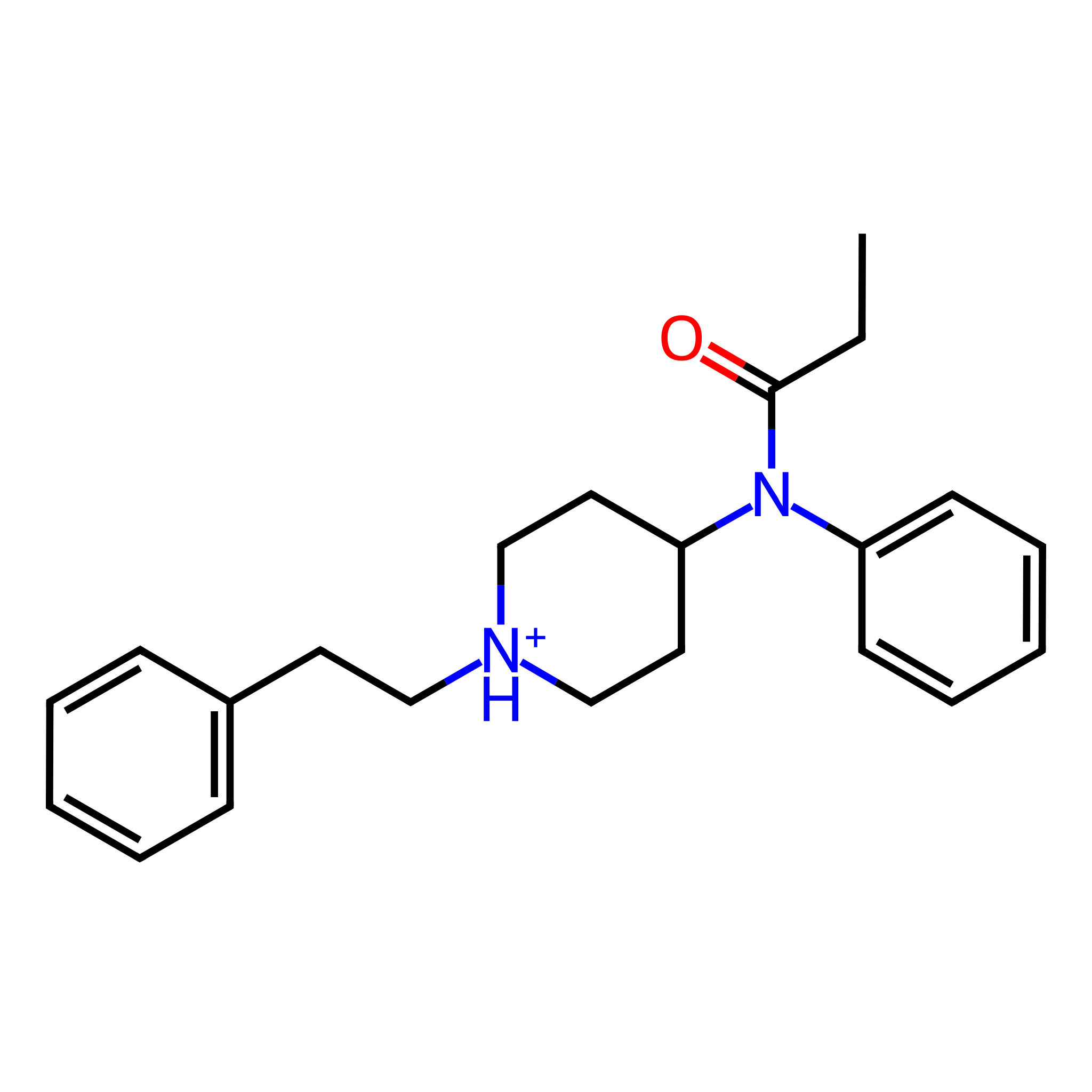}
    \caption{ }
    \label{fig:query_5}
\end{subfigure}%
\begin{subfigure}{.125\textwidth}
    \centering
    \includegraphics[width=\textwidth]{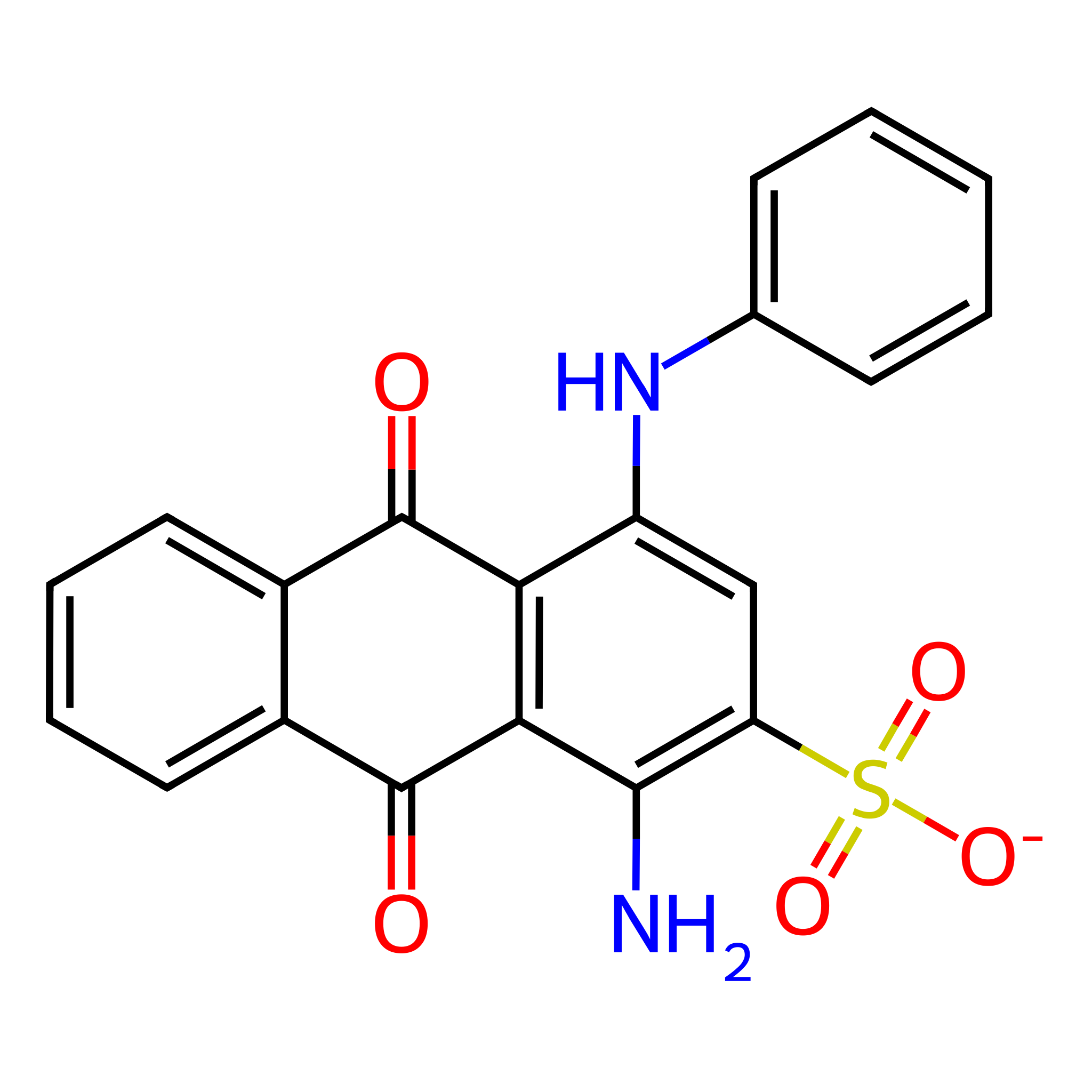}
    \caption{ }
    \label{fig:query_6}
\end{subfigure}%
\begin{subfigure}{.125\textwidth}
    \centering
    \includegraphics[width=\textwidth]{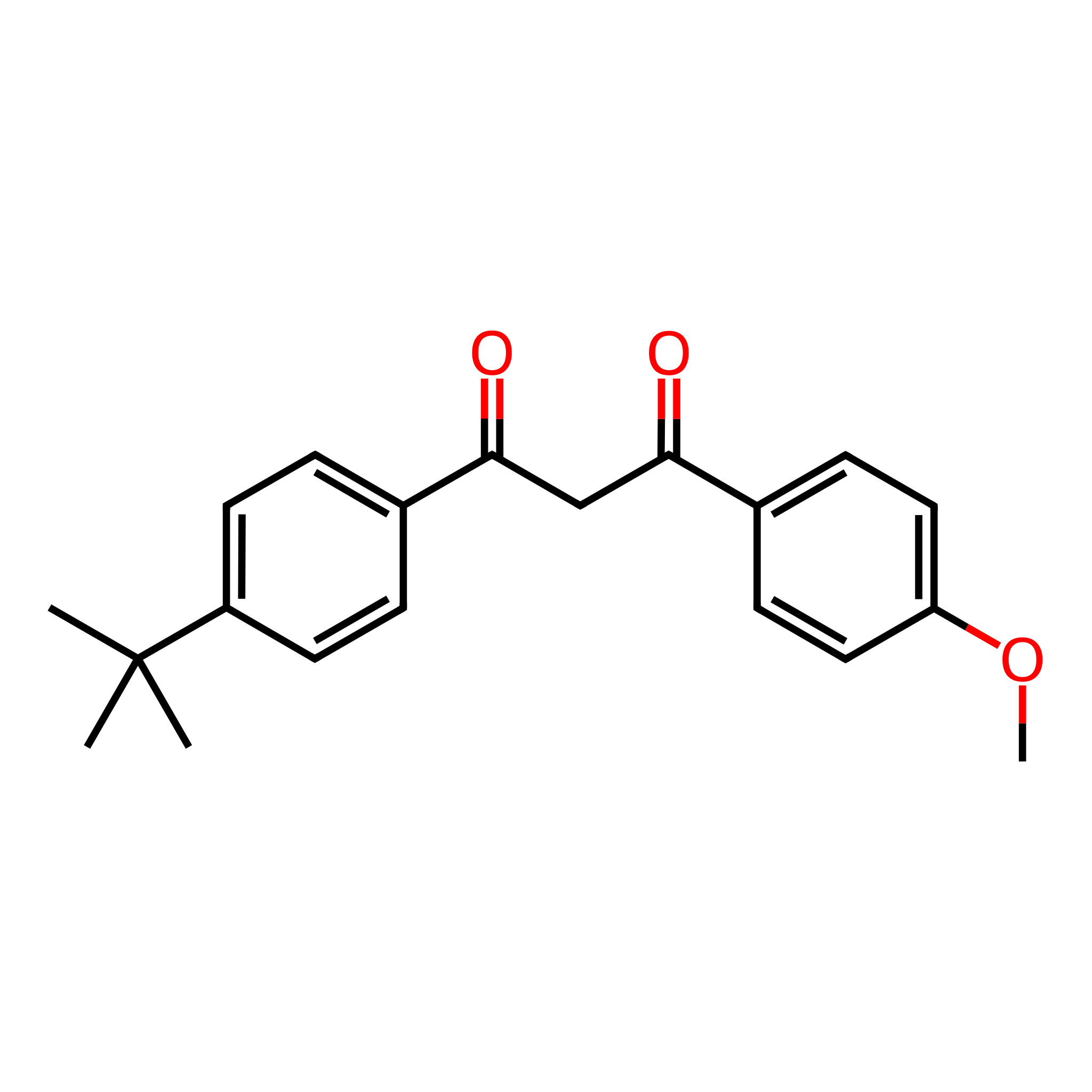}
    \caption{ }
    \label{fig:query_7}
\end{subfigure}%
\begin{subfigure}{.125\textwidth}
    \centering
    \includegraphics[width=\textwidth]{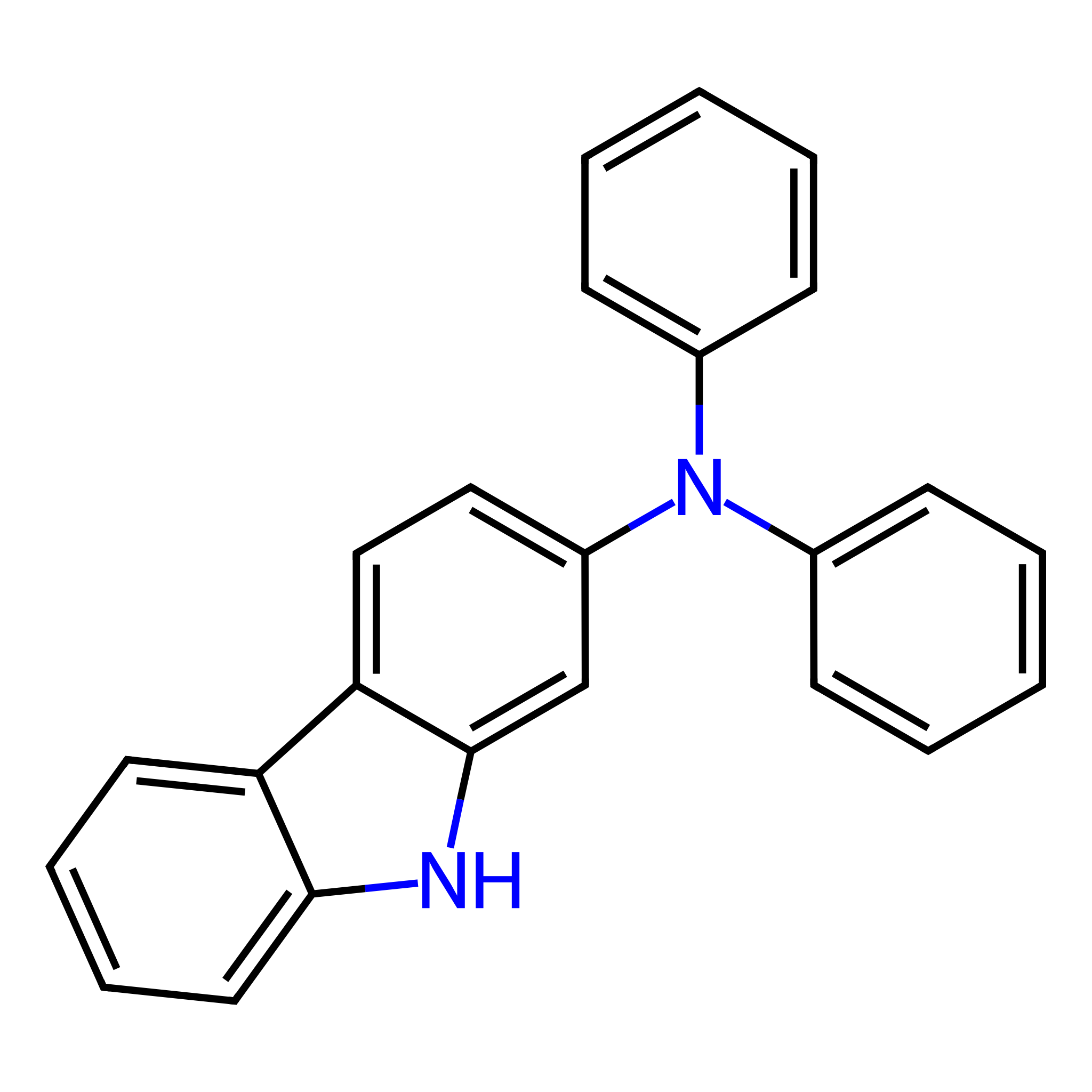}
    \caption{ }
    \label{fig:query_8}
\end{subfigure}\\
\begin{subfigure}{\textwidth}
    \centering
    \begin{tabular}{|c c c|}
    \hline
    \textbf{Query} & \textbf{Canon Alg.} & \textbf{SMILES} \\
    \hline
     & RDKit Atom 0 & CC1(C)SC2C(NC(=O)Cc3ccccc3)C(=O)N2C1C(=O)[O-]\\
    Penicillin G & RDKit Atom n & c1ccc(CC(=O)NC2C(=O)N3C2SC(C)(C)C3C(=O)[O-])cc1\\
     & OEChem & CC1(C(N2C(S1)C(C2=O)NC(=O)CC3=CC=CC=C3)C(=O)[O-])C\\
    \hline
    \end{tabular}
    \caption{ }
    \label{fig:query_peng_canonicalizations}
\end{subfigure}%
\captionsetup{singlelinecheck=false}
\caption{\textbf{Query molecules and canonical SMILES representations.} Query molecules made achiral during canonicalization. \textbf{(a).} Penicillin G; \textbf{(b).} Nirmatrelvir; \textbf{(c).} Zidovudine; \textbf{(d).} LSD; \textbf{(e).} Fentanyl; \textbf{(f).} Acid blue 25 FA; \textbf{(g).} Avobenzone; \textbf{(h).} 2-dPAC. \textbf{(i)}. Penicillin G SMILES strings for the three canonicalizations used herein. Unabridged SMILES for each query are listed in Table \ref{tab:canon_smiles_all_expanded}.}
\label{fig:queries}
\end{figure}

\begin{figure}[h]
\centering
\includegraphics{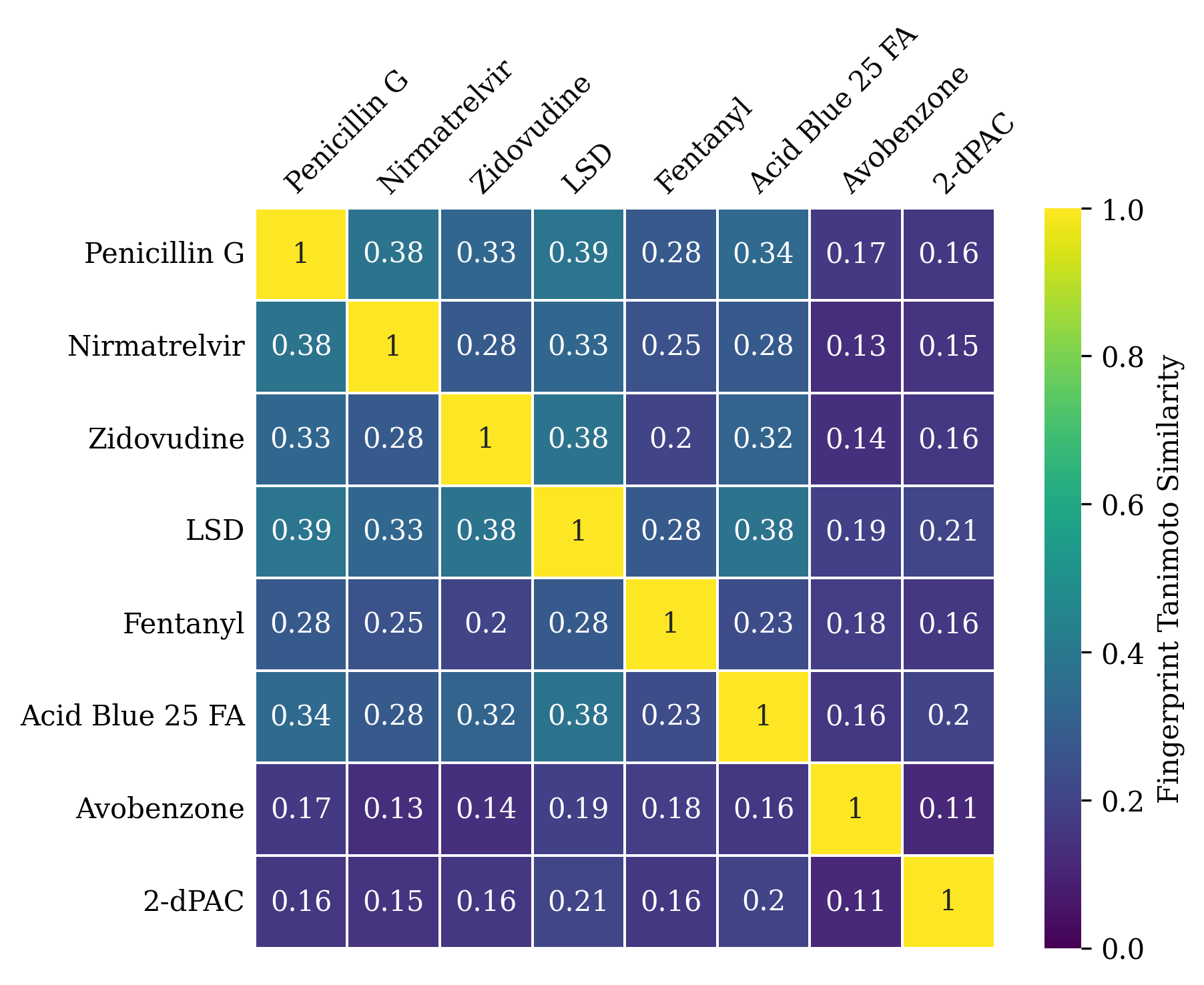}
\caption{\textbf{Fingerprint Tanimoto coefficients between each of the query molecules.} The drug-like molecules, as well as acid blue 25 FA, are more similar to one another than they are to avobenzone \& 2-dPAC. All of the molecules are fairly dissimilar to one another, with the highest similarity being 0.39 between LSD and penicillin G, and the lowest similarity being 0.11 between avobenzone and 2-dPAC.}
\label{fig:query_fprint_matrix}
\end{figure}

A CheSS search using each of three, different query canonicalizations was conducted on eight molecules of known function and roughly equal chemical complexity: penicillin G, nirmatrelvir, zidovudine, lysergic acid diethylamide (LSD), fentanyl, acid blue 25 free acid (acid blue 25 FA), avobenzone, and 2-diphenylaminocarbazole (2-dPAC) (Fig. \ref{fig:queries}). These molecules were chosen as they are of roughly equal complexity but otherwise represent two distinct classes of molecules: drug-like bioactive molecules and non-drug-like photochemical molecules (herein referred to as dye-like). In addition, the molecules were all sufficiently structurally dissimilar from one another, as determined by having a fingerprint Tanimoto coefficient (Tc) less than 0.60 (Fig \ref{fig:query_fprint_matrix}). That said, acid blue 25 FA was more similar to the drug-like molecules, whereas the dyes avobenzone and 2-dPAC were both highly dissimilar from all other query molecules (Fig. \ref{fig:query_fprint_matrix}).

\subsubsection{Statistical Analysis of Query Canonicalizations}

\begin{figure}[ht!]
    \centering
    \includegraphics[width=12cm]{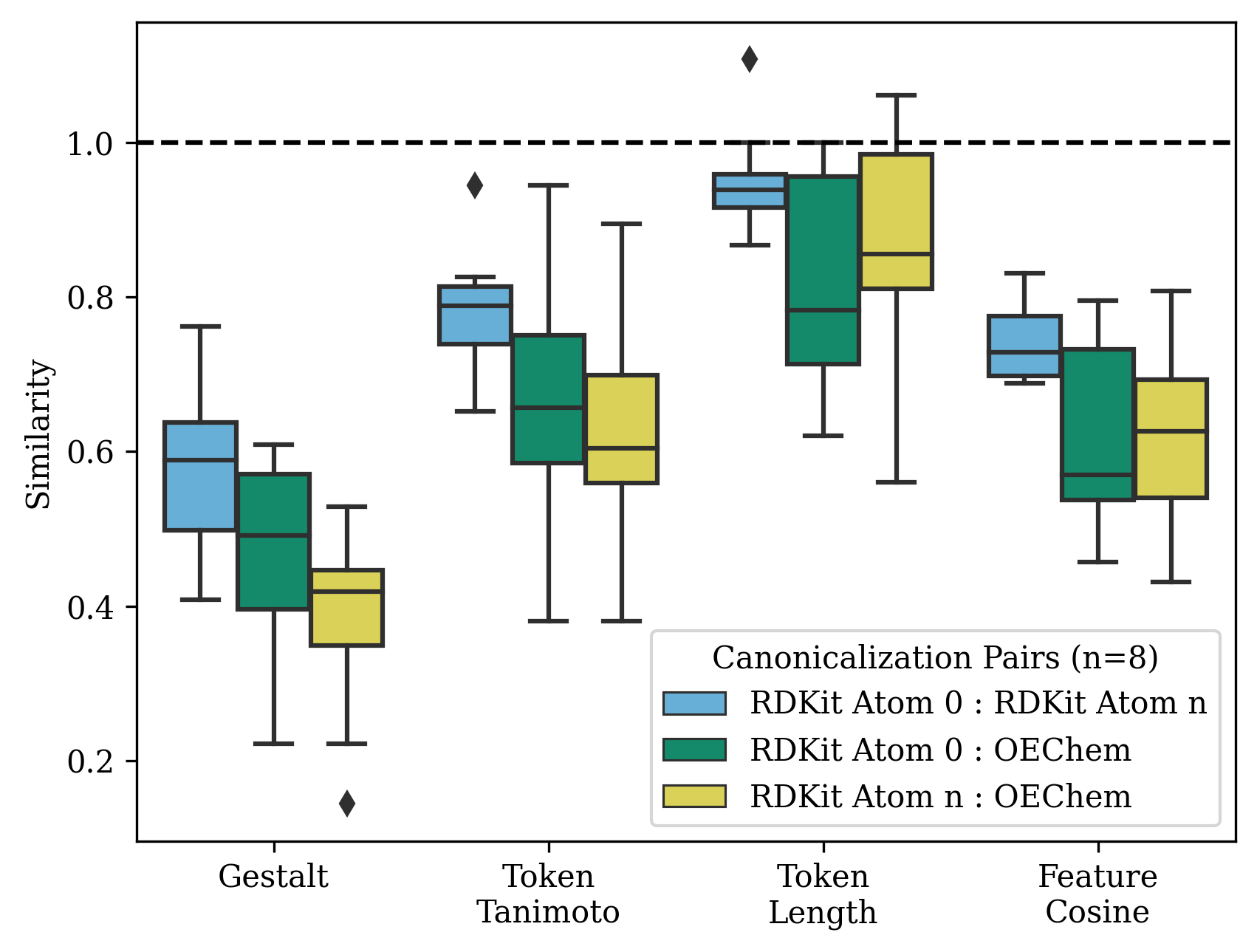}
    \captionsetup{singlelinecheck=false}
    \caption{\textbf{Similarity metrics between the three canonicalized representations for each query molecule.} Gestalt similarity demonstrates different canonicalizations result in markedly different strings. Token Tanimoto \& length ratios indicate these strings were tokenized into different inputs to the CLM. Feature cosine similarity between ChemBERTa-embedded vectors demonstrate that the differently canonicalized queries’ token vectors were interpreted differently by the model resulting in increased spread across feature space. Deviations from 1.0 for each metric represent divergence between canonicalizations.}
    \label{fig:query_similarity_metrics}
\end{figure}

For each query molecule, several similarity metrics were calculated between the three canonicalizations (pairwise comparisons, Figure \ref{fig:query_similarity_metrics}). Gestalt pattern matching, a string similarity metric, showed that each query canonicalized into different strings, with a mean pairwise value of 0.47 across canonicalizations (n=8) (Fig. \ref{fig:query_similarity_metrics}). Because the CLM does not directly receive strings as inputs, but instead receives the tokenized representations (integer-mapped subsections) of strings, the token vectors were analyzed to understand how these strings would be presented to the model. Tanimoto similarity, a metric comparing shared elements between two sets, was applied to the token vectors which indicated that the query strings were converted using markedly different input tokens (mean pairwise value of 0.69 across canonicalizations) (Fig. \ref{fig:query_similarity_metrics}). Similarly, token vector lengths were variable in length, with some queries differing by almost a factor of 2 depending on canonicalization (Fig. \ref{fig:query_similarity_metrics}). Changes in the token vectors cause differences in featurization, or the model’s interpretation of said input, and it was found that different embeddings were obtained depending on canonicalization (mean pairwise feature cosine similarity of 0.66), indicating that the model interpreted different canonicalizations of the same molecule as quite distinct inputs (Fig. \ref{fig:query_similarity_metrics}).

\subsubsection{Distribution of Top Hits}
\begin{figure}[ht!]
\centering
\begin{subfigure}{\textwidth}
    \centering
    \includegraphics[width=\textwidth]{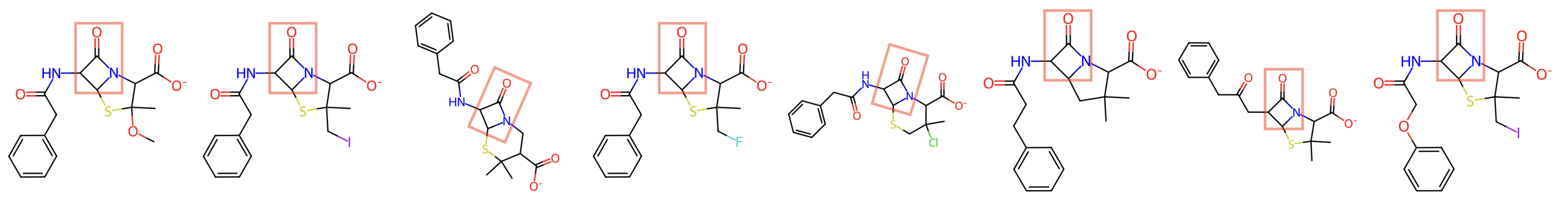}\\
    \captionsetup{width=8cm}
    \caption{ }
    \label{fig:rdkit_0_pen_g_top_5}
\end{subfigure}\\
\begin{subfigure}{\textwidth}
    \centering
    \includegraphics[width=\textwidth]{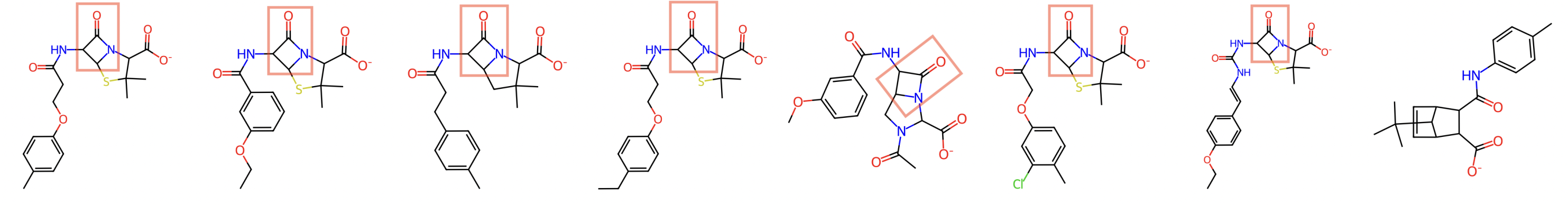}\\
    \caption{ }
    \label{fig:rdkit_n_pen_g_top_5}
\end{subfigure}\\
\begin{subfigure}{\textwidth}
    \centering
    \includegraphics[width=\textwidth]{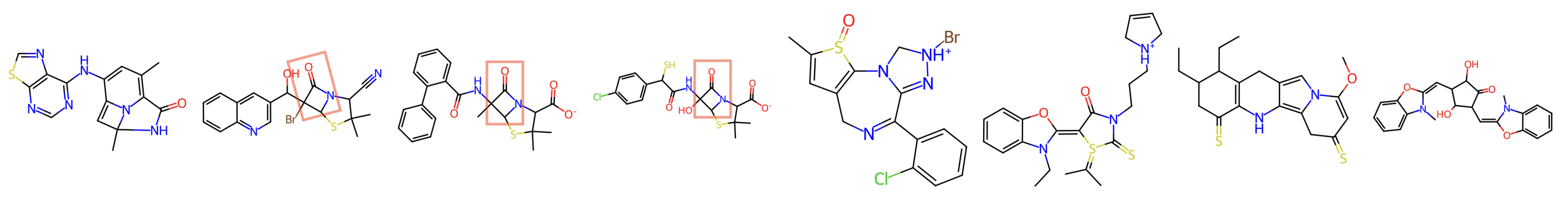}\\
    \caption{ }
    \label{fig:oechem_pen_g_top_5}
\end{subfigure}\\
\begin{subfigure}{0.5\textwidth}
    \centering
    \includegraphics[width=\textwidth]{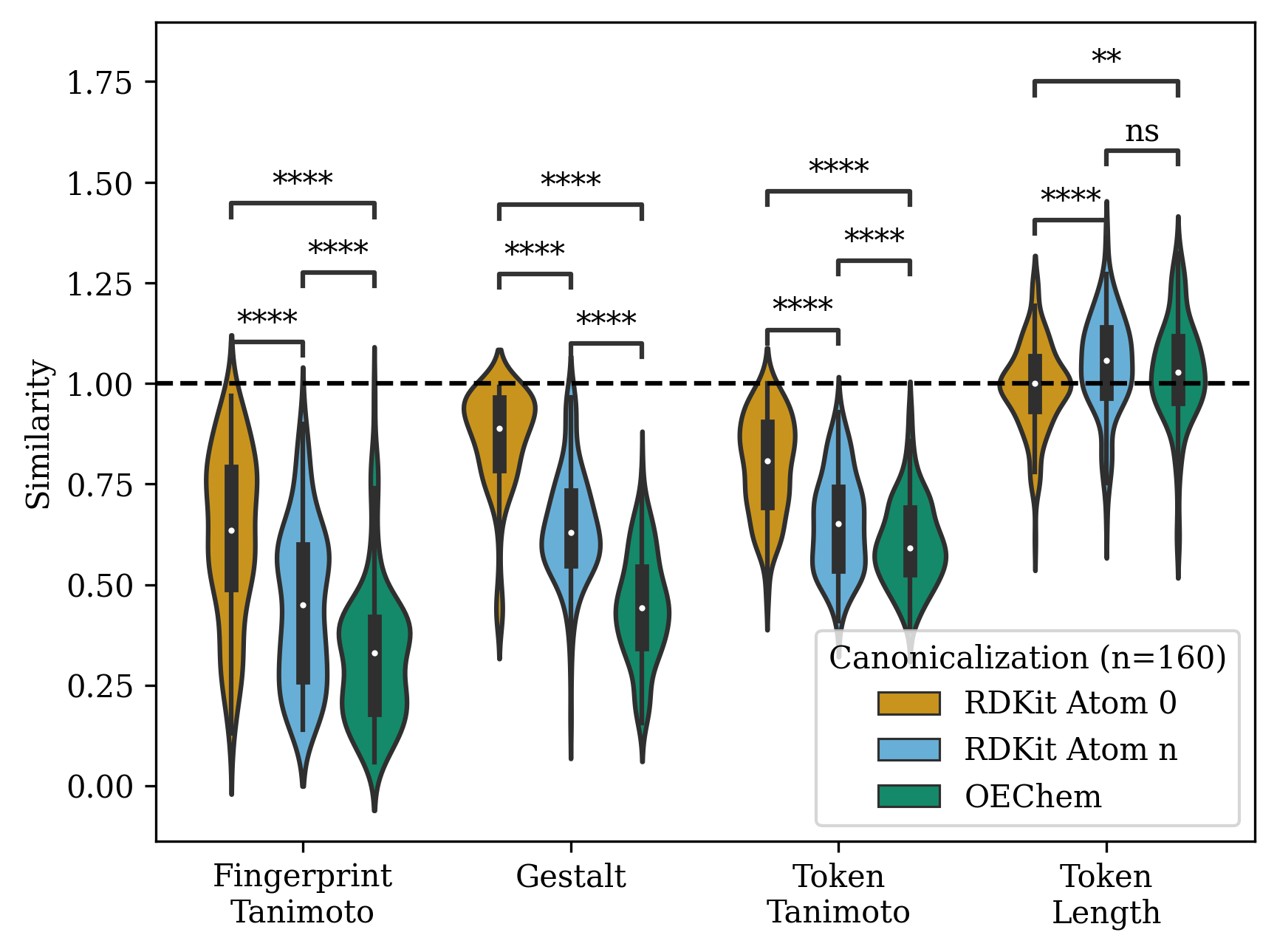}
    \caption{ }
    \label{fig:results_violin_all}
\end{subfigure}%
\begin{subfigure}{0.25\textwidth}
    \centering
    \includegraphics[width=\textwidth]{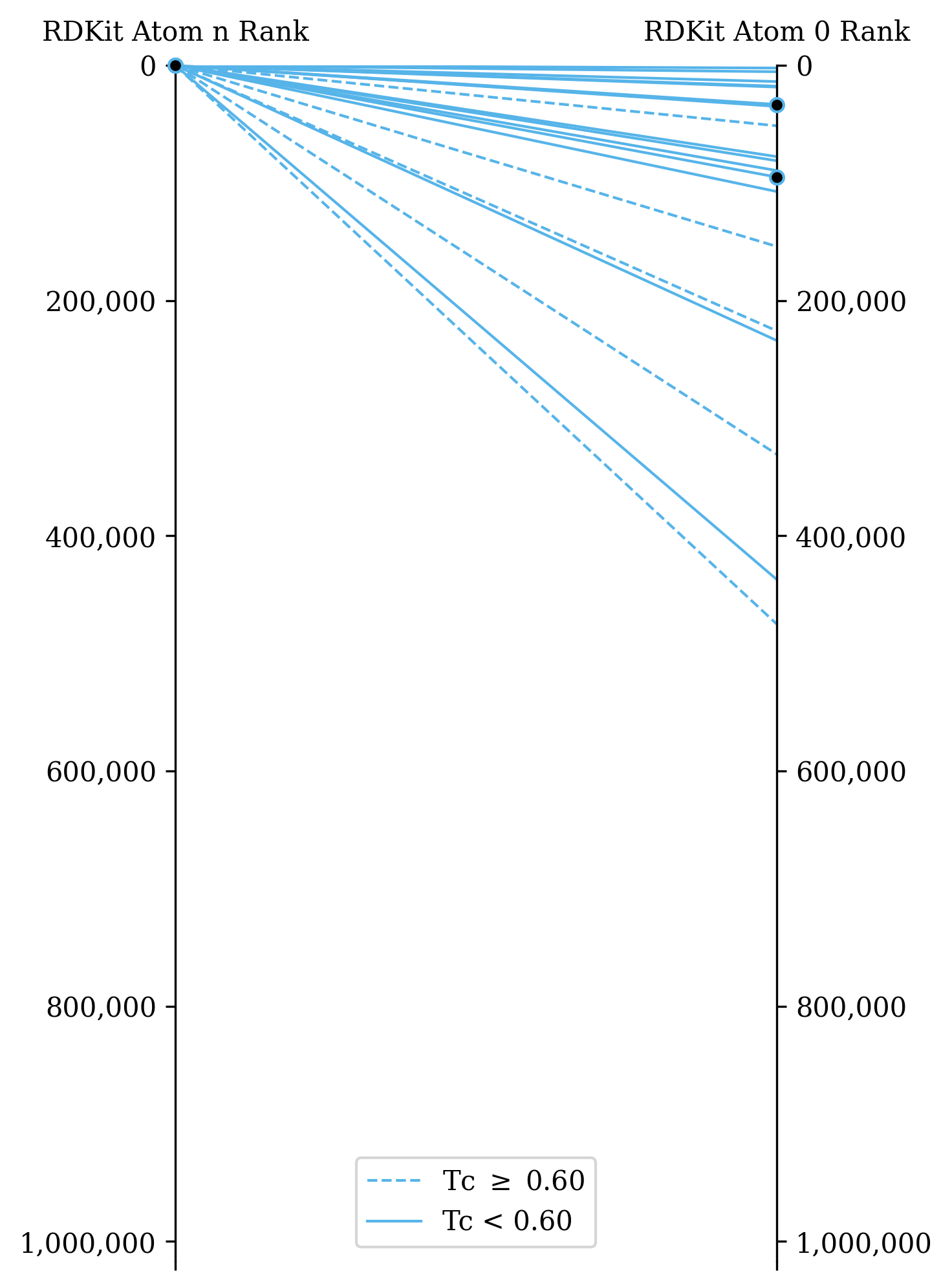}
    \caption{ }
    \label{fig:peng_slope_rdkit_n}
\end{subfigure}%
\begin{subfigure}{0.25\textwidth}
    \centering
    \includegraphics[width=\textwidth]{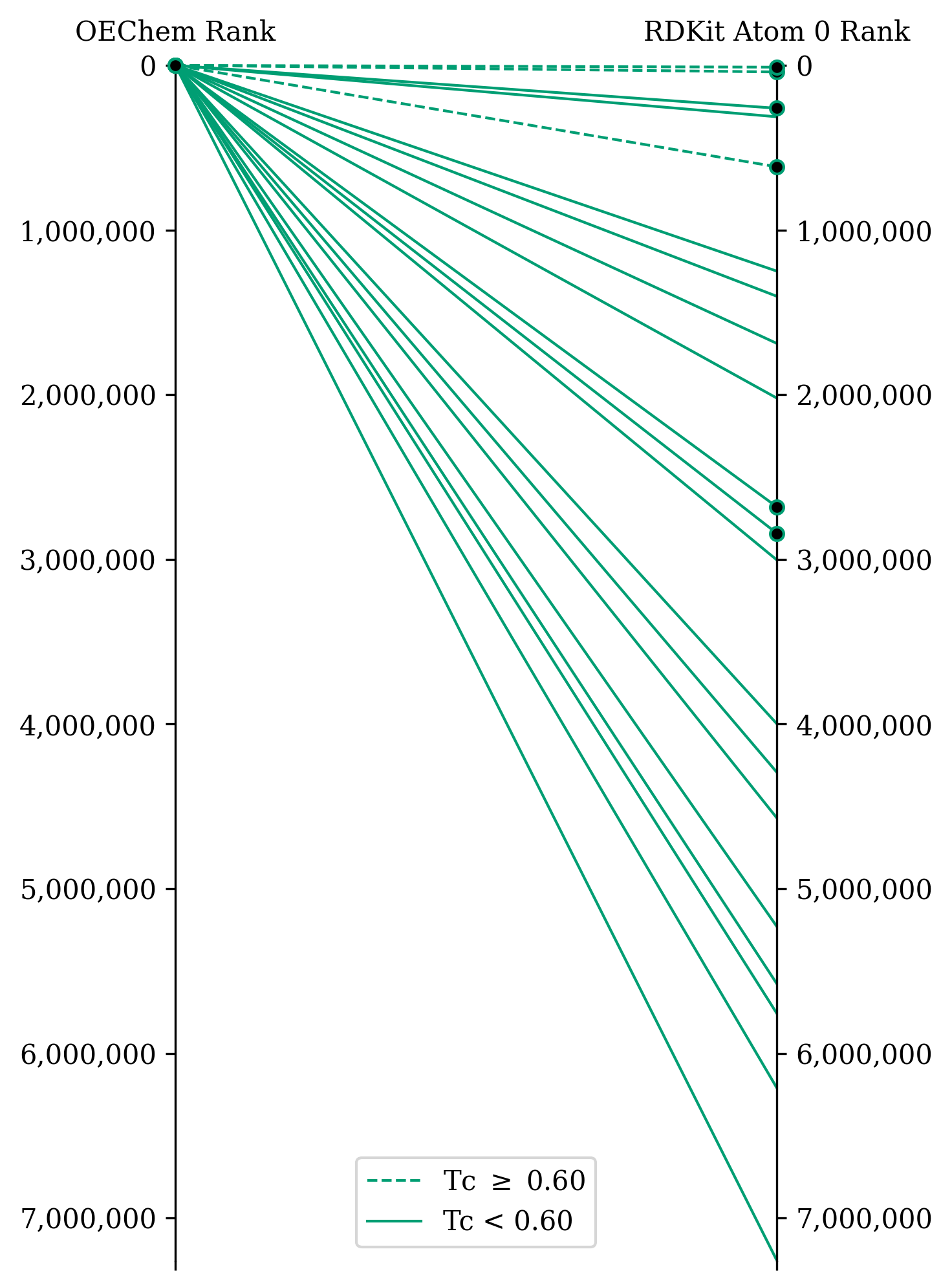}
    \caption{ }
    \label{fig:peng_slope_oechem}
\end{subfigure}
\captionsetup{singlelinecheck=false}
\caption{\textbf{Search behavior depends on canonicalization.} \textbf{(a-c).} Different canonicalizations return structurally distinct molecules, demonstrated by $\beta$-lactam ring-containing molecules in the top search results. \textbf{(a).} RDKit Atom 0 had 8/8 top results containing $\beta$-lactam rings. \textbf{(b).} RDKit Atom n had 7/8 top results containing $\beta$-lactam rings. \textbf{(c).} OEChem had 3/8 top results containing $\beta$-lactam rings. \textbf{(d).} Similarity metrics for all CheSS searches between each canonicalized query and respective top 20 results (n=160 for each canonicalization). Asterisks indicate the level of statistical significance for two-sided independent t-tests (ns, P$<$1.0; *, P$<$0.05; **, P$<$0.01; ***, P$<$0.001; **** P$<$0.0001). \textbf{(e-f).} The index rank of each alternate canonicalization's top 20 results for penicillin G compared to the index rank that these same molecules scored in the other canonicalizations' searches. Molecules functionally similar to the query indicated by a black dot, as determined by the patent search, and structurally similar to the query (Tc $\geq$ 0.60) indicated by a dashed line. Rank plots for each query and comparisons between RDKit Atom n and OEChem are listed in Fig. \ref{fig:slope_rank_1}. Queries with alternative canonicalizations were able to find molecules that would not have been found when the same canonicalization as the database was used, which were often functionally similar to the query.}
\label{fig:search_behavior_by_canon}
\end{figure}

In order to explore how different canonicalizations impact feature-based search behavior, similarity metrics were obtained comparing each canonicalized query to its respective top 20 CheSS search results. Queries canonicalized with RDKit Atom 0 yielded compounds high in structural similarity, as evidenced by fingerprint Tanimoto similarity, a measure of molecular substructure similarity, with a mean coefficient of 0.62 (n=160) (Fig. \ref{fig:results_violin_all}). In contrast, the mean fingerprint Tanimoto coefficients for RDKit Atom n and OEChem were 0.45 and 0.32 respectively. Another way to see the differences in these searches was that for RDKit Atom 0 canonicalized queries, 22\% could have been found from a fingerprint Tanimoto search using a cutoff as high as 0.80, indicating that nearly a quarter of the results were 1-2 atomic changes aways from the query molecule (Fig. \ref{fig:results_violin_all}). In contrast, only 6\% and 2\% of the top results for RDKit Atom n and OEChem, respectively, could have been found from this same search, indicating significant structural divergence (Fig. \ref{fig:results_violin_all}). 

At a more granular level, these structural differences are well-illustrated for a penicillin G query, in that there is a gradual diminishing of $\beta$-lactam-containing results as canonicalization diverges (Figs. \ref{fig:rdkit_0_pen_g_top_5}, \ref{fig:rdkit_n_pen_g_top_5}, \ref{fig:oechem_pen_g_top_5}): all of the top 8 hits for the RDKit Atom 0 canonicalizations contained $\beta$-lactams, while progressively fewer lactams were found for RDKit Atom n and OEChem. This trend in diverging structure was partially explained by Gestalt pattern matching similarity, in which the mean scores for RDKit Atom 0, RDKit Atom n, and OEChem were 0.86, 0.65, and 0.44 respectively, indicating that the average RDKit Atom 0 top result was a simple string permutation away from the original query, and thus also a simple structural modification away, but this was not the case for the alternate canonicalizations (Fig. \ref{fig:results_violin_all}). The token vector Tanimoto similarity demonstrated that RDKit Atom 0 canonicalization returned molecules with a high number of shared tokens to the query (mean of 0.80), whereas this was not the case with RDKit Atom n and OEChem (means 0.65, 0.60), indicating that the model’s ability to memorize tokens to determine feature similarity was reduced by different canonicalizations (Fig. \ref{fig:results_violin_all}). These results point to the possibility that the model utilizes non-obvious relationships to determine the alternative canonicalization’s location in feature space.

Interestingly, it was observed that the token vector length ratios for all canonicalizations’ results fell within about 20\% of each query’s token vector length, indicating that the model heavily utilized token vector length to determine feature space location and thus similarity (Fig. \ref{fig:results_violin_all}). This means that token vector length may constrain CLM-based similarity searches to confined regions of chemical space, with alternative canonicalizations acting as ways to bypass this predominant search criteria and thereby explore more distant regions of chemical space through variations in token vector length, ultimately allowing for more comprehensive and far-reaching similarity searches (Figs. \ref{fig:peng_slope_rdkit_n}, \ref{fig:peng_slope_oechem}).

\subsubsection{Patent Search Reveals Functionality of Molecules}

\begin{figure}[ht!]
\centering
\begin{subfigure}{.5\textwidth}
    \centering
    \includegraphics[width=\textwidth]{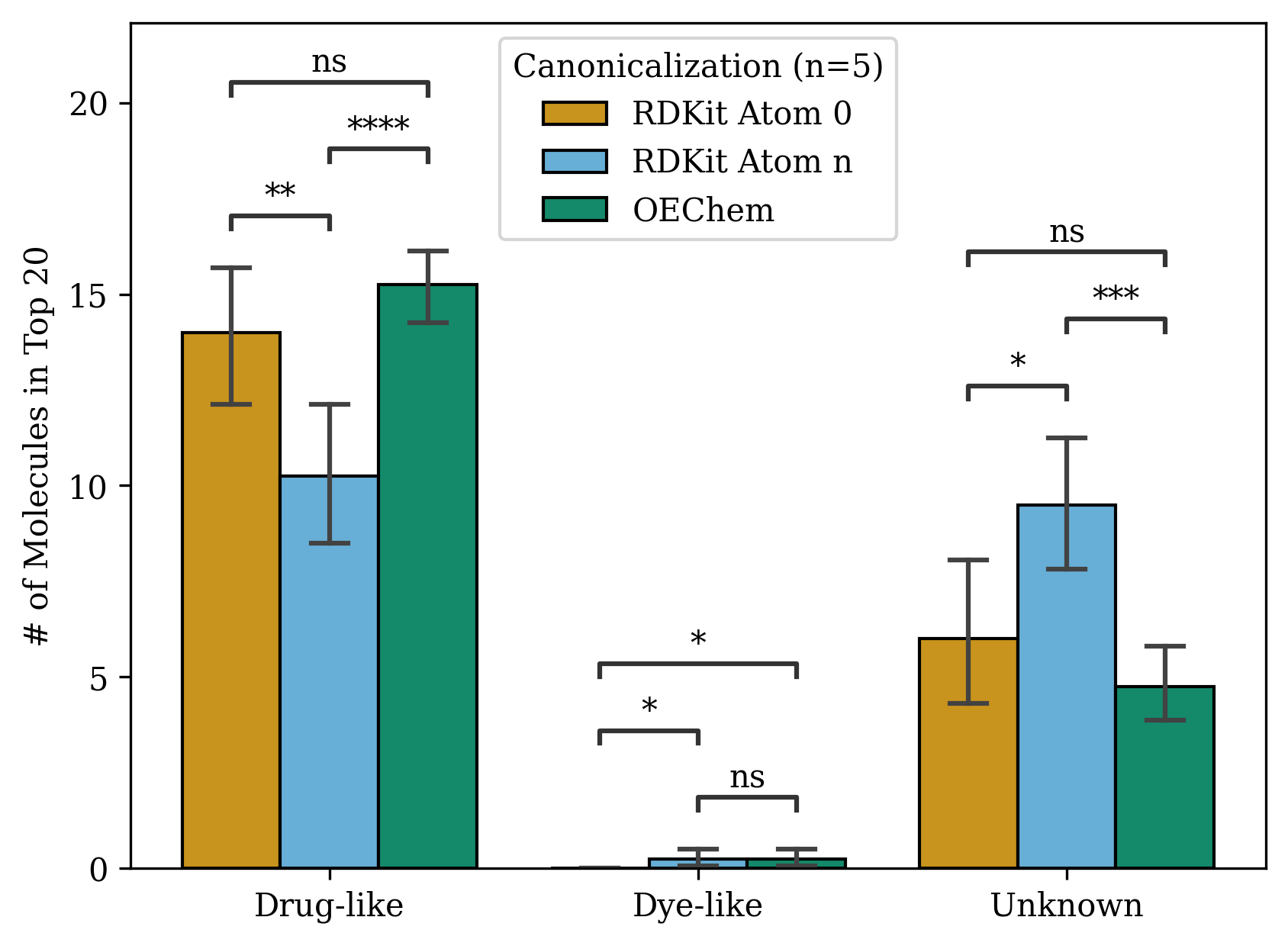}
    \caption{ }
    \label{fig:drug-like}
\end{subfigure}%
\begin{subfigure}{.5\textwidth}
    \centering
    \includegraphics[width=\textwidth]{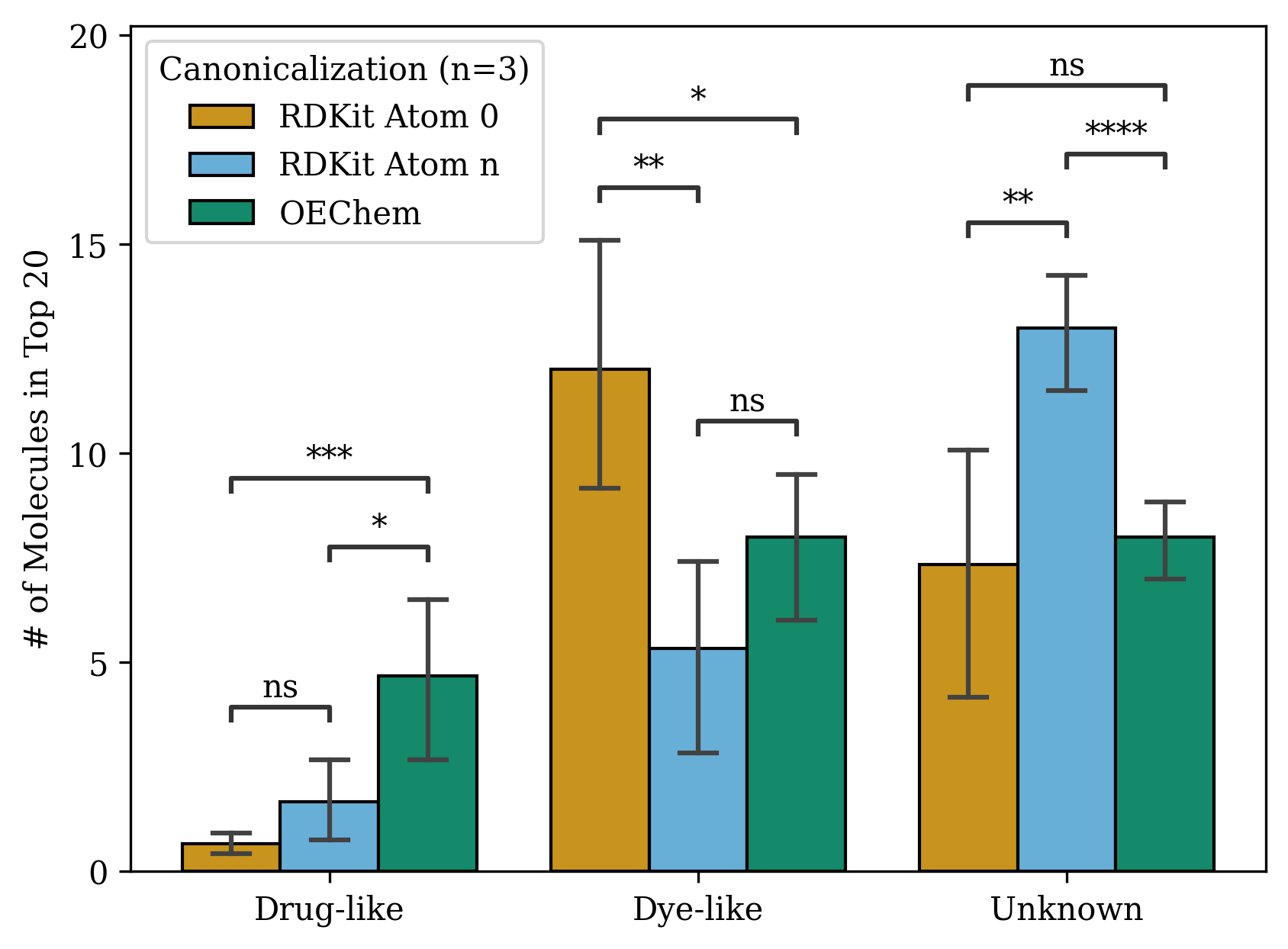}
    \caption{ }
    \label{fig:dye-like}
\end{subfigure}\\
\begin{subfigure}{.5\textwidth}
    \centering
    \includegraphics[width=\textwidth]{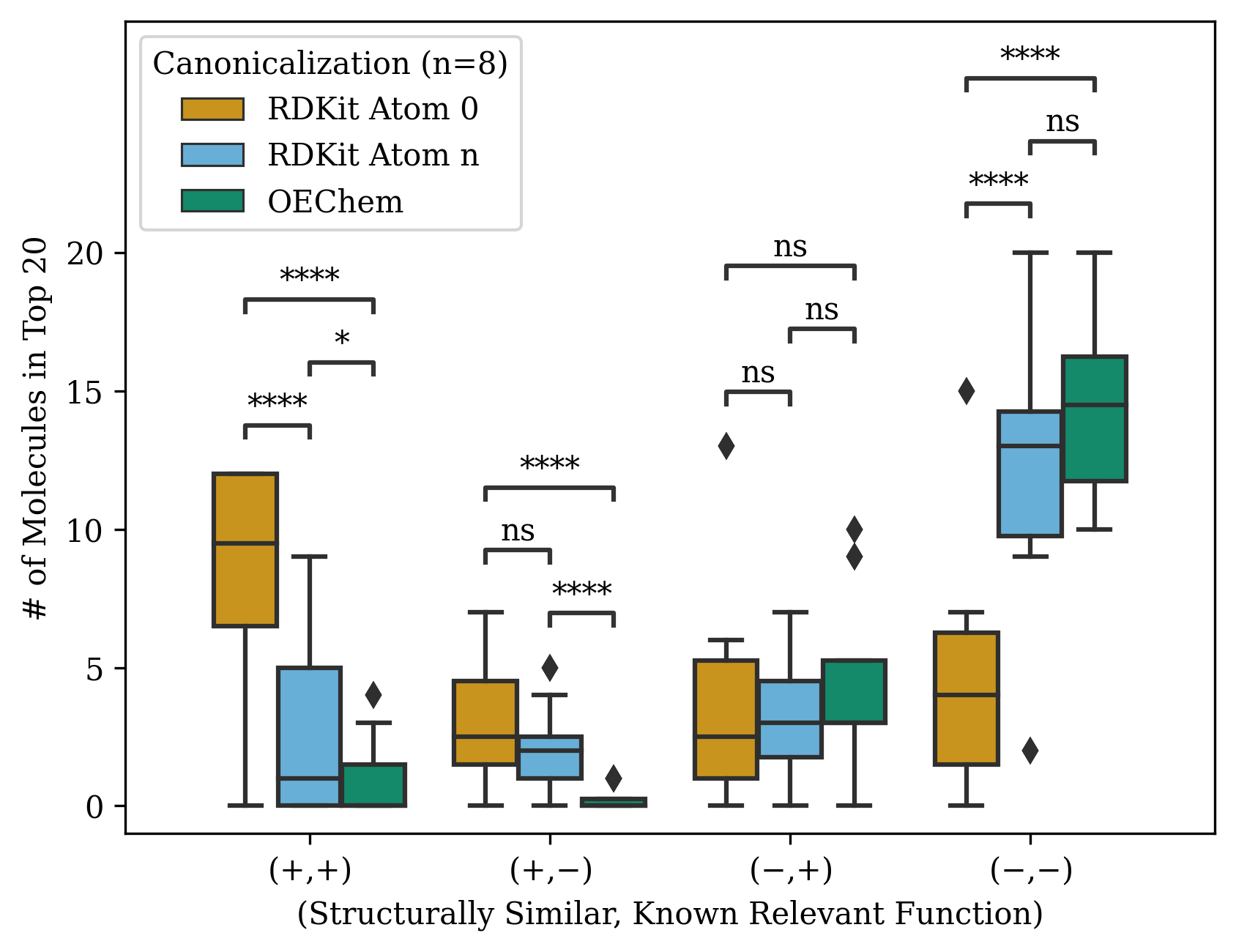}
    \caption{ }
    \label{fig:all-str-fun}
\end{subfigure}%
\begin{subfigure}{.5\textwidth}
    \centering
    \includegraphics[width=\textwidth]{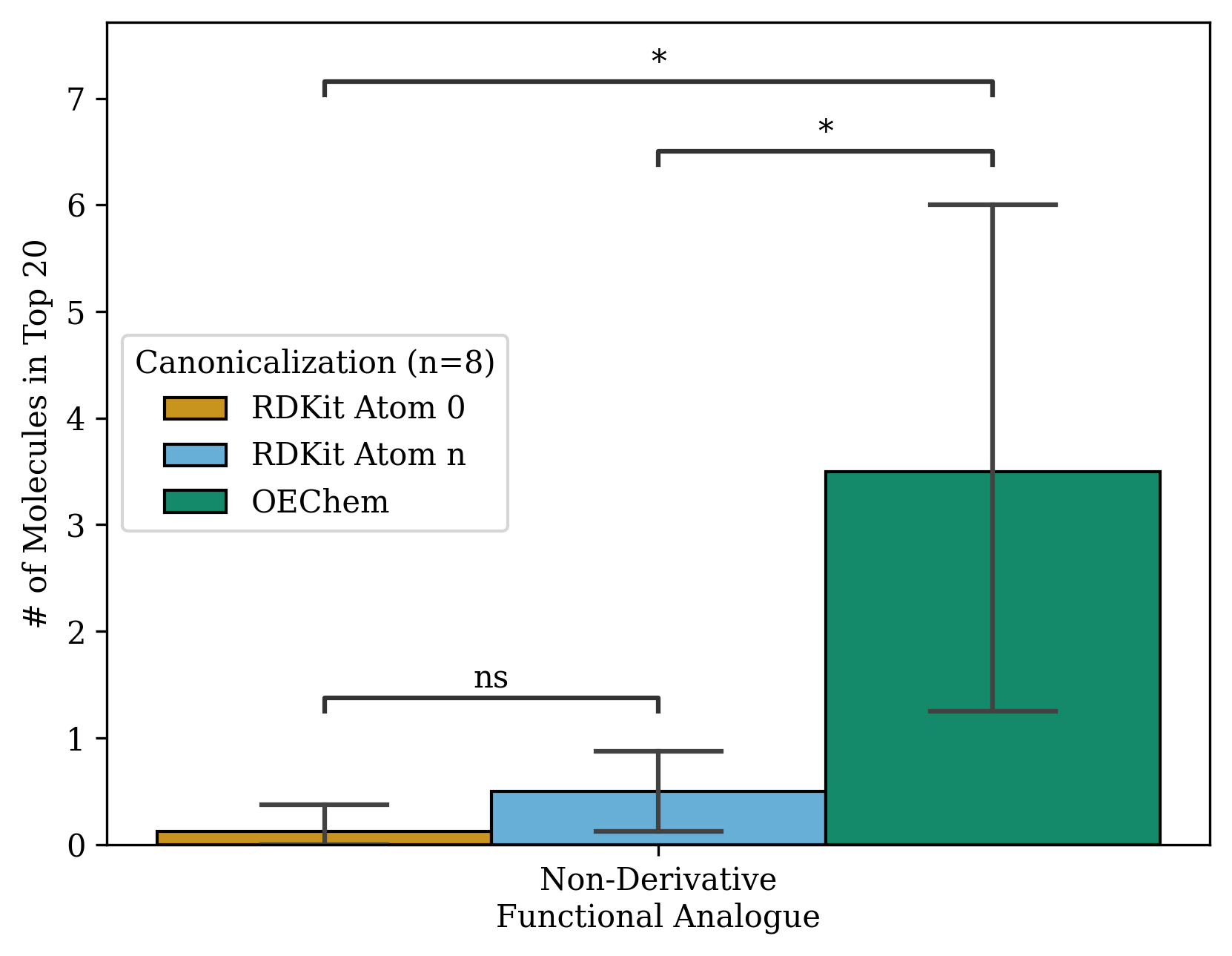}
    \caption{ }
    \label{fig:ndfa}
\end{subfigure}\\
\captionsetup{singlelinecheck=false}
\caption{\textbf{Patent-derived functional analysis for each canonicalization’s results.} \textbf{(a)}. Mean drug-like \& dye-like molecules returned in the top 20 results from drug-like queries (95\% CI, n=8). \textbf{(b)}. Mean drug-like \& dye-like molecules returned in the top 20 results from drug-like queries (95\% CI, n=8). \textbf{(c)}. Structure-function categorization across all queries for each canonicalization (n=8 for each canonicalization). Structural similarity determined by fingerprint Tanimoto similarity ($+$ indicates Tc $\geq$ 0.60, and $-$ indicates Tc $<$ 0.60). Functional similarity determined by patent search ($+$ indicates similar function, $-$ indicates no known relevant function to query). Criteria for similar function for each molecule was as follows: Penicillin G: antibiotic \cite{fleming1941penicillin}; nirmatrelvir: protease inhibitor or antiviral \cite{nirmatrelvir_covid}; zidovudine: antiviral \cite{zidovudine_hiv}; LSD: 5-HT receptor agonist or dopaminergic agonist \cite{lsd_5ht, lsd_dopamine}; fentanyl: opioid analgesic or muscarinic receptor agonist \cite{fentanyl_opioid, fentanyl_muscarinic_heart, fentanyl_muscarinic_brain}; acid blue 25 FA: dye or electroluminescent; avobenzone: UV-Absorption, electroluminescent \cite{avobenzone_sunscreen, avobenzone_oled}; 2-dPAC: electroluminescent \cite{2dpac_electrochem}. \textbf{(d).} Mean non-derivative functional analogues returned in the top 20 results (95\% CI, n=8). Asterisks indicate the level of statistical significance for two-sided independent t-tests (ns, P$<$1.0; *, P$<$0.05; **, P$<$0.01; ***, P$<$0.001; **** P$<$0.0001).}
\label{fig:patent_figs}
\end{figure}

In order to begin to determine the functional significance of the search results, patent and literature searches were conducted on the top 20 results from each search. In general, functionally drug-like queries returned high levels of drug-like molecules and few dye-like molecules, and conversely, queries on functionally dye-like molecules returned more dye-like molecules, and many fewer drug-like molecules (Figs. \ref{fig:drug-like}, \ref{fig:dye-like}). An exception to the latter statement were molecules identified using OEChem inquiries, but this skew was due almost solely to results from the avobenzone search (which in turn has known biological activity \cite{avobenzone_bio_activity_1, avobenzone_bio_activity_2}). In general, the baseline of random drug-like molecules returned for drug-based queries exists, but is relatively low.

\begin{figure}[h]
\centering
\begin{subfigure}{.125\textwidth}
    \centering
    \includegraphics[width=\textwidth]{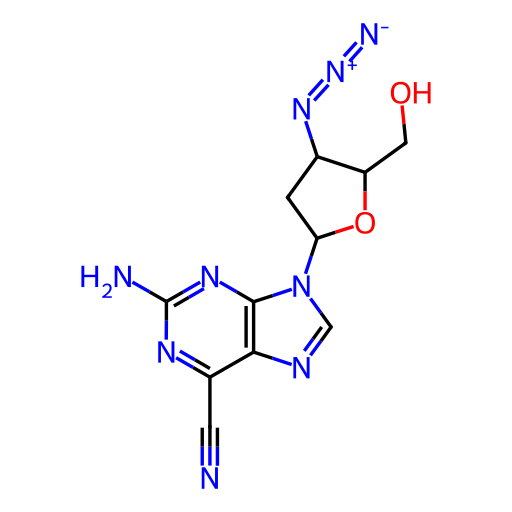}
    \caption{ }
    \label{fig:mol_of_interest_1}
\end{subfigure}%
\begin{subfigure}{.125\textwidth}
    \centering
    \includegraphics[width=\textwidth]{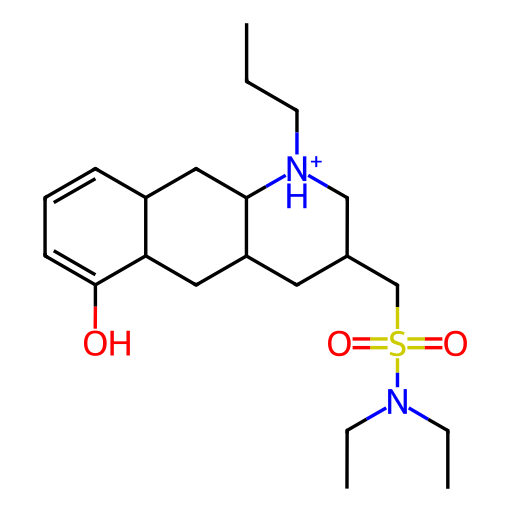}
    \caption{ }
    \label{fig:mol_of_interest_2}
\end{subfigure}%
\begin{subfigure}{.125\textwidth}
    \centering
    \includegraphics[width=\textwidth]{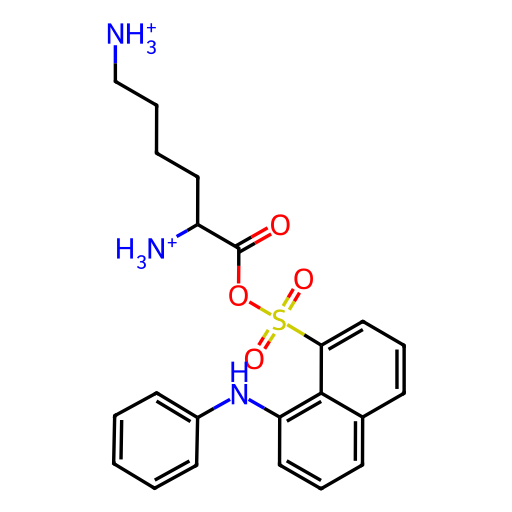}
    \caption{ }
    \label{fig:mol_of_interest_3}
\end{subfigure}%
\begin{subfigure}{.125\textwidth}
    \centering
    \includegraphics[width=\textwidth]{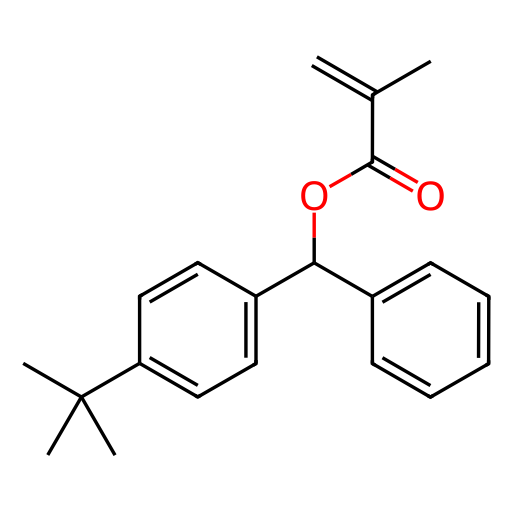}
    \caption{ }
    \label{fig:mol_of_interest_4}
\end{subfigure}%
\begin{subfigure}{.125\textwidth}
    \centering
    \includegraphics[width=\textwidth]{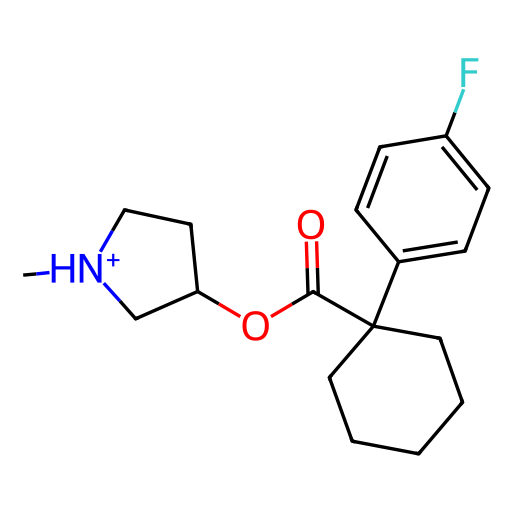}
    \caption{ }
    \label{fig:mol_of_interest_5}
\end{subfigure}%
\begin{subfigure}{.125\textwidth}
    \centering
    \includegraphics[width=\textwidth]{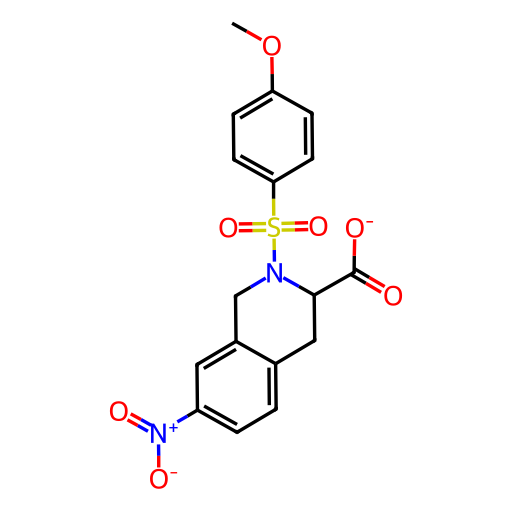}
    \caption{ }
    \label{fig:mol_of_interest_6}
\end{subfigure}%
\begin{subfigure}{.125\textwidth}
    \centering
    \includegraphics[width=\textwidth]{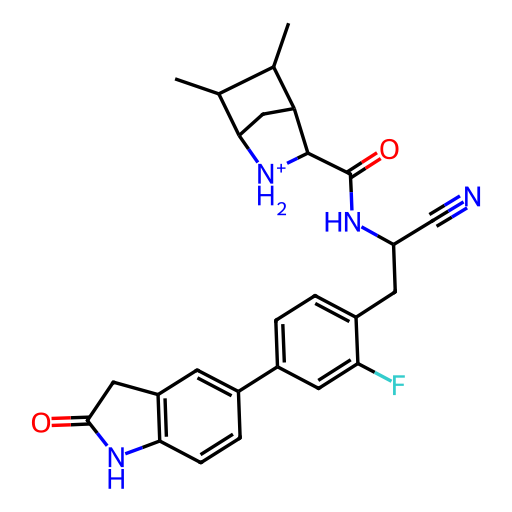}
    \caption{ }
    \label{fig:mol_of_interest_7}
\end{subfigure}\\
\begin{subfigure}{.125\textwidth}
    \centering
    \includegraphics[width=\textwidth]{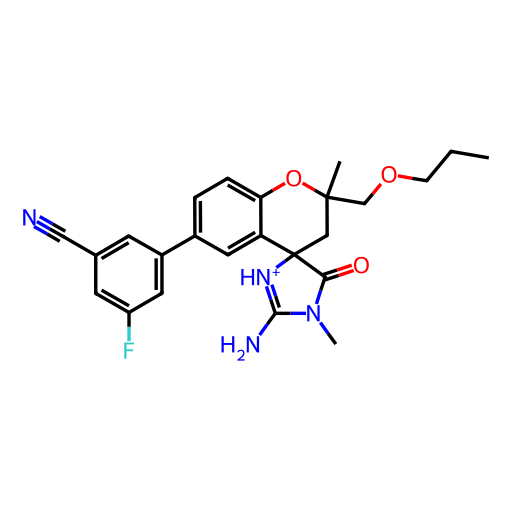}
    \caption{ }
    \label{fig:mol_of_interest_8}
\end{subfigure}%
\begin{subfigure}{.125\textwidth}
    \centering
    \includegraphics[width=\textwidth]{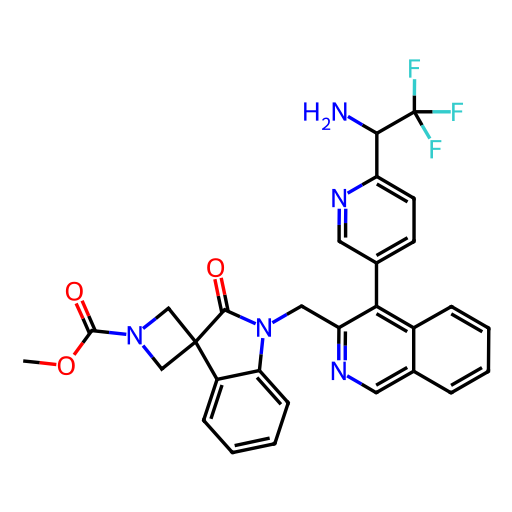}
    \caption{ }
    \label{fig:mol_of_interest_9}
\end{subfigure}%
\begin{subfigure}{.125\textwidth}
    \centering
    \includegraphics[width=\textwidth]{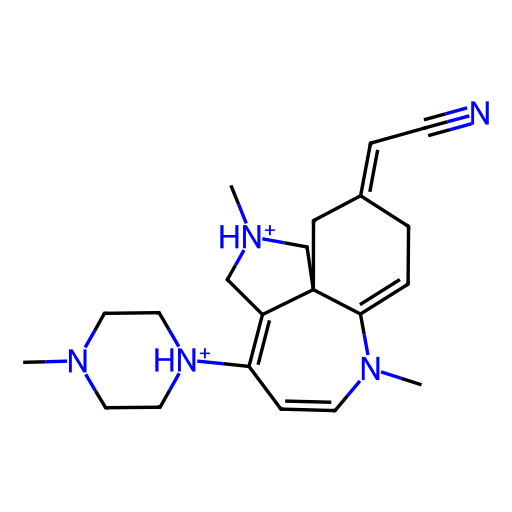}
    \caption{ }
    \label{fig:mol_of_interest_10}
\end{subfigure}%
\begin{subfigure}{.125\textwidth}
    \centering
    \includegraphics[width=\textwidth]{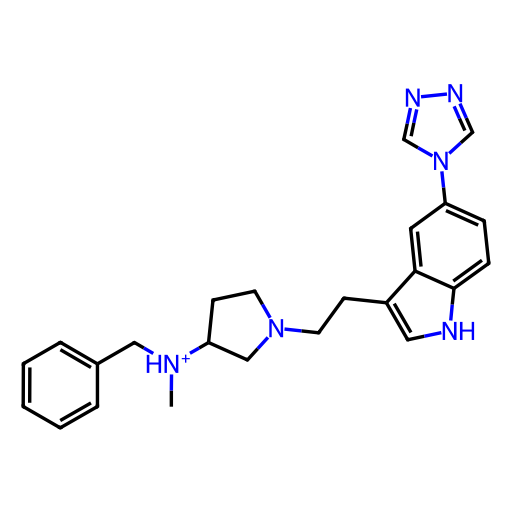}
    \caption{ }
    \label{fig:mol_of_interest_11}
\end{subfigure}%
\begin{subfigure}{.125\textwidth}
    \centering
    \includegraphics[width=\textwidth]{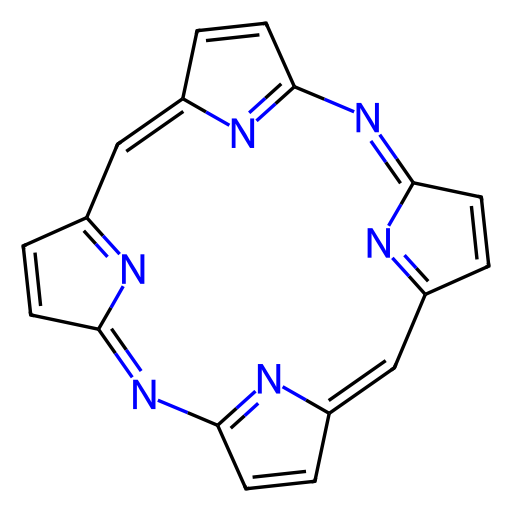}
    \caption{ }
    \label{fig:mol_of_interest_12}
\end{subfigure}%
\begin{subfigure}{.125\textwidth}
    \centering
    \includegraphics[width=\textwidth]{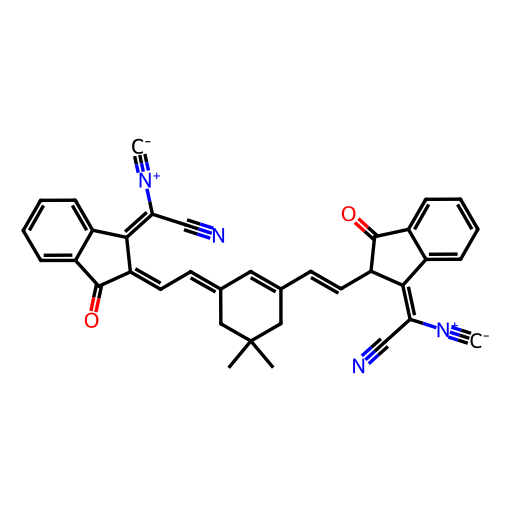}
    \caption{ }
    \label{fig:mol_of_interest_13}
\end{subfigure}%
\begin{subfigure}{.125\textwidth}
    \centering
    \includegraphics[width=\textwidth]{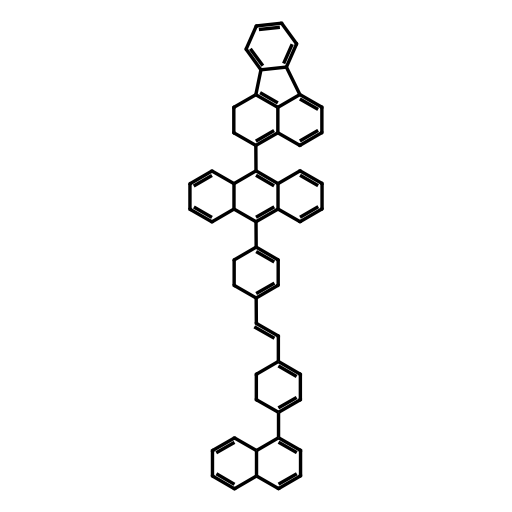}
    \caption{ }
    \label{fig:mol_of_interest_14}
\end{subfigure}%
\captionsetup{singlelinecheck=false}
\caption{\textbf{Structures of molecules discussed herein.} \textbf{(a).} RDKit Atom n zidovudine top result \#2; \textbf{(b).} RDKit Atom n LSD top result \#7; \textbf{(c).} RDKit Atom n acid blue 25 FA top result \#9; \textbf{(d).} RDKit Atom n avobenzone top result \#13; \textbf{(e).} RDKit Atom n fentanyl top result \#20; \textbf{(f).} RDKit Atom n penicillin G top result \#9; \textbf{(g).} OEChem nirmatrelvir top result \#16; \textbf{(h).} OEChem nirmatrelvir top result \#17; \textbf{(i).} OEChem nirmatrelvir top result \#8; \textbf{(j).} OEChem LSD top result \#8; \textbf{(k).} OEChem LSD top result \#14; \textbf{(l).} OEChem acid blue 25 FA top result \#5; \textbf{(m).} OEChem acid blue 25 FA top result \#20; \textbf{(n).} OEChem 2-dPAC top result \#15.}
\label{fig:mols_of_interest_2}
\end{figure}

The functional similarity of query results was contrasted with fingerprint Tanimoto similarity (Fig. \ref{fig:all-str-fun}). The categorization of functionality was either positive (known relevant functionality to the query) or negative (unknown relevant functionality), and the structural similarity was either positive (Tc $\geq$ 0.60) or negative (Tc $<$ 0.60). Criteria / ontologies for similar functionality for each molecule were as follows: Penicillin G: antibiotic \cite{fleming1941penicillin}; nirmatrelvir: protease inhibitor or antiviral \cite{nirmatrelvir_covid}; zidovudine: antiviral \cite{zidovudine_hiv}; LSD: 5-HT receptor agonist or dopaminergic agonist \cite{lsd_5ht, lsd_dopamine}; fentanyl: opioid analgesic or muscarinic receptor agonist \cite{fentanyl_opioid, fentanyl_muscarinic_heart, fentanyl_muscarinic_brain}; acid blue 25 FA: dye or electroluminescent; avobenzone: UV-Absorption, electroluminescent \cite{avobenzone_sunscreen, avobenzone_oled}; 2-dPAC: electroluminescent \cite{2dpac_electrochem}. To illustrate, for OEChem-canonicalized nirmatrelvir (a SARS-CoV-2 main protease inhibitor) several top results (7, 8, 16, and 17) were classified as positive, as these compounds were known protease inhibitors and / or antivirals (Figs. \ref{fig:mol_of_interest_8} \ref{fig:mol_of_interest_9}, \ref{fig:mol_of_interest_10}). A table of the top 20 results for each search, complete with links to their PubChem pages, relevant patents and functional descriptors is listed in Table \ref{tab:patents_list}.

As expected, CheSS searches with the query molecules canonicalized with RDKit Atom 0 resulted in the identification of molecules with similar structures and functions. There were 31 structurally dissimilar molecules (Tc $<$ 0.60) with shared functionality to the query, 30 of which were nonetheless obvious structural derivatives. Penicillin G returned $\beta$-lactam antibiotics, nirmatrelvir returned Hepatitis C Virus (HCV) protease inhibitors, zidovudine returned antiviral pyrimidine nucleosides, LSD returned psychoactive ergolines, fentanyl returned narcotic piperidine analogues, acid blue 25 FA returned anthraquinone dyes, avobenzone returned dibenzoylmethane permutants, and 2-dPAC returned electroluminescent carbazole and triphenylamine derivatives.

In contrast to RDKit Atom 0 queries, RDKit Atom n queries returned molecules that had greater structural diversity, but that still contained many structural analogues. There were 26 structurally dissimilar molecules (Tc $<$ 0.60) with shared functionality to the query, some 22 of which were relatively obvious structural derivatives, including purine and pyrimidine analogs of zidovudine (Fig. \ref{fig:mol_of_interest_1}). However, these hits also included a quite distinct dopaminergic agonist for LSD (Fig. \ref{fig:mol_of_interest_2}), a hydrophobic fluorescence probe for acid blue 25 FA (Fig. \ref{fig:mol_of_interest_3}), a refractive copolymer for avobenzone (Fig. \ref{fig:mol_of_interest_4}), and a fentanyl-like agonist for its known coronary muscarinic receptor (Fig. \ref{fig:mol_of_interest_5}) \cite{fentanyl_muscarinic_heart, fentanyl_muscarinic_brain}. 2-dPAC did not return any molecules with known relevant functionality using this canonicalization, though the results had similar aromaticity to the query. Interestingly, tetracycline antibiotics have been used to inhibit metalloproteases, and the penicillin G query returned two non-tetracycline metalloprotease inhibitors (Fig. \ref{fig:mol_of_interest_6}) \cite{metalloprotease_tetra}.

OEChem generally returned molecules that were structurally highly dissimilar to the query. There were 35 structurally dissimilar molecules (Tc $<$ 0.60) with shared functionality to the query, and in contrast to the previous two canonicalizations, only 7 of these were obvious structural derivatives. The more diverse compounds with shared functionality included two non-HCV protease inhibitors (Figs. \ref{fig:mol_of_interest_7}, \ref{fig:mol_of_interest_8}); a Respiratory Syncytial Virus antiviral for nirmatrelvir (Fig. \ref{fig:mol_of_interest_9}); a dopaminergic and serotonergic agonist (Fig. \ref{fig:mol_of_interest_10}) and a 5-HT1 receptor agonist for LSD (Fig. \ref{fig:mol_of_interest_11}); porphyrins (Fig. \ref{fig:mol_of_interest_12}) and other conjugated dyes for acid blue 25 FA (Fig. \ref{fig:mol_of_interest_13}); photovoltaic and electroluminescent molecules for avobenzone; and highly conjugated electroluminescent molecules for 2-dPAC (Fig. \ref{fig:mol_of_interest_14}). Taken together, these hits provide anecdotal proof for the hypothesis that changing query canonicalizations can lead to the discovery of novel, but functional, chemical compounds. These insights may provide interesting avenues for drug repurposing, and that molecules with previously unknown functions may serve as leads for novel drug discovery.

\subsection{Explanation of Search Behavior}

We find that alternative canonicalizations influence the search behavior of CheSS through changes in tokenization, which causes the CLM to weight higher-order relationships more importantly when creating embeddings. There are stark differences between RDKit and OEChem canonicalizations, notably their differences in the representation of aromatic rings. OEChem prefers the Kekul\'e form (C1=CC=CC=C1), while RDKit prefers to use lowercase with assumed aromaticity (c1ccccc1), and these differences, among others, result in markedly different tokenization, both in the composition and the length of the tokenized vectors (Fig. \ref{fig:search_behavior_by_canon}). Since the CLM has demonstrated a bias toward embedding molecules with similar token vector lengths and (to a lesser extent) token composition to the query (Fig. \ref{fig:search_behavior_by_canon}), a query with a different canonicalization will tokenize into a radically different token vector and thereby make the CLM more likely to return molecules with SMILES representations that are highly dissimilar from the original same-canonicalized query. Despite this behavior, and potentially because of it, CheSS searches with alternative canonicalizations found diverse chemical structures with similar functional properties, as demonstrated by patent and literature searches (Fig. \ref{fig:patent_figs}).

When CLMs are forced to go beyond simple token patterns to determine similarity, more nuanced relationships may appear. Given the nature of transformers, it is indiscernible what these relationships are, but based on our analysis we find it possible that the CLM may, for example, key on the apposition of functional groups in space, in a way similar to how receptors perceive ligands.


\subsection{Drawbacks, Future Improvements, and Potential for Misuse}

The database used by CheSS consisted of the $\sim$10M molecules used as a training set for ChemBERTa, and thus it is difficult to predict the behavior of queries that differ greatly from the molecules in this dataset. In addition, very small molecules may not differ in their canonicalized representations, leading to more homogeneity between queries. Nonetheless, the method itself is extensible for use with any dataset, and may invite discussions regarding what datasets and CLMs are most useful for moving between different canonicalizations to identify functional analogues. We did not at this time explore whether non-canonicalized, yet valid, variations of SMILES strings will lead to similar results. 

We also note that the threat of dual use for chemical machine learning models has been a topic of discussion amongst researchers \cite{urbina2022dual}. While the model used in our implementation of CheSS was unsupervised and not trained for identifying toxic molecules, a successful chemical similarity search tool carries inherent risks. We therefore advise caution in considering public implementations of these tools and recommend restricting searches to avoid queries with the potential for malicious use. 


\section{Conclusion}
In this study, we created a chemical similarity search pipeline utilizing a transformer encoder-based chemical language model to generate embeddings upon which similarity scores can be computed. From this, we designed a prompt engineering strategy that expands upon existing chemical semantic searches by creating a method able to identify structurally dissimilar molecules with similar function. We demonstrate the utility of this search method to identify non-obvious functional compounds related to multiple different query molecules. This method may aid repurposing known compounds or in discovering new structural classes of molecules that have desirable functionality. Despite potential drawbacks, we believe that CheSS and the canonicalization prompt engineering method discussed herein will be of broad interest to the chemical community, as it begins to explore how machine learning can be used outside of staid similarity queries.


\section{Acknowledgements}

The authors acknowledge the Texas Advanced Computing Center at The University of Texas at Austin for providing high-performance computing resources. This work was supported by the Welch Foundation (C.W.K.), the Blumberg Centennial Professorship in Molecular Evolution (C.O.W.), and the Reeder Centennial Fellowship in Systematic and Evolutionary Biology at The University of Texas at Austin (C.O.W.). 

\section{Author Contributions}
Conceptualization, C.W.K. and A.L.F.; Methodology, C.W.K.; Software, C.W.K. and A.L.F.; Validation, C.W.K. and A.L.F; Formal Analysis, C.W.K.; Investigation, C.W.K.; Resources, A.D.E., C.O.W.; Data Curation, C.W.K. and A.L.F.; Writing - Original Draft, C.W.K. and A.L.F.; Writing - Review \& Editing, C.W.K., A.L.F., A.D.E., and C.O.W.; Visualization, C.W.K and A.L.F.; Supervision, A.D.E. and C.O.W.; Funding Acquisition, A.D.E. and C.O.W.

\section{Declaration of Interests}
The authors declare no competing interests.

\section {Data and Code Availability}
Project code and discussed search results can be found at https://github.com/kosonocky/CheSS.


\bibliographystyle{unsrtnat}
\bibliography{references}  


\clearpage
\section*{Supplementary Figures}
\beginsupplement

\begin{table}[h]
\captionsetup{singlelinecheck=false}
\caption{\textbf{Different canonical SMILES string representations for each molecule query molecule.}}
\label{tab:canon_smiles_all_expanded}
\centering
\begin{adjustbox}{center}
\begin{tabular}{|c c c|}
\hline
\textbf{Query} & \textbf{Canon Alg.} & \textbf{SMILES} \\
\hline
 & RDKit Atom 0 & CC1(C)SC2C(NC(=O)Cc3ccccc3)C(=O)N2C1C(=O)[O-]\\
Penicillin G & RDKit Atom 13 & c1ccc(CC(=O)NC2C(=O)N3C2SC(C)(C)C3C(=O)[O-])cc1\\
 & OEChem & CC1(C(N2C(S1)C(C2=O)NC(=O)CC3=CC=CC=C3)C(=O)[O-])C\\
\hline
& RDKit Atom 0 & CC(C)(C)C(NC(=O)C(F)(F)F)C(=O)N1CC2C(C1C(=O)NC(C\#N)CC1CCNC1=O)C2(C)C\\
Nirmatrelvir & RDKit Atom 21 & N(C(=O)C1C2C(CN1C(=O)C(NC(=O)C(F)(F)F)C(C)(C)C)C2(C)C)C(C\#N)CC1CCNC1=O\\
& OEChem & CC1(C2C1C(N(C2)C(=O)C(C(C)(C)C)NC(=O)C(F)(F)F)C(=O)NC(CC3CCNC3=O)C\#N)C\\
\hline
& RDKit Atom 0 & Cc1cn(C2CC(N=[N+]=[N-])C(CO)O2)c(=O)[nH]c1=O\\
Zidovudine & RDKit Atom 15 & O=c1[nH]c(=O)c(C)cn1C1CC(N=[N+]=[N-])C(CO)O1\\
& OEChem & CC1=CN(C(=O)NC1=O)C2CC(C(O2)CO)N=[N+]=[N-]\\
\hline
& RDKit Atom 0 & CCN(CC)C(=O)C1C=C2c3cccc4[nH]cc(c34)CC2[NH+](C)C1\\
LSD & RDKit Atom 9 & C12=CC(C(=O)N(CC)CC)C[NH+](C)C1Cc1c[nH]c3cccc2c13\\
& OEChem & CCN(CC)C(=O)C1C[NH+](C2CC3=CNC4=CC=CC(=C34)C2=C1)C\\
\hline
& RDKit Atom 0 & CCC(=O)N(c1ccccc1)C1CC[NH+](CCc2ccccc2)CC1\\
Fentanyl & RDKit Atom 17 & c1(CC[NH+]2CCC(N(C(=O)CC)c3ccccc3)CC2)ccccc1\\
& OEChem & CCC(=O)N(C1CC[NH+](CC1)CCC2=CC=CC=C2)C3=CC=CC=C3\\
\hline
& RDKit Atom 0 & Nc1c(S(=O)(=O)[O-])cc(Nc2ccccc2)c2c1C(=O)c1ccccc1C2=O\\
Acid Blue 25 FA & RDKit Atom 26 & C1(=O)c2ccccc2C(=O)c2c(N)c(S(=O)(=O)[O-])cc(Nc3ccccc3)c21\\
& OEChem & C1=CC=C(C=C1)NC2=CC(=C(C3=C2C(=O)C4=CC=CC=C4C3=O)N)S(=O)(=O)[O-]\\
\hline
& RDKit Atom 0 & COc1ccc(C(=O)CC(=O)c2ccc(C(C)(C)C)cc2)cc1\\
Avobenzone & RDKit Atom 8 & C(C(=O)c1ccc(OC)cc1)C(=O)c1ccc(C(C)(C)C)cc1\\
& OEChem & CC(C)(C)C1=CC=C(C=C1)C(=O)CC(=O)C2=CC=C(C=C2)OC\\
\hline
& RDKit Atom 0 & c1ccc(N(c2ccccc2)c2ccc3c(c2)[nH]c2ccccc23)cc1\\
2-dPAC & RDKit Atom 18 & c12ccccc1c1ccc(N(c3ccccc3)c3ccccc3)cc1[nH]2\\
& OEChem & C1=CC=C(C=C1)N(C2=CC=CC=C2)C3=CC4=C(C=C3)C5=CC=CC=C5N4\\
\hline
\end{tabular}
\end{adjustbox}
\end{table}


\newpage
\FloatBarrier

\newpage
\FloatBarrier


\newpage
\begin{figure}[h]
\centering
\begin{subfigure}{\textwidth}
    \centering
    \includegraphics[width=\textwidth]{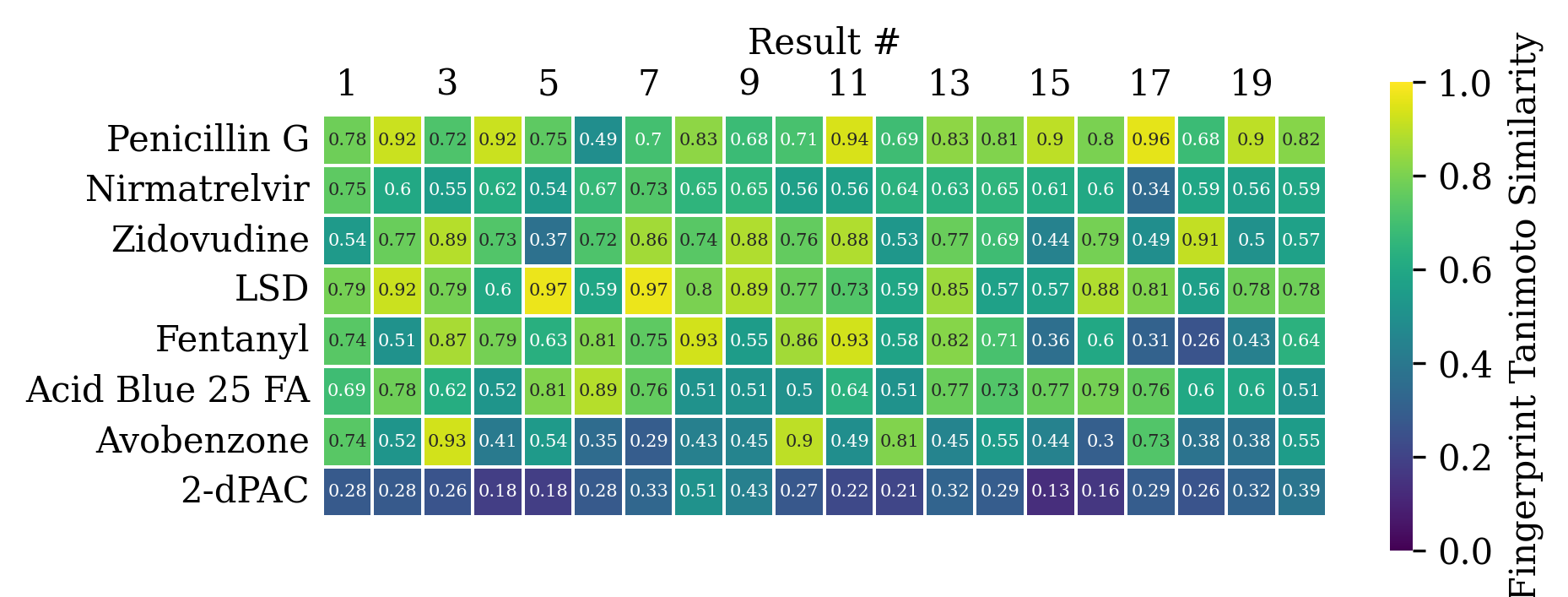}
    \captionsetup{width=6cm}
    \caption{ }
    \label{fig:rdkit_atom_0_results_fingerprint}
\end{subfigure}\\
\begin{subfigure}{\textwidth}
    \centering
    \includegraphics[width=\textwidth]{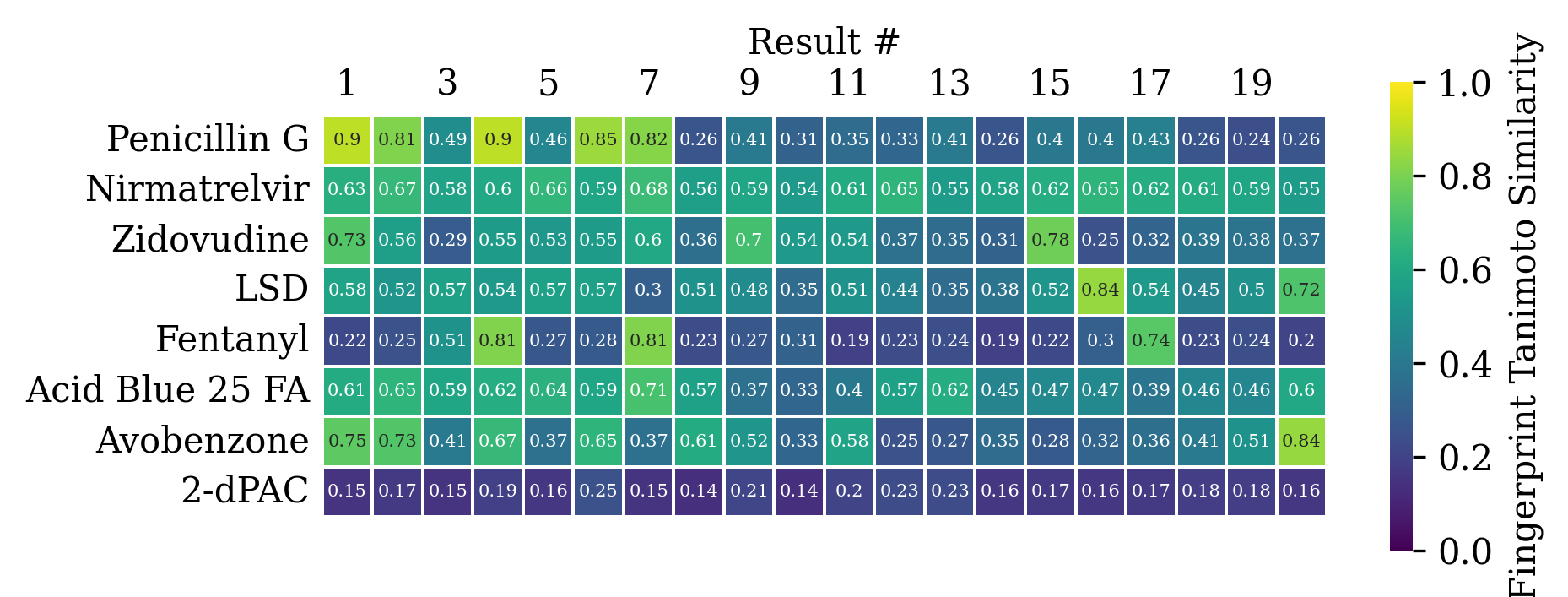}
    \captionsetup{width=6cm}
    \caption{ }
    \label{fig:rdkit_atom_n_results_fingerprint}
\end{subfigure}\\
\begin{subfigure}{\textwidth}
    \centering
    \includegraphics[width=\textwidth]{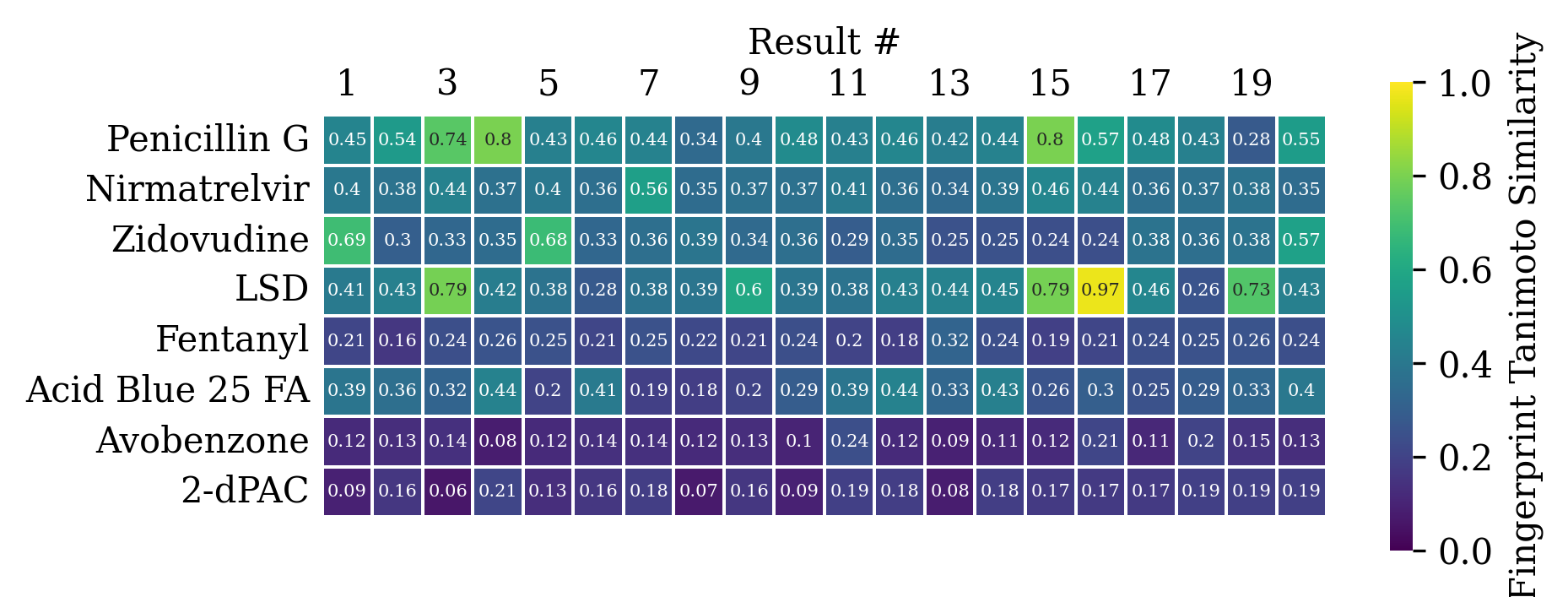}
    \captionsetup{width=6cm}
    \caption{ }
    \label{fig:oechem_results_fingerprint}
\end{subfigure}\\
\caption{\textbf{Fingerprint Tanimoto coefficients between the query molecule and the top 20 most similar molecules to the query (by feature cosine similarity) for each canonicalization}. \textbf{(a-c).} In order: RDKit Atom 0, RDKit Atom n, OEChem. These demonstrate that the RDKit Atom 0 search is providing results similar to a fingerprint structural search, whereas this is less so the case in RDKit Atom n, and even less so in the OEChem search. The exception to this is 2-dPAC, in which none of the molecules would have reasonably been found with a fingerprint search.}
\label{fig:results_fingerprint}
\end{figure}
\FloatBarrier


\newpage
\begin{figure}[h]
\centering
\begin{subfigure}{\textwidth}
    \centering
    \includegraphics[width=\textwidth]{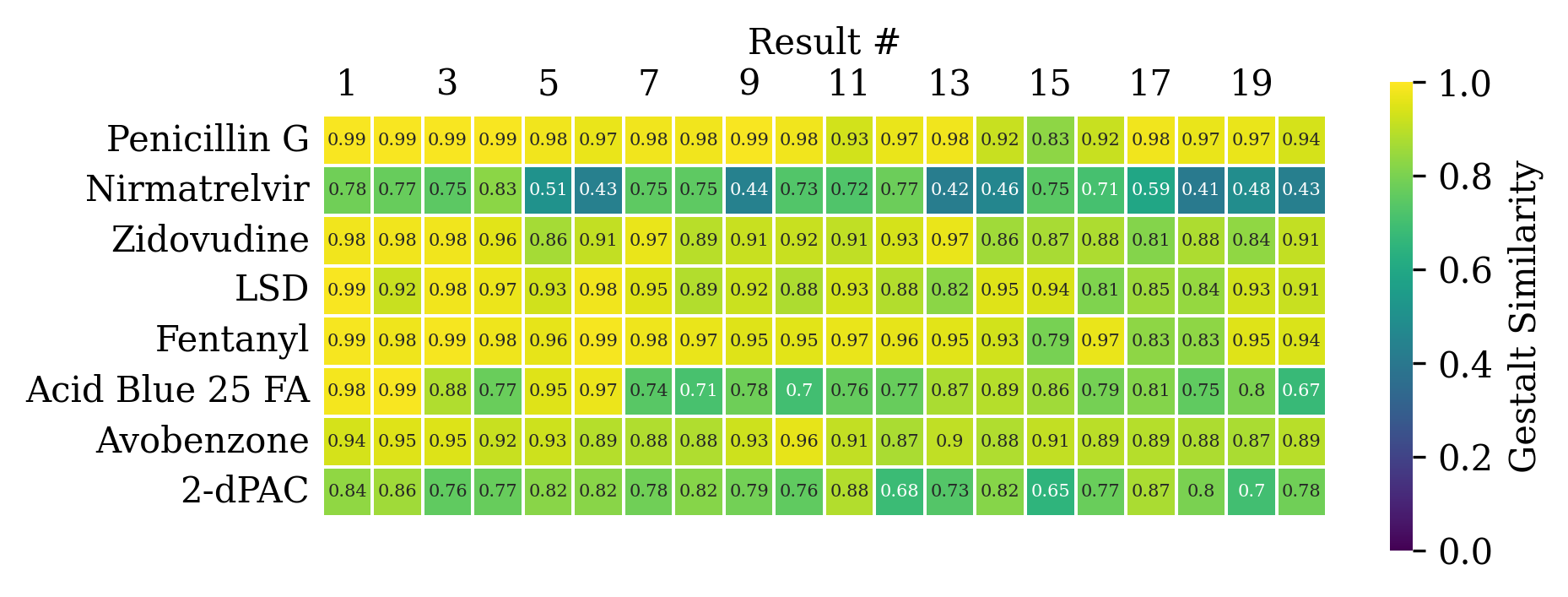}
    \captionsetup{width=6cm}
    \caption{ }
    \label{fig:rdkit_atom_0_results_gestalt}
\end{subfigure}\\
\begin{subfigure}{\textwidth}
    \centering
    \includegraphics[width=\textwidth]{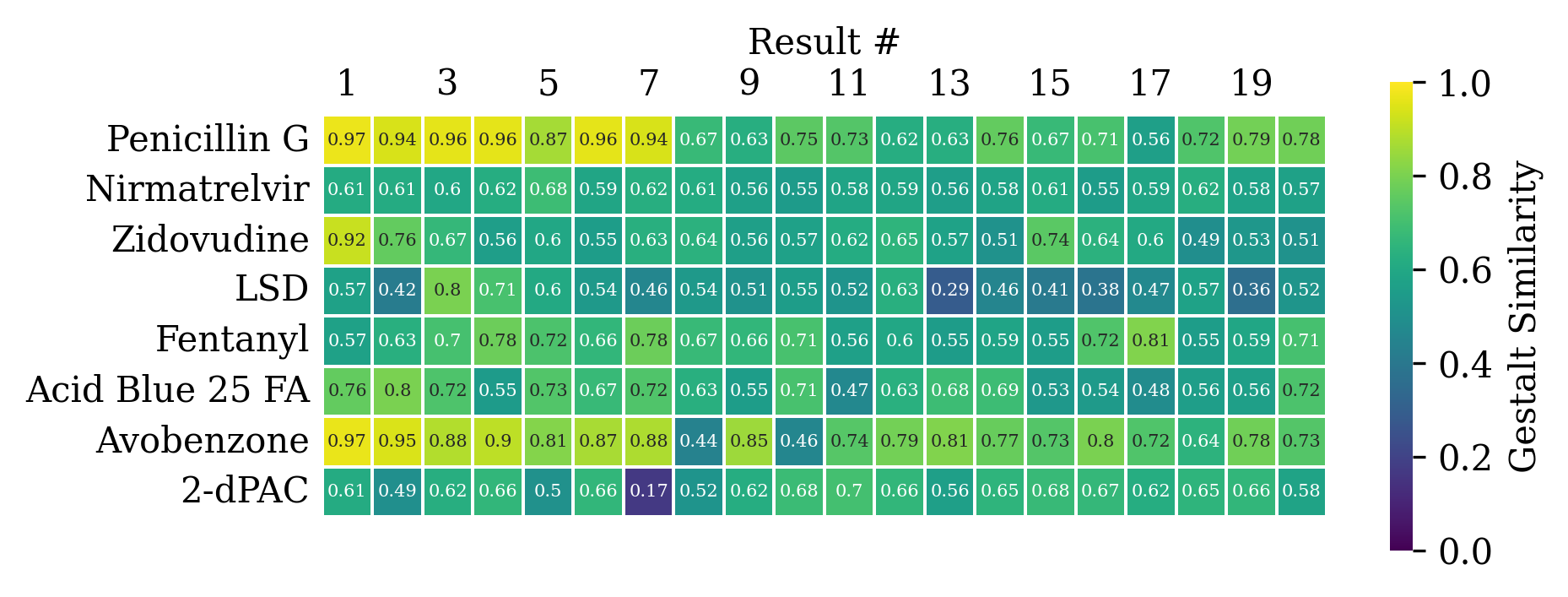}
    \captionsetup{width=6cm}
    \caption{ }
    \label{fig:rdkit_atom_n_results_gestalt}
\end{subfigure}\\
\begin{subfigure}{\textwidth}
    \centering
    \includegraphics[width=\textwidth]{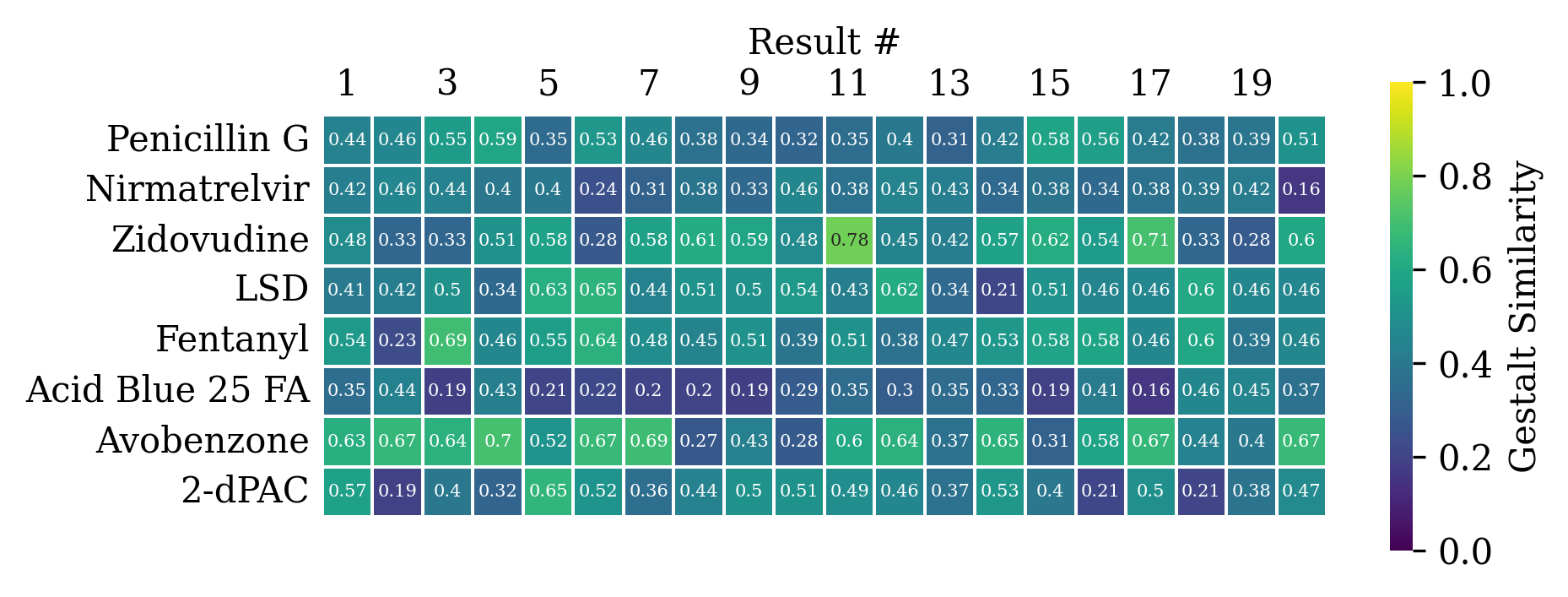}
    \captionsetup{width=6cm}
    \caption{ }
    \label{fig:oechem_results_gestalt}
\end{subfigure}\\
\caption{\textbf{Gestalt similarity between the strings of the top 20 most similar molecules to the query (by feature cosine similarity) and the canonicalized query string of each respective canonicalization}. \textbf{(a-c).} In order: RDKit Atom 0, RDKit Atom n, OEChem. These demonstrate that the RDKit Atom 0 search is providing results very similar to a simple string similarity search, whereas this is less so the case in RDKit Atom n, and even less so in the OEChem search.}
\label{fig:results_gestalt}
\end{figure}
\FloatBarrier


\newpage
\begin{figure}[h]
\centering
\begin{subfigure}{\textwidth}
    \centering
    \includegraphics[width=\textwidth]{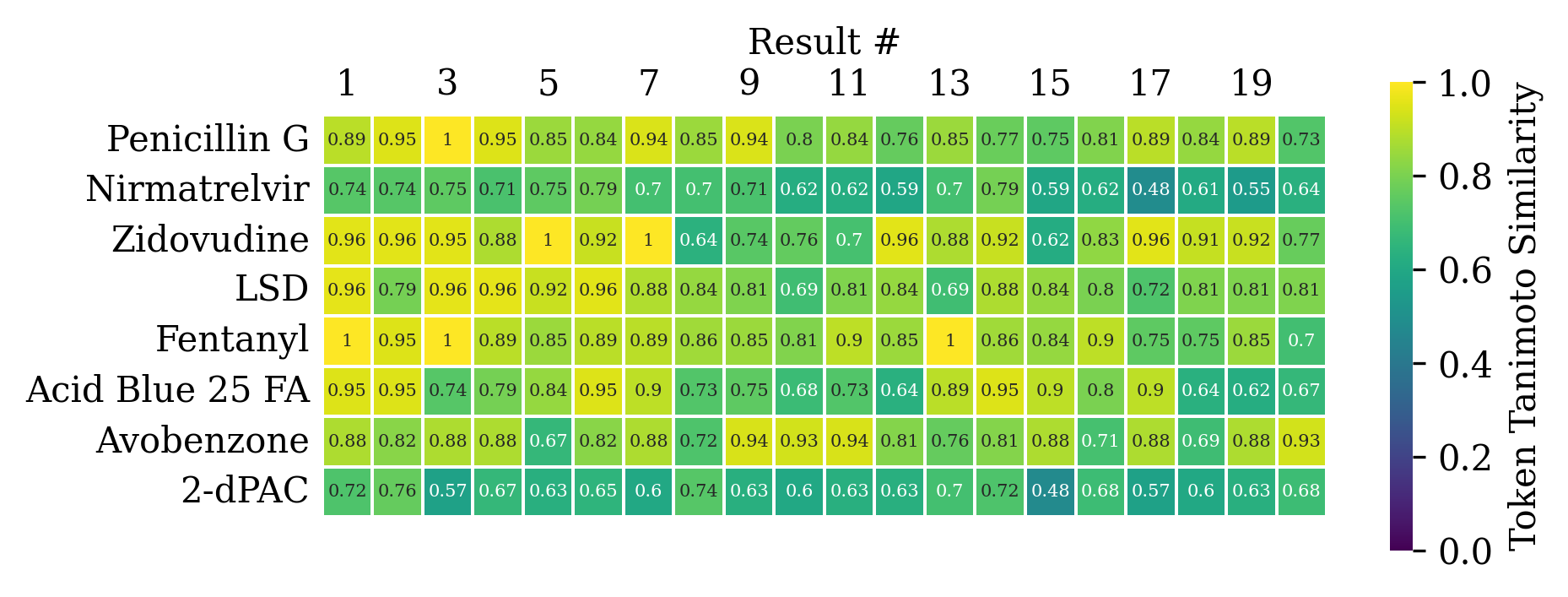}
    \captionsetup{width=6cm}
    \caption{ }
    \label{fig:rdkit_atom_0_results_token_similarity}
\end{subfigure}\\
\begin{subfigure}{\textwidth}
    \centering
    \includegraphics[width=\textwidth]{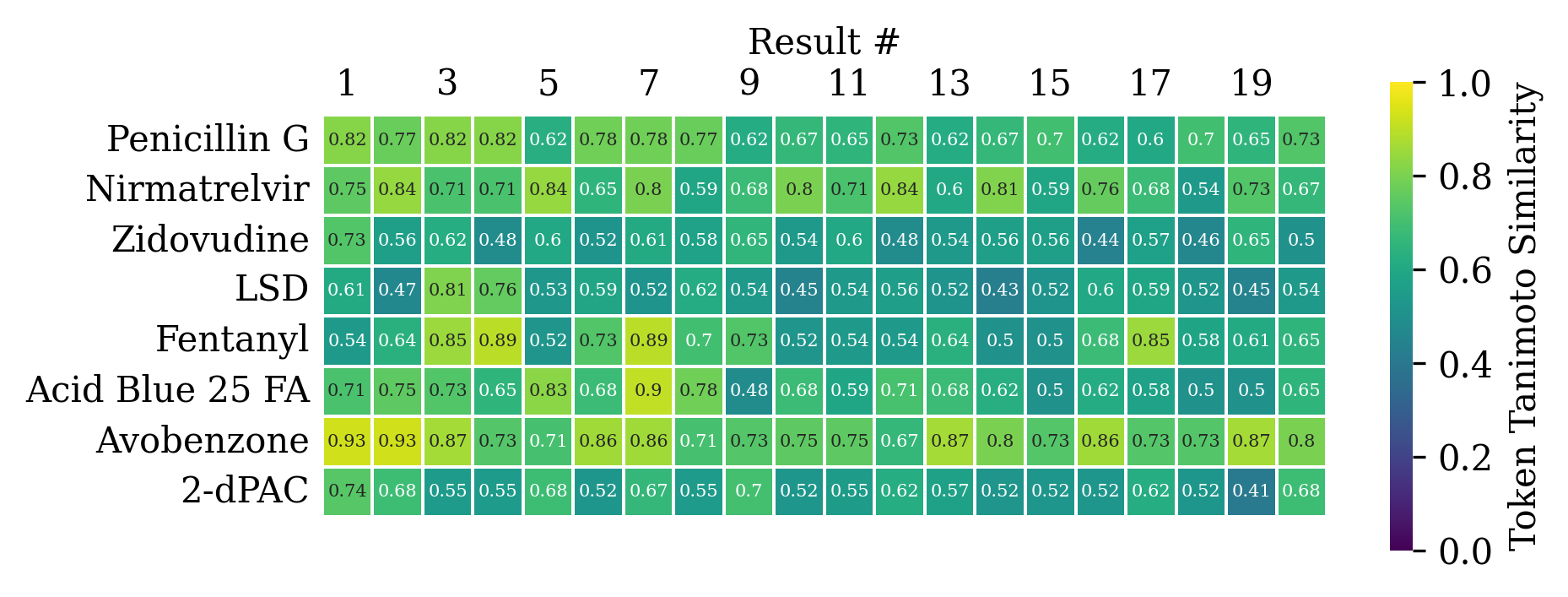}
    \captionsetup{width=6cm}
    \caption{ }
    \label{fig:rdkit_atom_n_results_token_similarity}
\end{subfigure}\\
\begin{subfigure}{\textwidth}
    \centering
    \includegraphics[width=\textwidth]{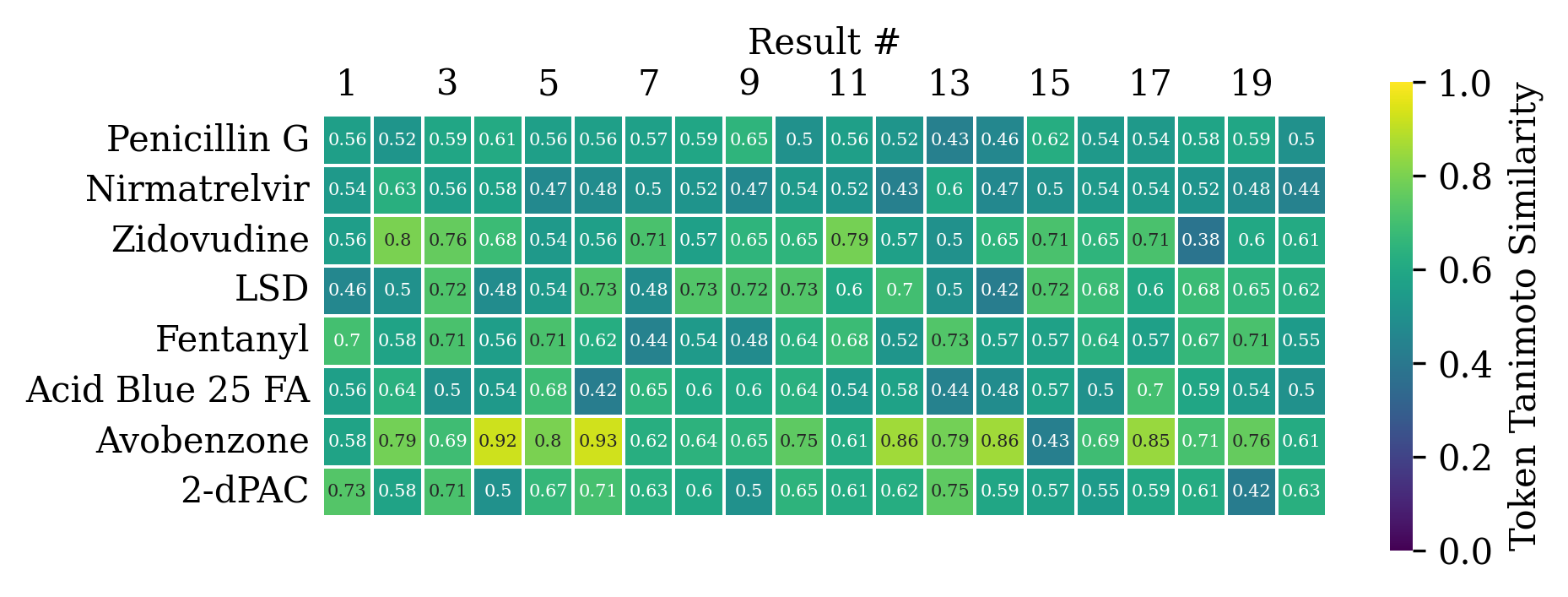}
    \captionsetup{width=6cm}
    \caption{ }
    \label{fig:oechem_results_token_similarity}
\end{subfigure}\\
\caption{\textbf{Token vector Tanimoto ratios between the query molecule's tokenized SMILES vector and the tokenized SMILES vectors of the top 20 most similar molecules to the query (by feature cosine similarity) for each canonicalization}. \textbf{(a-c).} In order: RDKit Atom 0, RDKit Atom n, OEChem. These demonstrate that searches using RDKit Atom 0 to canonicalize the SMILES string will return molecules with a high number of shared tokens to the query, whereas this is less so the case with RDKit Atom n and OEChem. Despite these differences, nearly all of the top results share at least 50\% of the tokens with the query.}
\label{fig:results_token_similarity}
\end{figure}
\FloatBarrier


\newpage
\begin{figure}[h]
\centering
\begin{subfigure}{\textwidth}
    \centering
    \includegraphics[width=\textwidth]{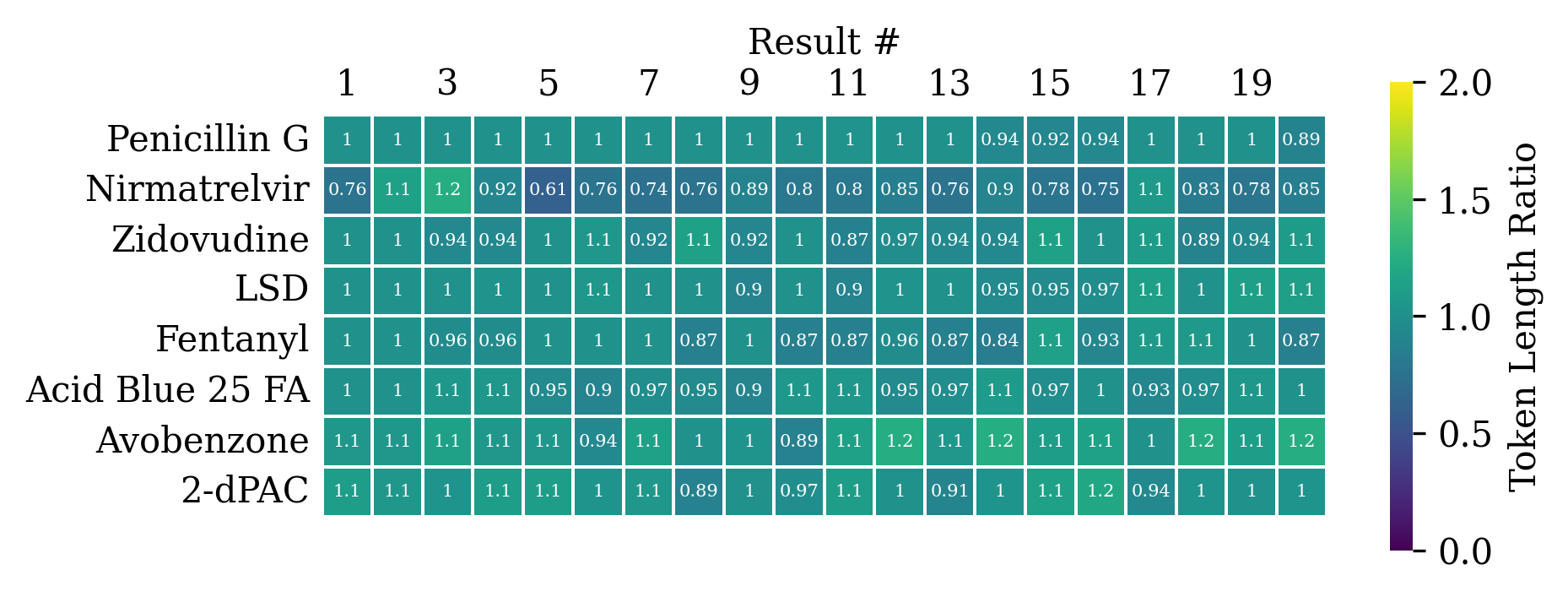}
    \captionsetup{width=6cm}
    \caption{ }
    \label{fig:rdkit_atom_0_results_token_len_ratio}
\end{subfigure}\\
\begin{subfigure}{\textwidth}
    \centering
    \includegraphics[width=\textwidth]{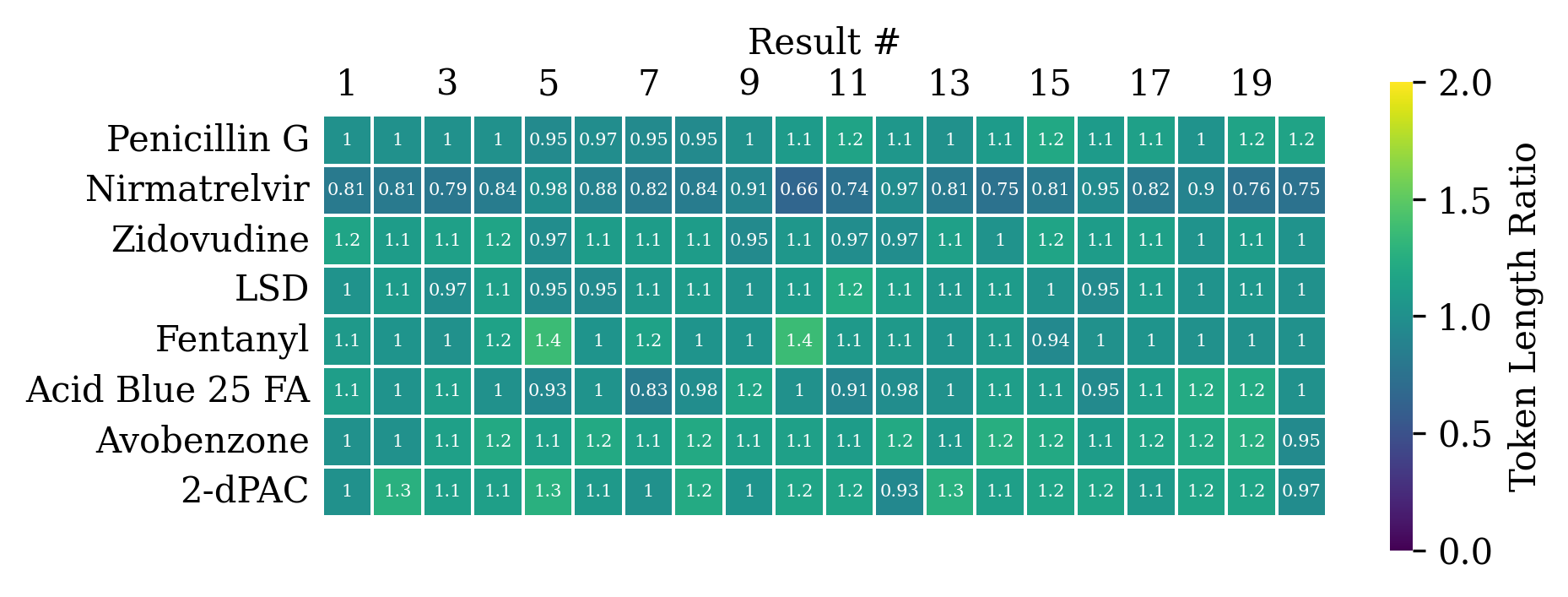}
    \captionsetup{width=6cm}
    \caption{ }
    \label{fig:rdkit_atom_n_results_token_len_ratio}
\end{subfigure}\\
\begin{subfigure}{\textwidth}
    \centering
    \includegraphics[width=\textwidth]{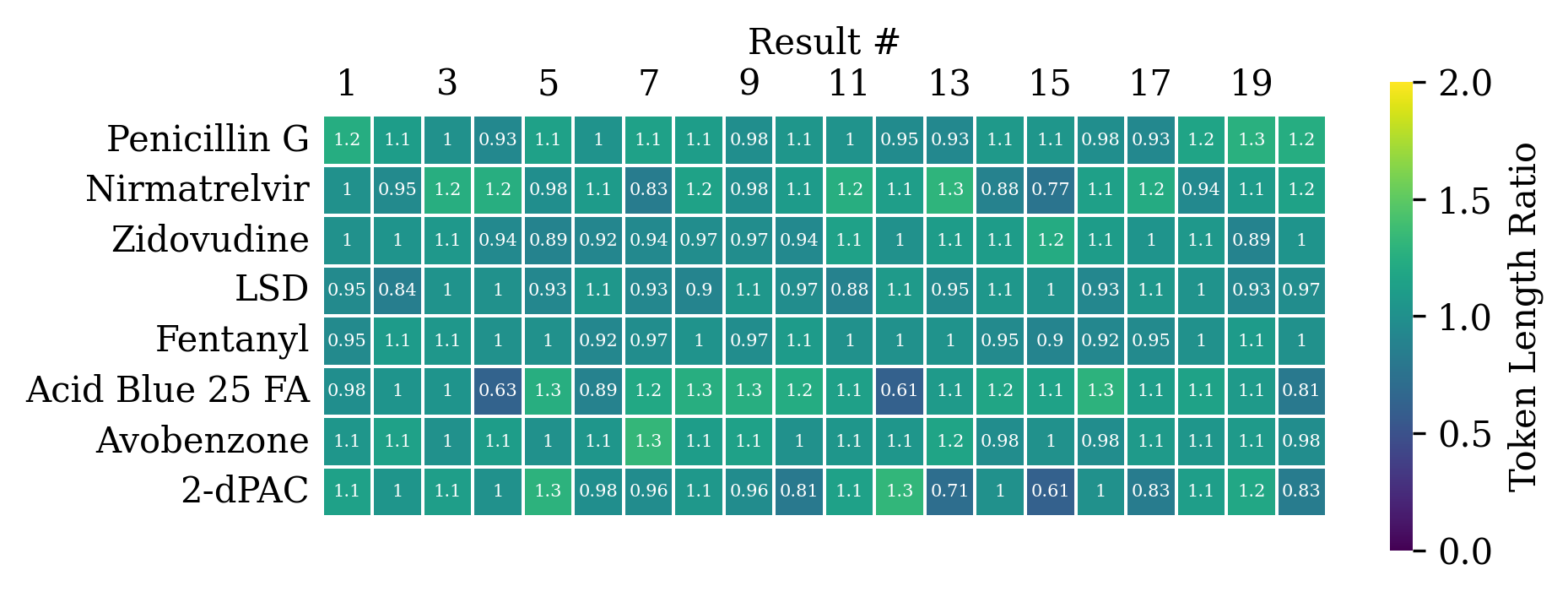}
    \captionsetup{width=6cm}
    \caption{ }
    \label{fig:oechem_results_token_len_ratio}
\end{subfigure}\\
\caption{\textbf{Token vector length ratio between the query molecule's tokenized SMILES vector and the tokenized SMILES vectors of the top 20 most similar molecules to the query (by feature cosine similarity) for each canonicalization}. These demonstrate that the length of the tokenized vectors for almost all of the results fall within 20\% of the query's length, indicating that the length of the tokenized SMILES vector is a significant factor in how the top results are determined.}
\label{fig:results_token_length}
\end{figure}
\FloatBarrier




\newpage
\begin{figure}[h]
\centering
\begin{subfigure}{\textwidth}
    \centering
    \includegraphics[width=0.4\textwidth]{figs/rank_slope_rdkit_atom_n_vs_rdkit_atom_0_penicillin.png}\includegraphics[width=0.4\textwidth]{figs/rank_slope_oechem_vs_rdkit_atom_0_penicillin.png}
    \captionsetup{width=4cm}
    \caption{ }
    \label{fig:slope_rank_penicillin}
\end{subfigure}\\
\begin{subfigure}{\textwidth}
    \centering
    \includegraphics[width=0.4\textwidth]{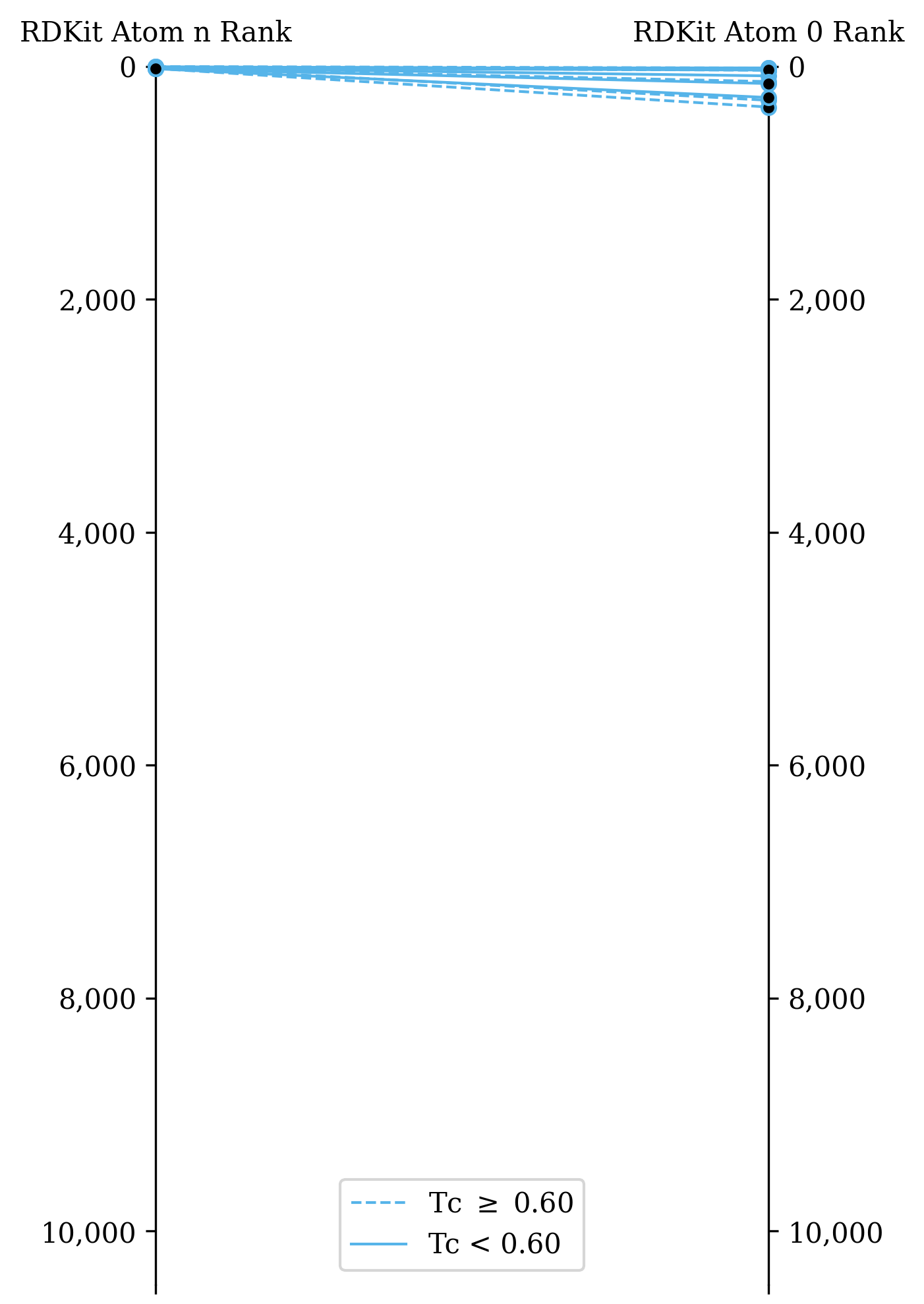}\includegraphics[width=0.4\textwidth]{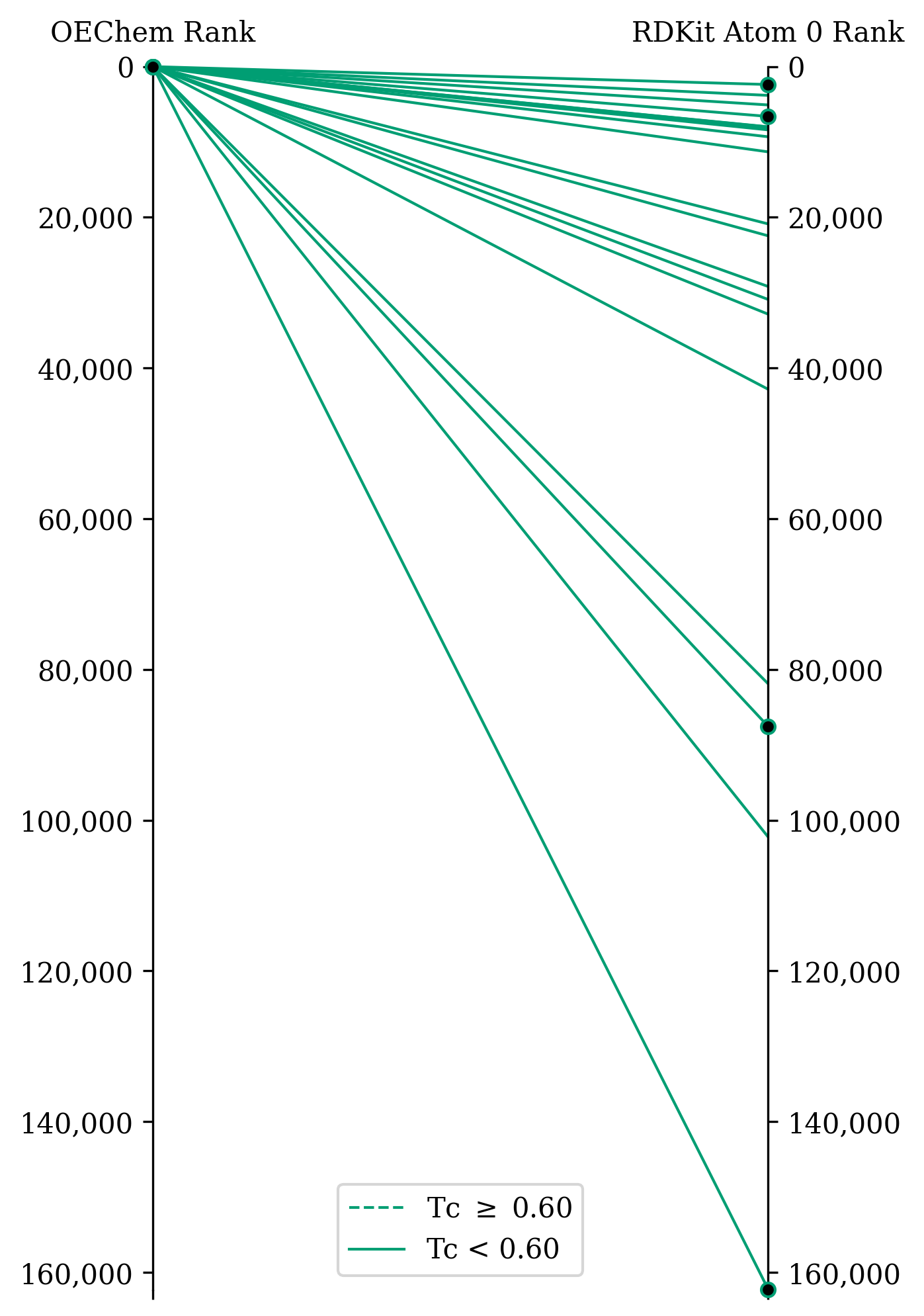}
    \captionsetup{width=4cm}
    \caption{ }
    \label{fig:slope_rank_nirmatrelvir}
\end{subfigure}\\
\caption{\textbf{The index rank of each alternate canonicalization’s top 20 results for each query compared to the index rank that these same molecules scored in the other canonicalizations’ searches.} \textbf{(a-h).} In order: penicillin G, nirmatrelvir, zidovudine, LSD, fentanyl, acid blue 25 FA, avobenzone, 2-dPAC. Molecules functionally similar to the query indicated by a black dot, as determined by the patent search, and structurally similar to the query (Tc $\geq$ 0.60) indicated by a dashed line. These demonstrate that queries that underwent alternative canonicalization were able to identify functional molecules that would have been impractical to find using the standard canonicalization.}
\label{fig:slope_rank_1}
\end{figure}%
\begin{figure}[h]\ContinuedFloat
\centering
\begin{subfigure}{\textwidth}
    \centering
    \includegraphics[width=0.4\textwidth]{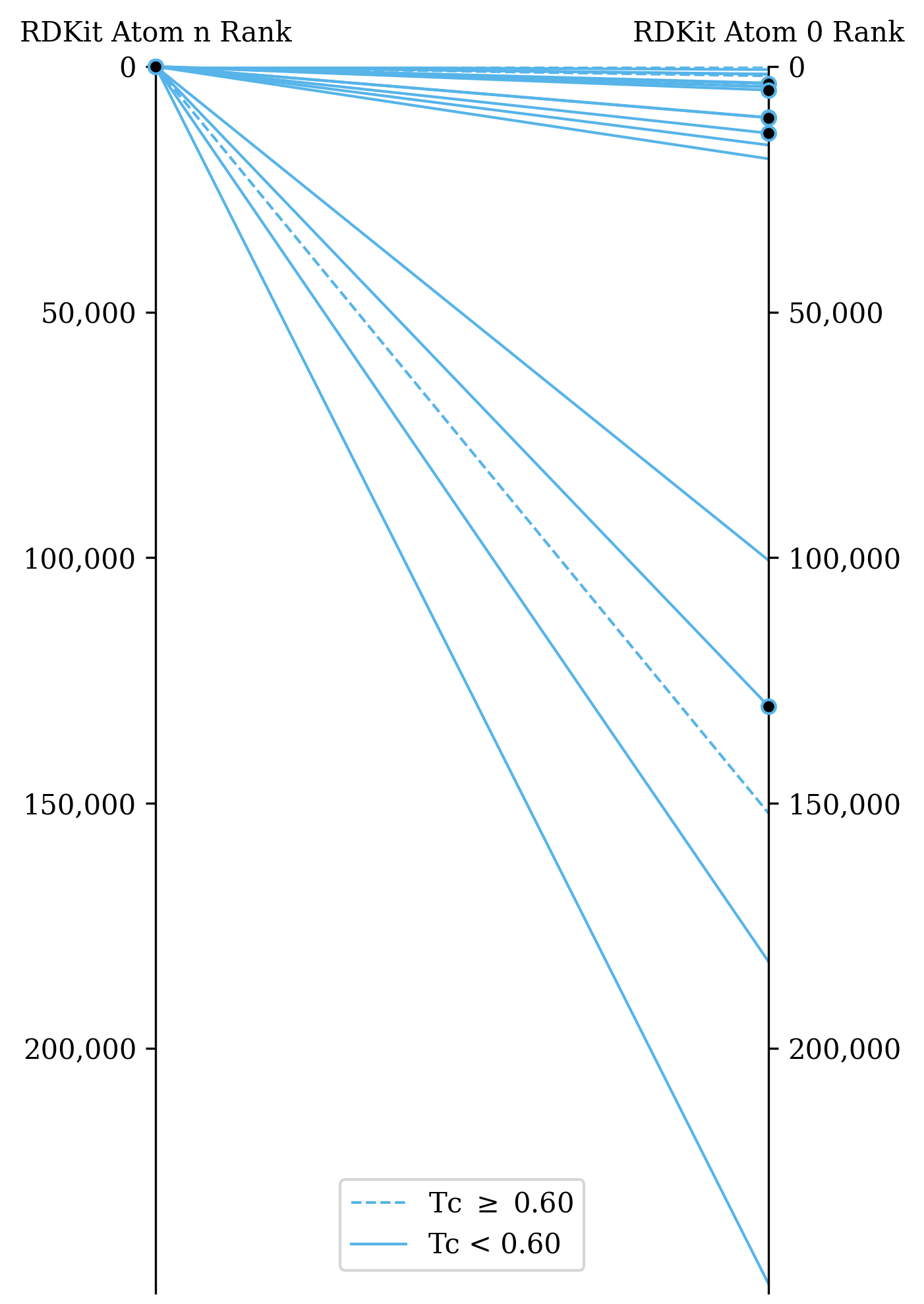}\includegraphics[width=0.4\textwidth]{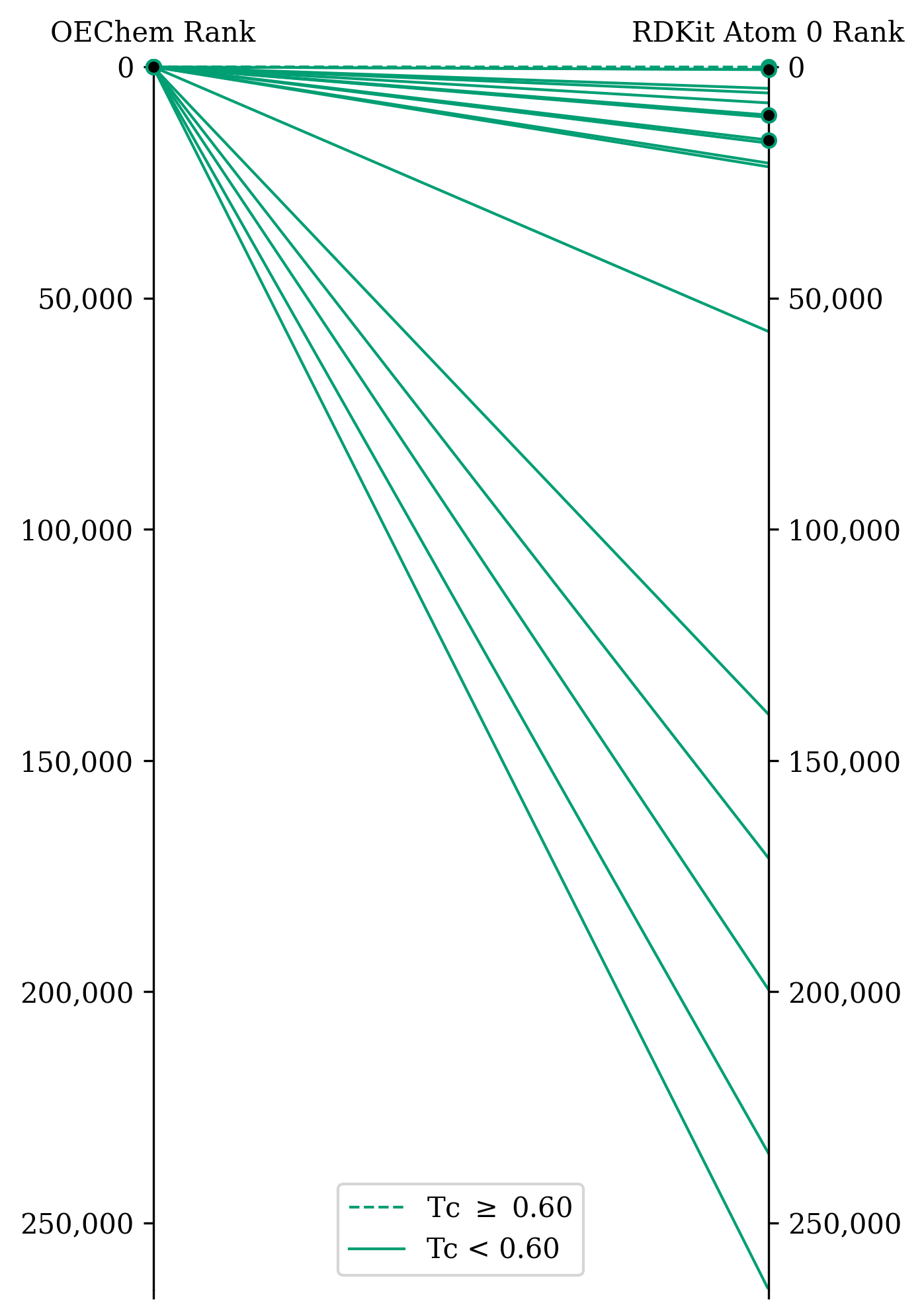}
    \captionsetup{width=4cm}
    \caption{ }
    \label{fig:slope_rank_zidovudine}
\end{subfigure}\\
\begin{subfigure}{\textwidth}
    \centering
    \includegraphics[width=0.4\textwidth]{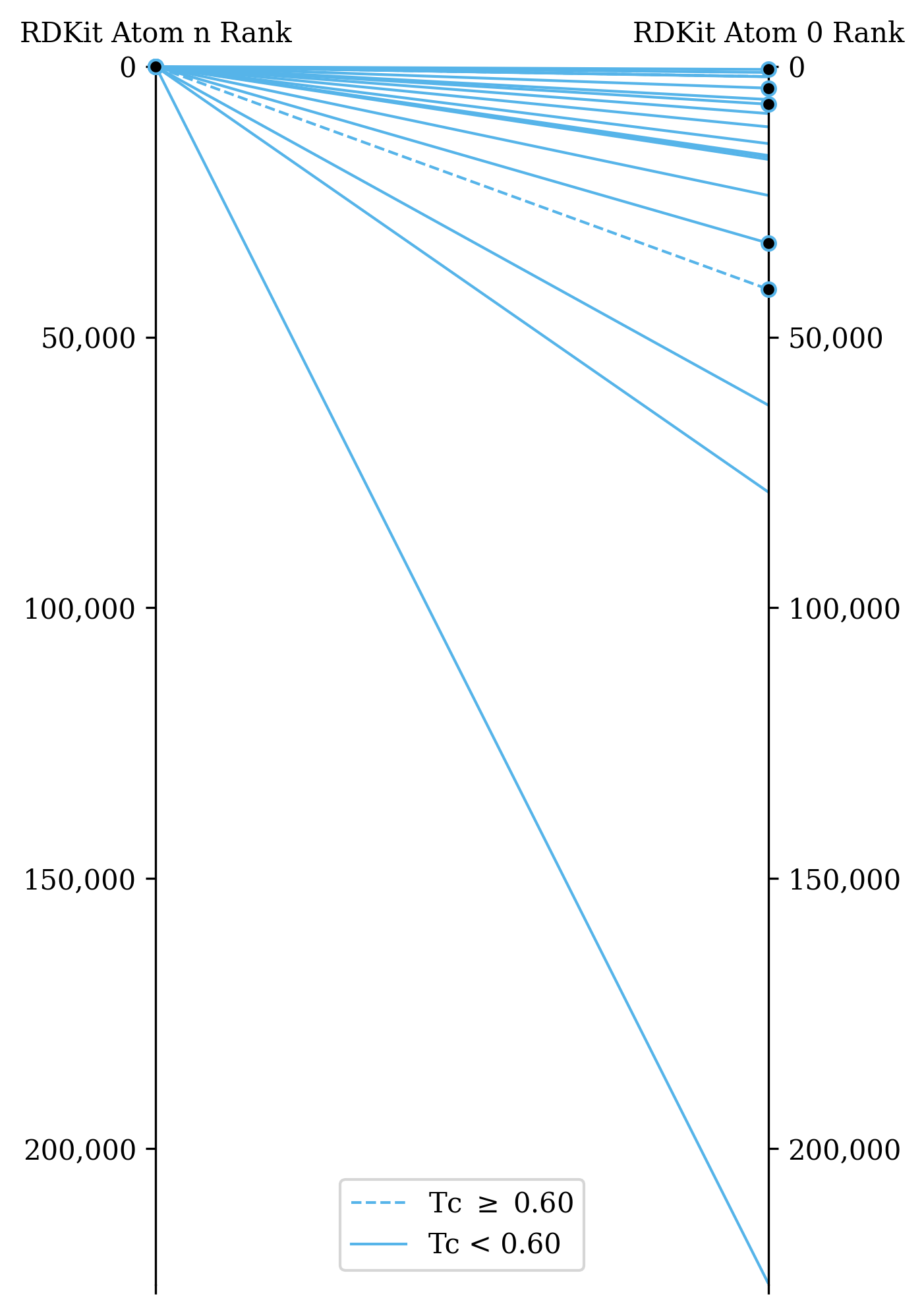}\includegraphics[width=0.4\textwidth]{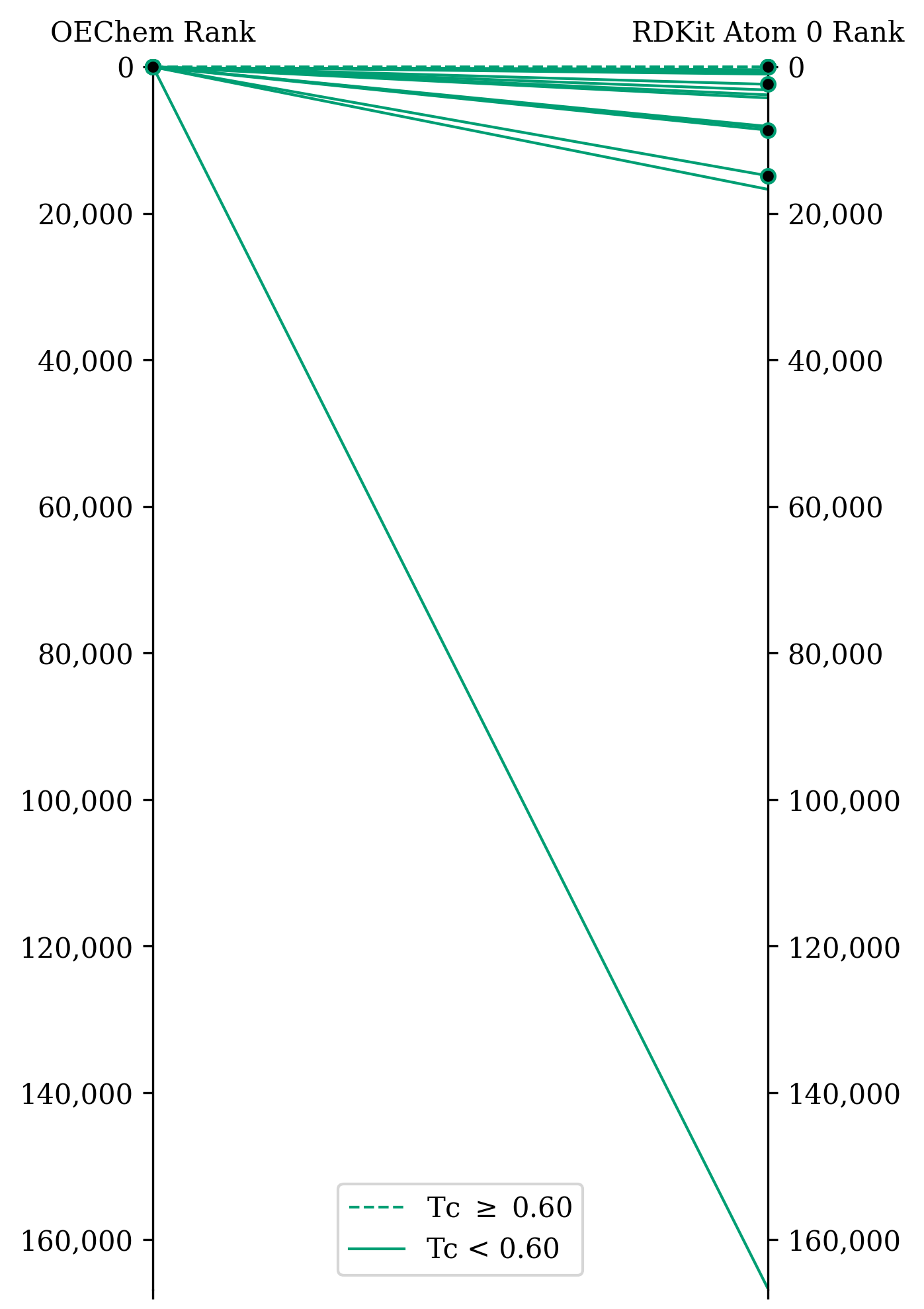}
    \captionsetup{width=4cm}
    \caption{ }
    \label{fig:slope_rank_lsd}
\end{subfigure}\\
\caption{\textbf{The index rank of each alternate canonicalization’s top 20 results for each query compared to the index rank that these same molecules scored in the other canonicalizations’ searches.} \textbf{(a-h).} In order: penicillin G, nirmatrelvir, zidovudine, LSD, fentanyl, acid blue 25 FA, avobenzone, 2-dPAC. Molecules functionally similar to the query indicated by a black dot, as determined by the patent search, and structurally similar to the query (Tc $\geq$ 0.60) indicated by a dashed line. These demonstrate that queries that underwent alternative canonicalization were able to identify functional molecules that would have been impractical to find using the standard canonicalization.}
\label{fig:slope_rank_2}
\end{figure}%
\begin{figure}[h]\ContinuedFloat
\centering
\begin{subfigure}{\textwidth}
    \centering
    \includegraphics[width=0.4\textwidth]{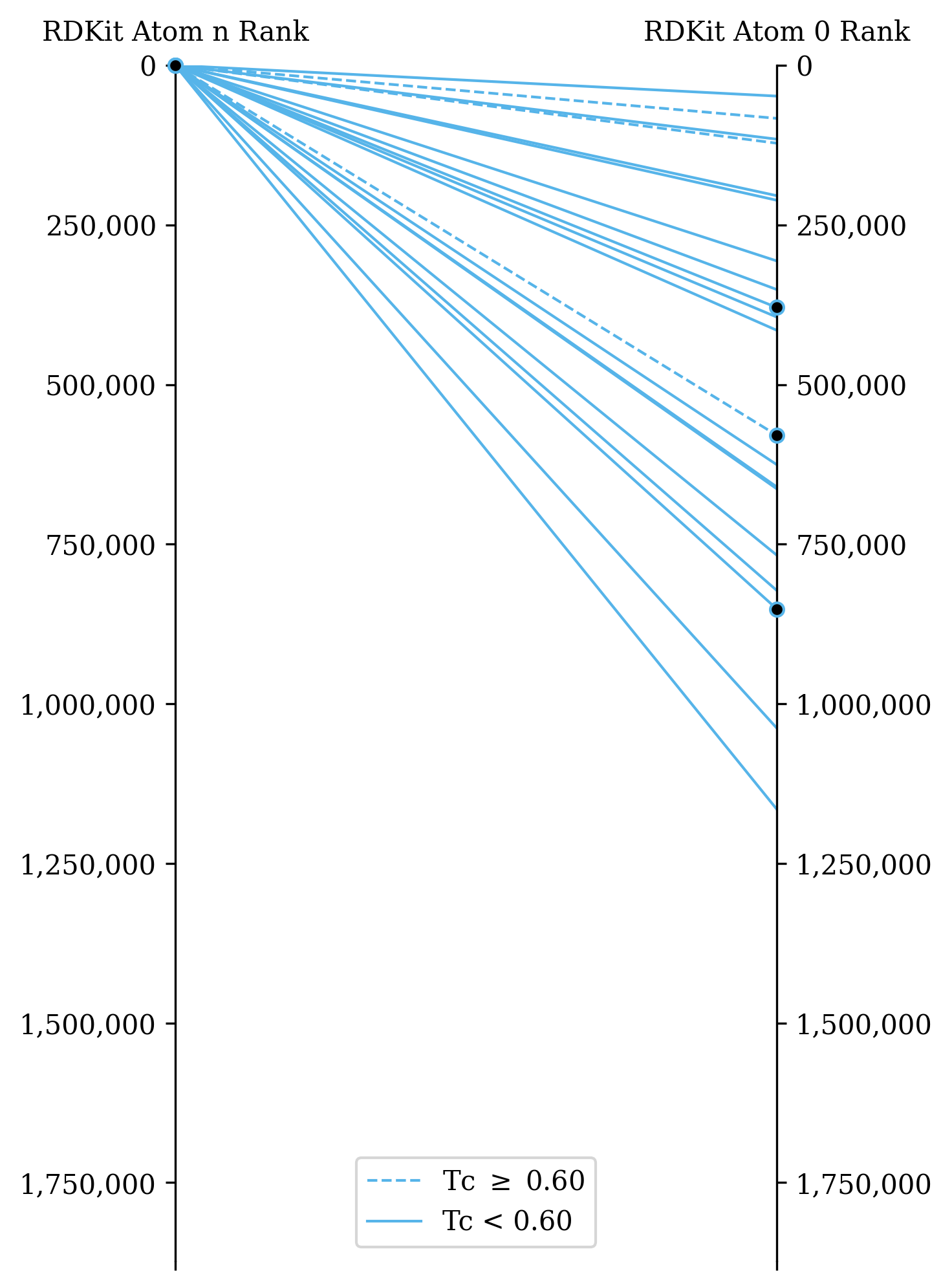}\includegraphics[width=0.4\textwidth]{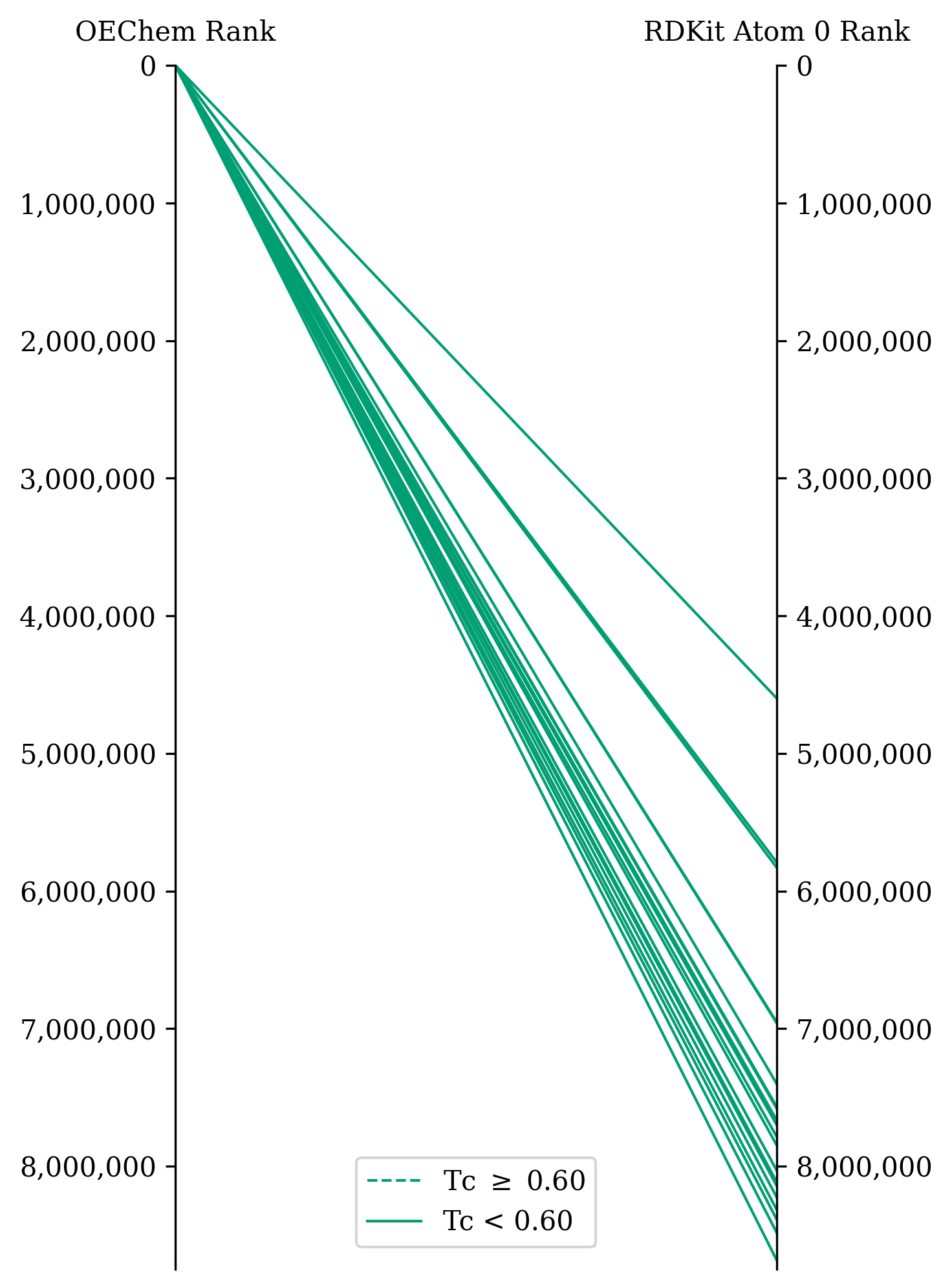}
    \captionsetup{width=4cm}
    \caption{ }
    \label{fig:slope_rank_fentanyl}
\end{subfigure}\\
\begin{subfigure}{\textwidth}
    \centering
    \includegraphics[width=0.4\textwidth]{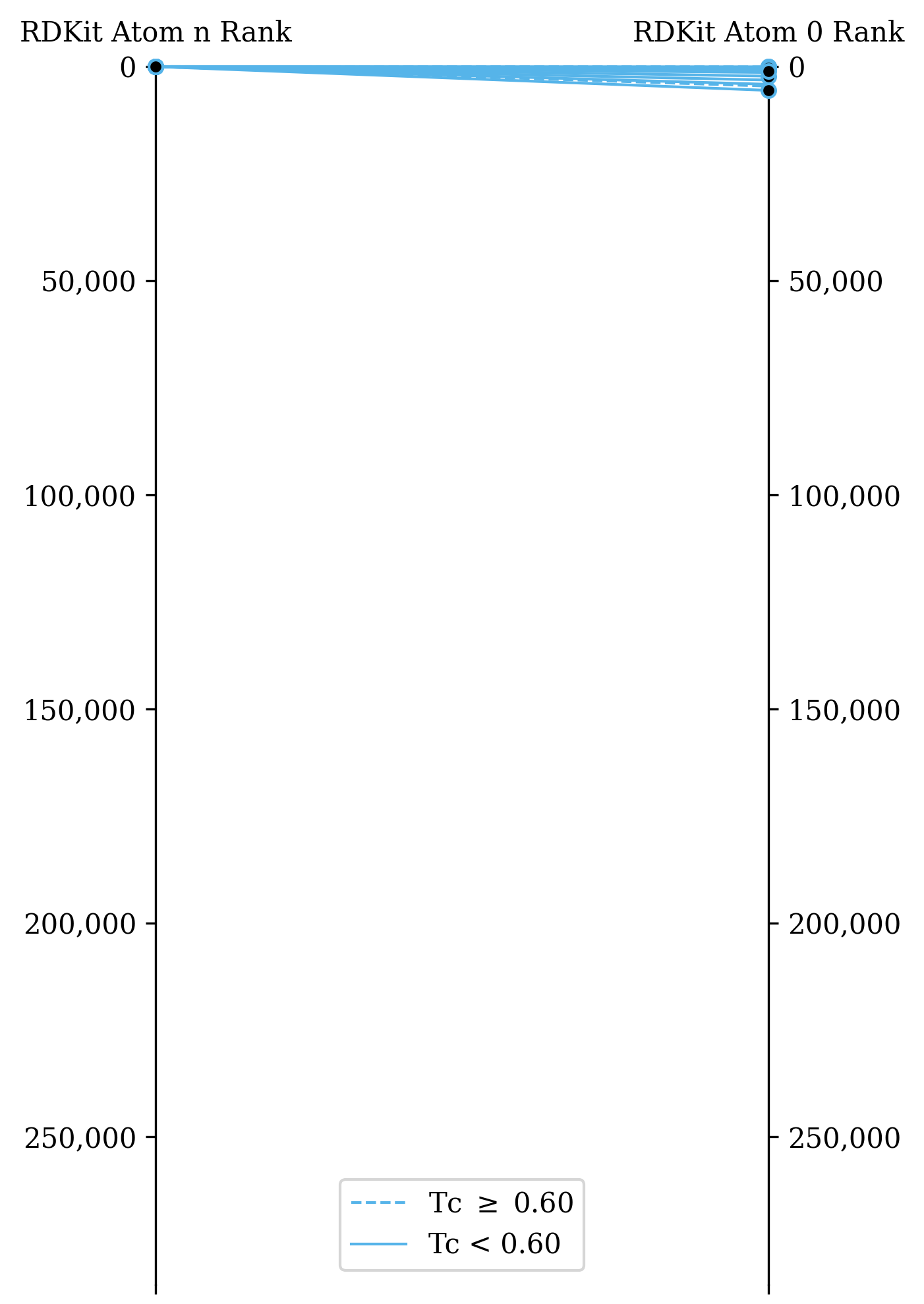}\includegraphics[width=0.4\textwidth]{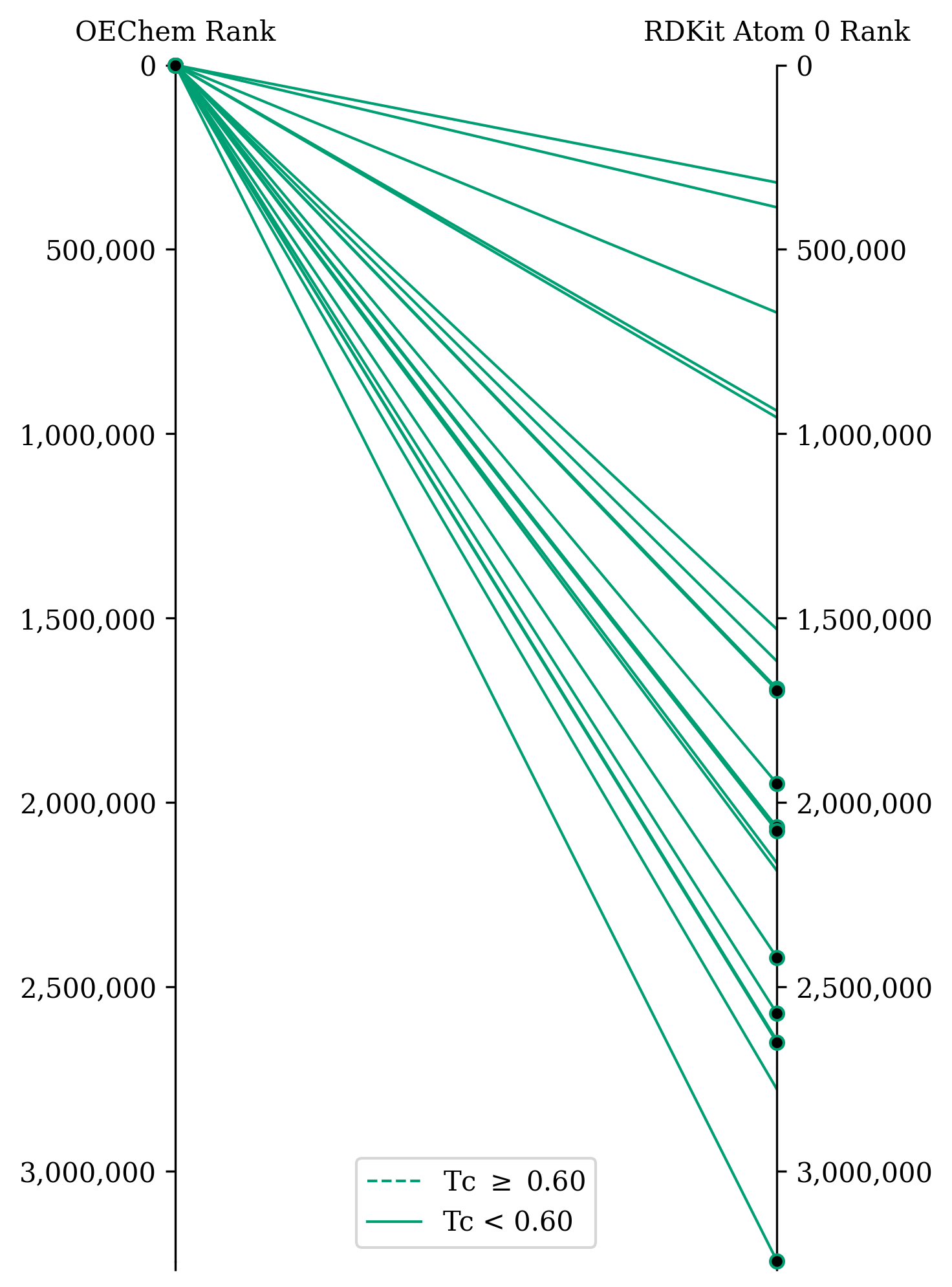}
    \captionsetup{width=4cm}
    \caption{ }
    \label{fig:slope_rank_acid_blue_25_fa}
\end{subfigure}\\
\caption{\textbf{The index rank of each alternate canonicalization’s top 20 results for each query compared to the index rank that these same molecules scored in the other canonicalizations’ searches.} \textbf{(a-h).} In order: penicillin G, nirmatrelvir, zidovudine, LSD, fentanyl, acid blue 25 FA, avobenzone, 2-dPAC. Molecules functionally similar to the query indicated by a black dot, as determined by the patent search, and structurally similar to the query (Tc $\geq$ 0.60) indicated by a dashed line. These demonstrate that queries that underwent alternative canonicalization were able to identify functional molecules that would have been impractical to find using the standard canonicalization.}
\label{fig:slope_rank_3}
\end{figure}%
\begin{figure}[h]\ContinuedFloat
\centering
\begin{subfigure}{\textwidth}
    \centering
    \includegraphics[width=0.4\textwidth]{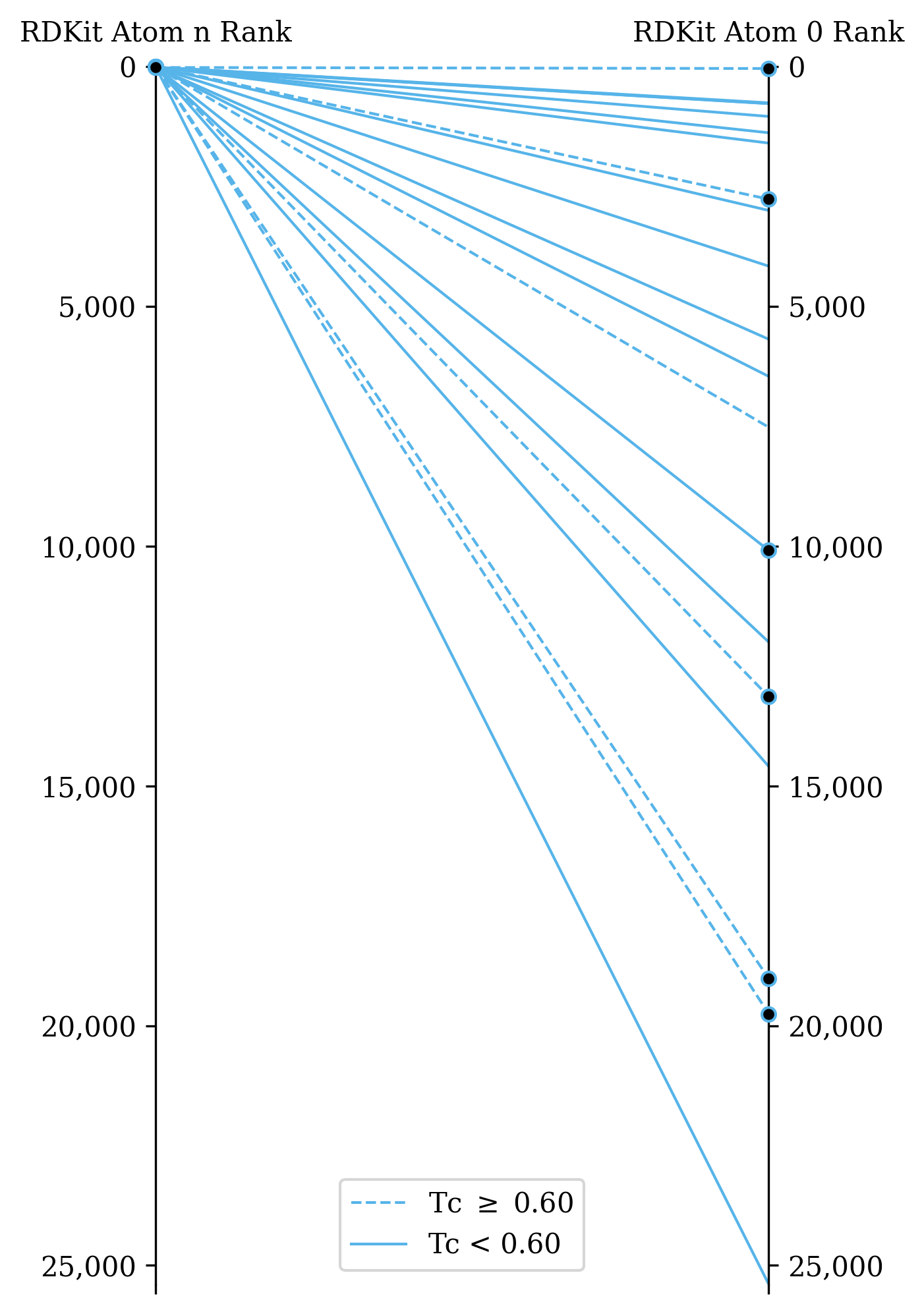}\includegraphics[width=0.4\textwidth]{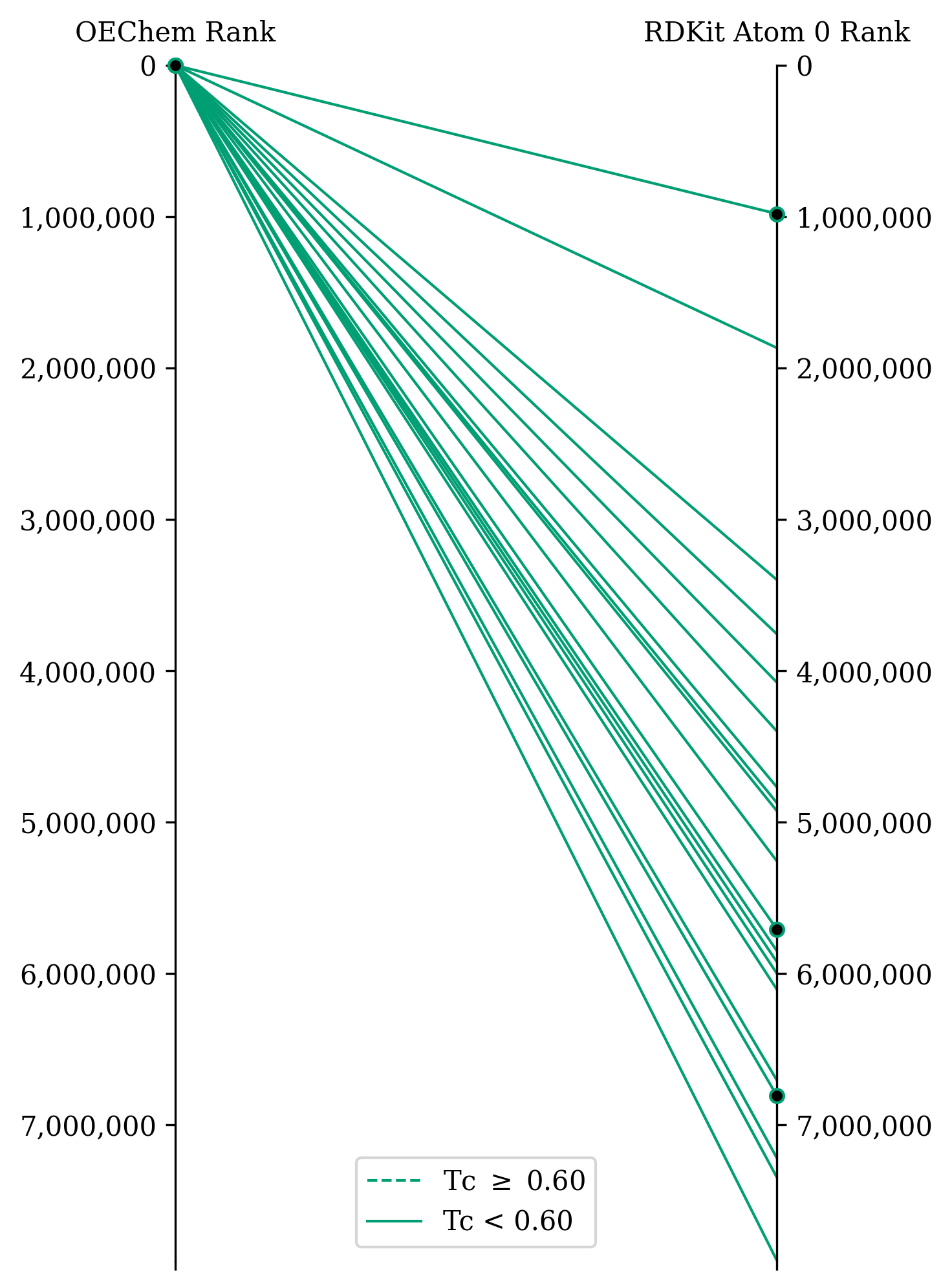}
    \captionsetup{width=4cm}
    \caption{Avobenzone}
    \label{fig:slope_rank_avobenzone}
\end{subfigure}\\
\begin{subfigure}{\textwidth}
    \centering
    \includegraphics[width=0.4\textwidth]{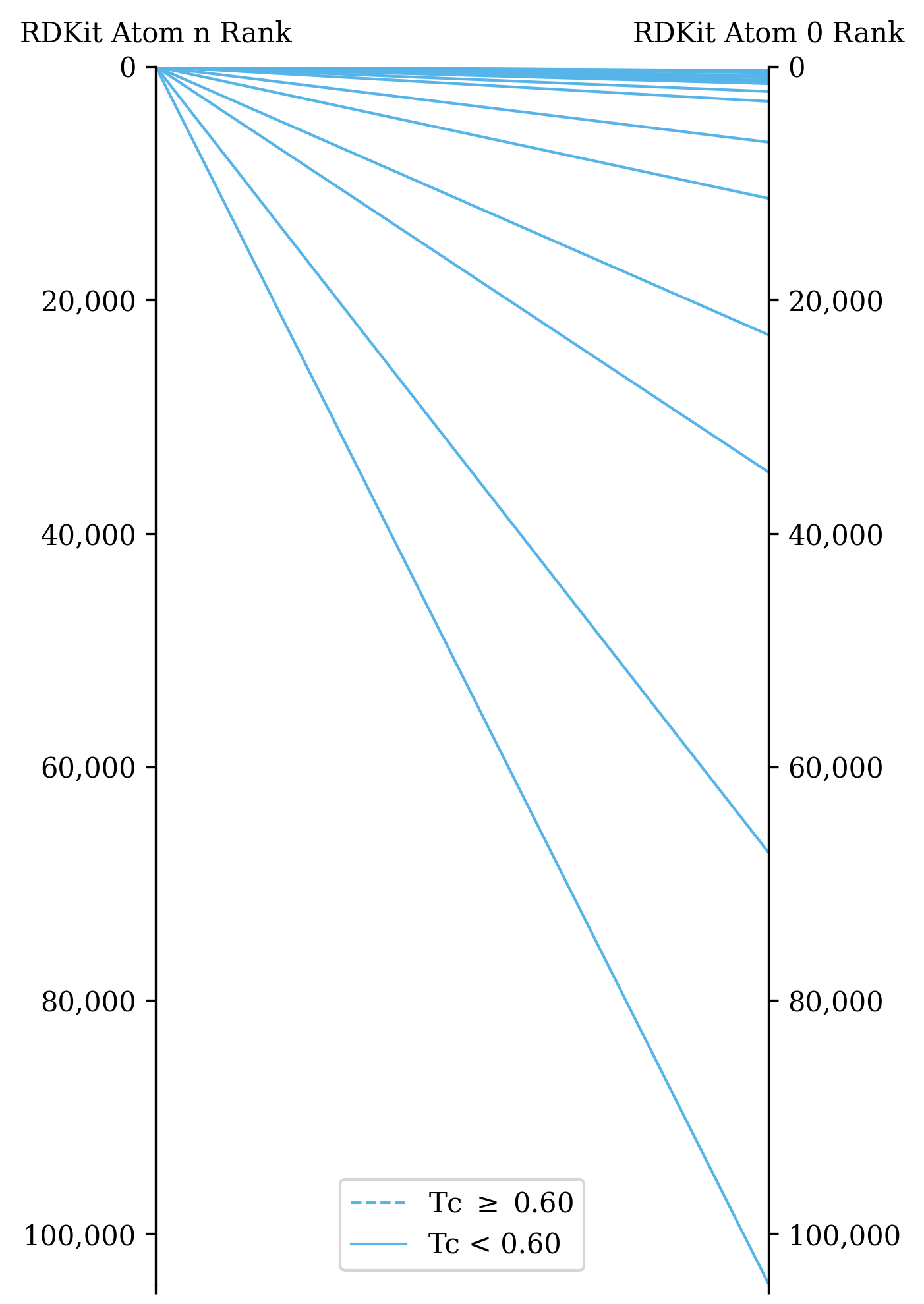}\includegraphics[width=0.4\textwidth]{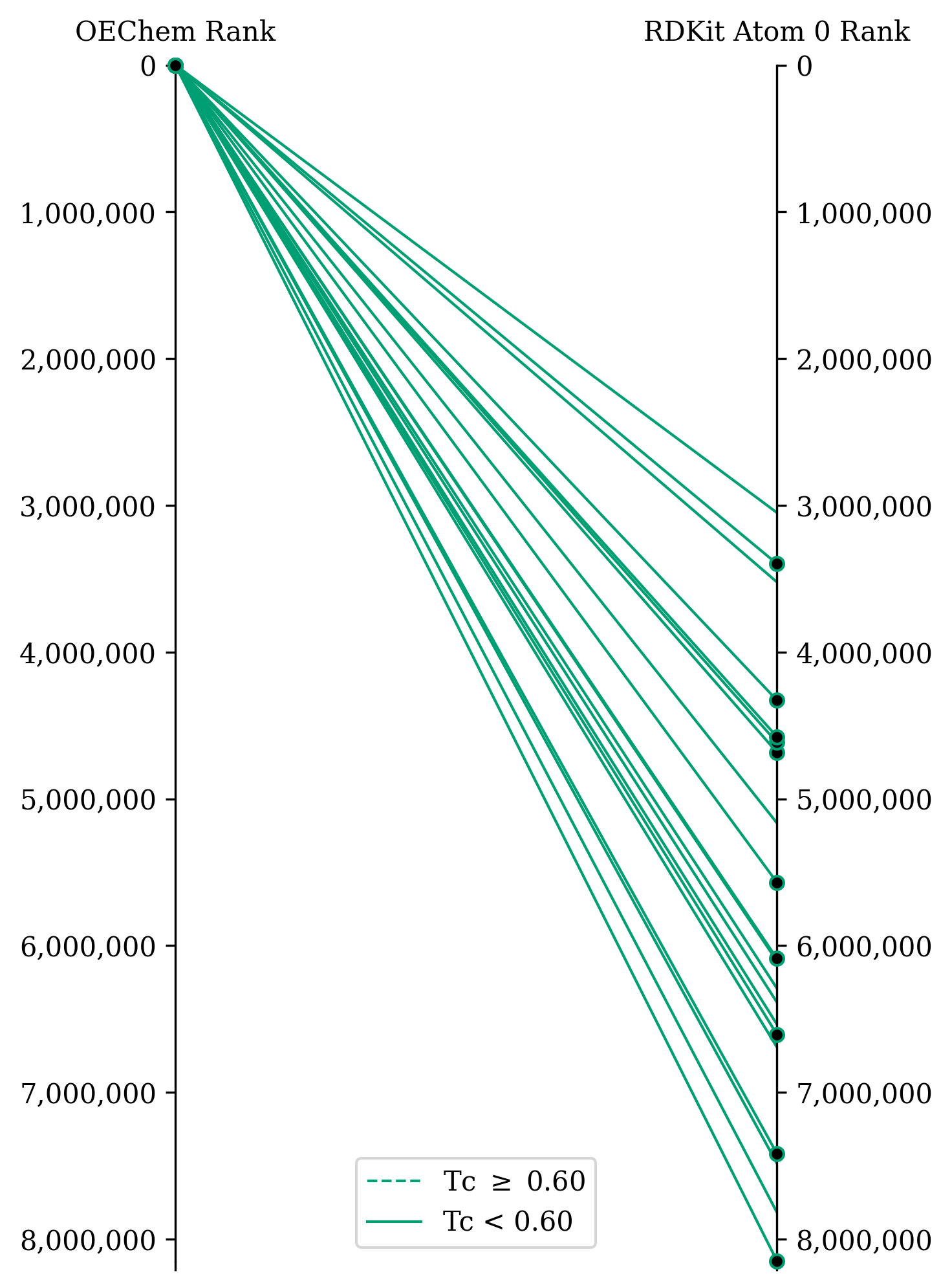}
    \captionsetup{width=4cm}
    \caption{2-dPAC}
    \label{fig:slope_rank_2-diphenylaminocarbazole}
\end{subfigure}\\
\caption{\textbf{The index rank of each alternate canonicalization’s top 20 results for each query compared to the index rank that these same molecules scored in the other canonicalizations’ searches.} \textbf{(a-h).} In order: penicillin G, nirmatrelvir, zidovudine, LSD, fentanyl, acid blue 25 FA, avobenzone, 2-dPAC. Molecules functionally similar to the query indicated by a black dot, as determined by the patent search, and structurally similar to the query (Tc $\geq$ 0.60) indicated by a dashed line. These demonstrate that queries that underwent alternative canonicalization were able to identify functional molecules that would have been impractical to find using the standard canonicalization.}
\label{fig:slope_rank_4}
\end{figure}
\FloatBarrier


\newpage
\begin{figure}[h]
\centering
\begin{subfigure}{\textwidth}
    \centering
    \includegraphics[width=\textwidth]{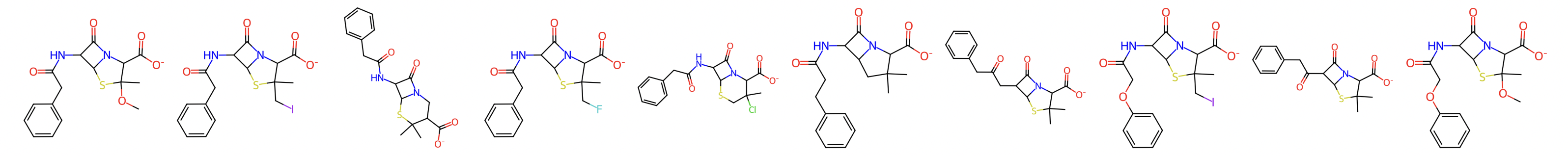}\\
    \captionsetup{width=8cm}
    \caption{Penicillin G RDKit Atom 0 (1-10)}
    \label{fig:mols_pen_rdkit_atom_0_1-10}
\end{subfigure}\\
\begin{subfigure}{\textwidth}
    \centering
    \includegraphics[width=\textwidth]{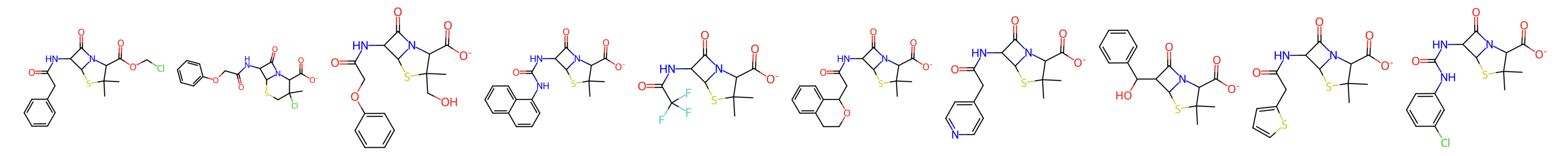}\\
    \captionsetup{width=8cm}
    \caption{Penicillin G RDKit Atom 0 (11-20)}
    \label{fig:mols_pen_rdkit_atom_0_11-20}
\end{subfigure}\\
\begin{subfigure}{\textwidth}
    \centering
    \includegraphics[width=\textwidth]{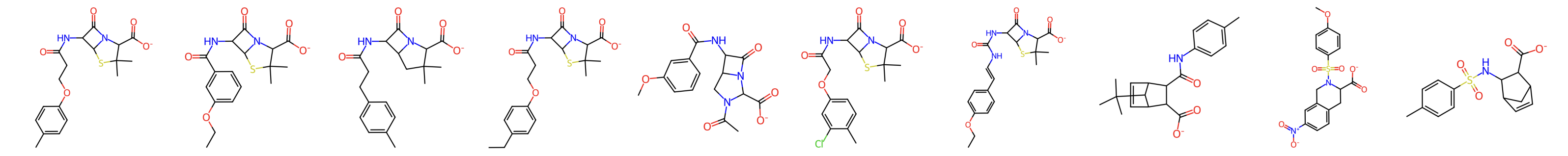}\\
    \captionsetup{width=8cm}
    \caption{Penicillin G RDKit Atom n (1-10)}
    \label{fig:mols_pen_rdkit_atom_n_1-10}
\end{subfigure}\\
\begin{subfigure}{\textwidth}
    \centering
    \includegraphics[width=\textwidth]{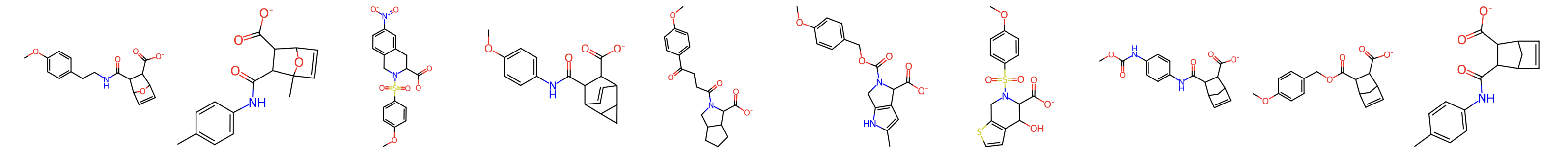}\\
    \captionsetup{width=8cm}
    \caption{Penicillin G RDKit Atom n (11-20)}
    \label{fig:mols_pen_rdkit_atom_n_11-20}
\end{subfigure}\\
\begin{subfigure}{\textwidth}
    \centering
    \includegraphics[width=\textwidth]{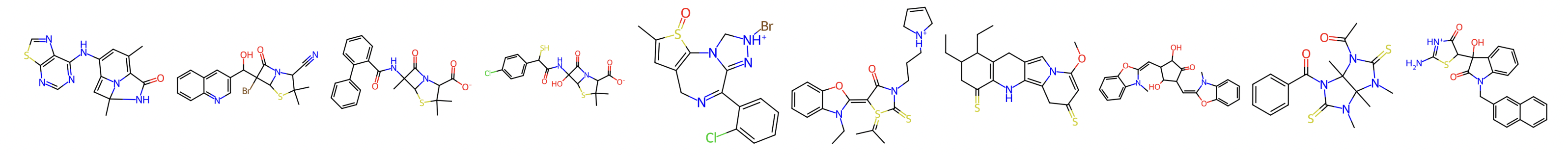}\\
    \captionsetup{width=8cm}
    \caption{Penicillin G OEChem (1-10)}
    \label{fig:mols_pen_oechem_1-10}
\end{subfigure}\\
\begin{subfigure}{\textwidth}
    \centering
    \includegraphics[width=\textwidth]{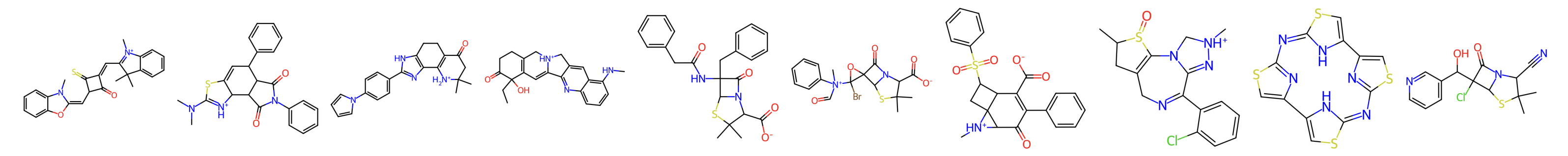}\\
    \captionsetup{width=8cm}
    \caption{Penicillin G OEChem (11-20)}
    \label{fig:mols_pen_oechem_11-20}
\end{subfigure}\\
\begin{subfigure}{\textwidth}
    \centering
    \includegraphics[width=\textwidth]{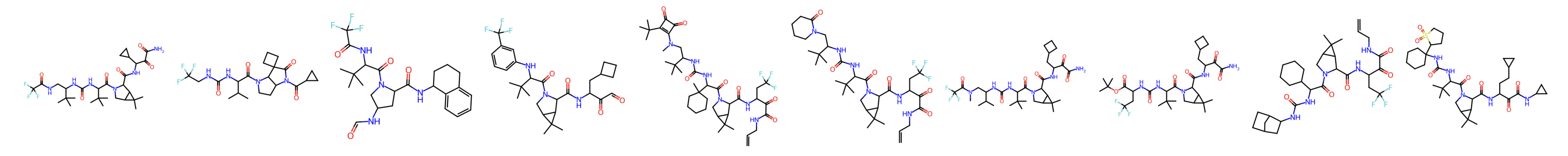}\\
    \captionsetup{width=8cm}
    \caption{Nirmatrelvir RDKit Atom 0 (1-10)}
    \label{fig:mols_pax_rdkit_atom_0_1-10}
\end{subfigure}\\
\begin{subfigure}{\textwidth}
    \centering
    \includegraphics[width=\textwidth]{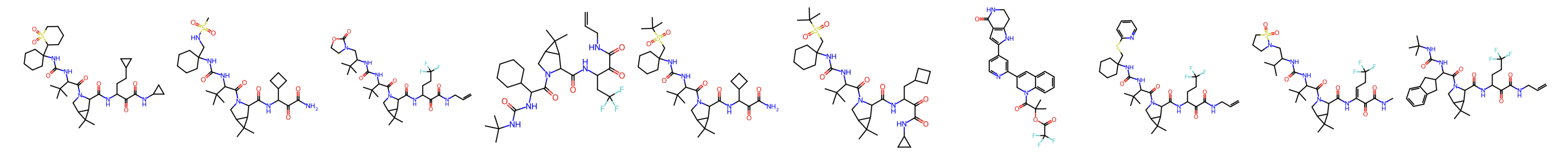}\\
    \captionsetup{width=8cm}
    \caption{Nirmatrelvir RDKit Atom 0 (11-20)}
    \label{fig:mols_pax_rdkit_atom_0_11-20}
\end{subfigure}\\
\begin{subfigure}{\textwidth}
    \centering
    \includegraphics[width=\textwidth]{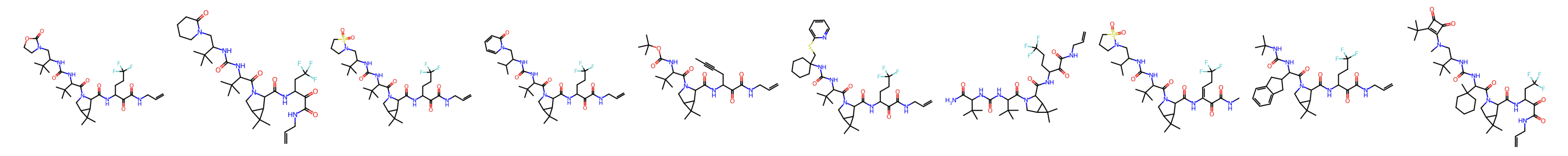}\\
    \captionsetup{width=8cm}
    \caption{Nirmatrelvir RDKit Atom n (1-10)}
    \label{fig:mols_pax_rdkit_atom_n_1-10}
\end{subfigure}\\
\begin{subfigure}{\textwidth}
    \centering
    \includegraphics[width=\textwidth]{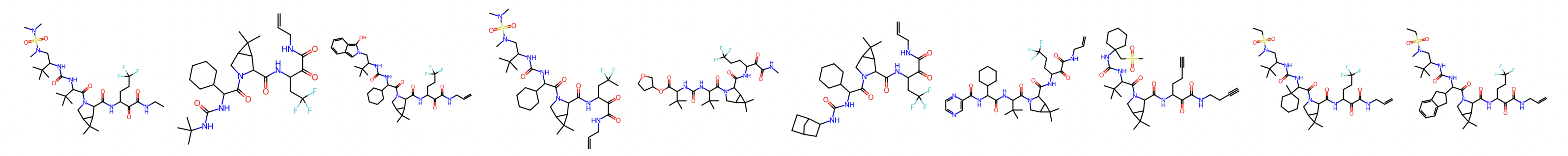}\\
    \captionsetup{width=8cm}
    \caption{Nirmatrelvir RDKit Atom n (11-20)}
    \label{fig:mols_pax_rdkit_atom_n_11-20}
\end{subfigure}\\
\captionsetup{singlelinecheck=false}
\caption{\textbf{Structures of top 20 results for each query.}}
\label{fig:mols_supp_1}
\end{figure}%
\begin{figure}[h]\ContinuedFloat
\centering
\begin{subfigure}{\textwidth}
    \centering
    \includegraphics[width=\textwidth]{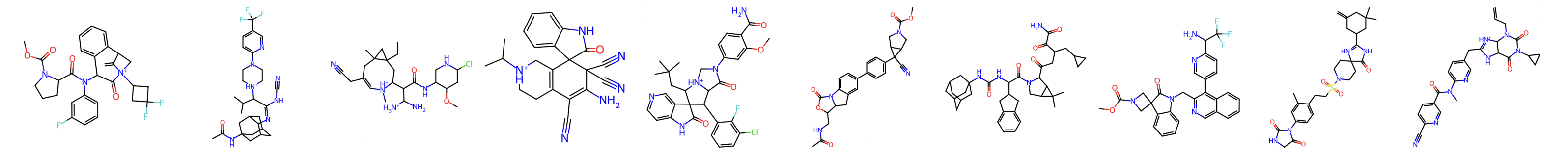}\\
    \captionsetup{width=8cm}
    \caption{Nirmatrelvir OEChem (1-10)}
    \label{fig:mols_pax_oechem_1-10}
\end{subfigure}\\
\begin{subfigure}{\textwidth}
    \centering
    \includegraphics[width=\textwidth]{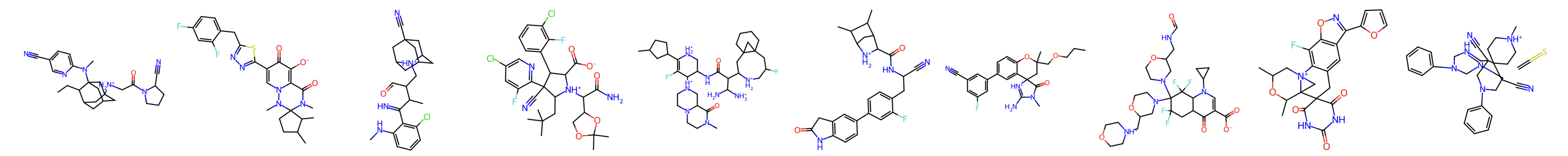}\\
    \captionsetup{width=8cm}
    \caption{Nirmatrelvir OEChem (11-20)}
    \label{fig:mols_pax_oechem_11-20}
\end{subfigure}\\
\begin{subfigure}{\textwidth}
    \centering
    \includegraphics[width=\textwidth]{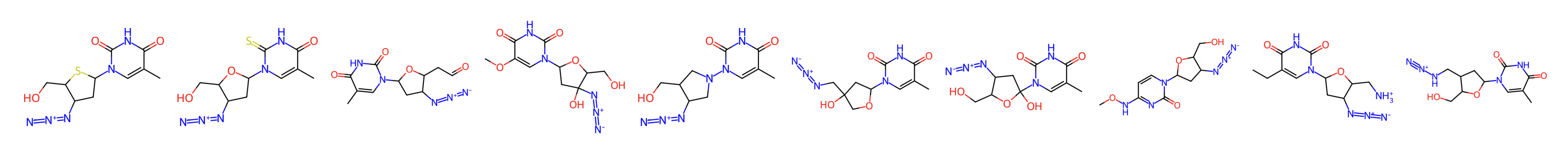}\\
    \captionsetup{width=8cm}
    \caption{Zidovudine RDKit Atom 0 (1-10)}
    \label{fig:mols_azt_rdkit_atom_0_1-10}
\end{subfigure}\\
\begin{subfigure}{\textwidth}
    \centering
    \includegraphics[width=\textwidth]{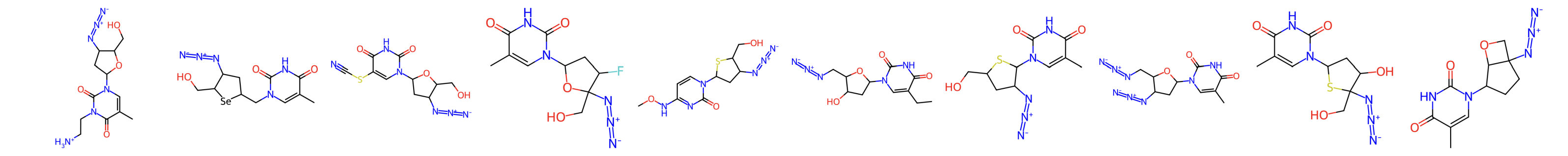}\\
    \captionsetup{width=8cm}
    \caption{Zidovudine RDKit Atom 0 (11-20)}
    \label{fig:mols_azt_rdkit_atom_0_11-20}
\end{subfigure}\\
\begin{subfigure}{\textwidth}
    \centering
    \includegraphics[width=\textwidth]{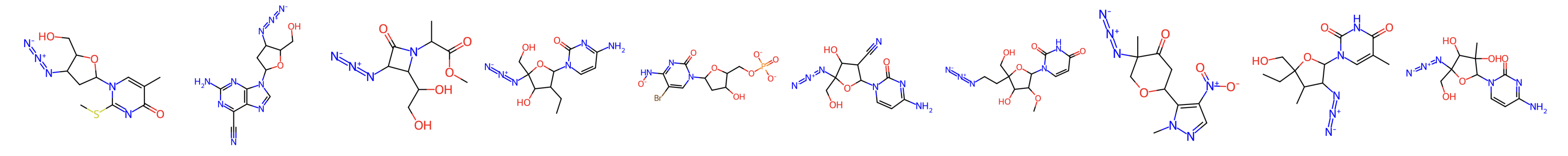}\\
    \captionsetup{width=8cm}
    \caption{Zidovudine RDKit Atom n (1-10)}
    \label{fig:mols_azt_rdkit_atom_n_1-10}
\end{subfigure}\\
\begin{subfigure}{\textwidth}
    \centering
    \includegraphics[width=\textwidth]{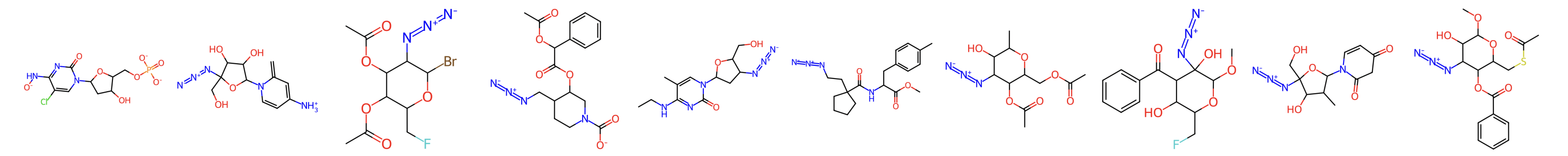}\\
    \captionsetup{width=8cm}
    \caption{Zidovudine RDKit Atom n (11-20)}
    \label{fig:mols_azt_rdkit_atom_n_11-20}
\end{subfigure}\\
\begin{subfigure}{\textwidth}
    \centering
    \includegraphics[width=\textwidth]{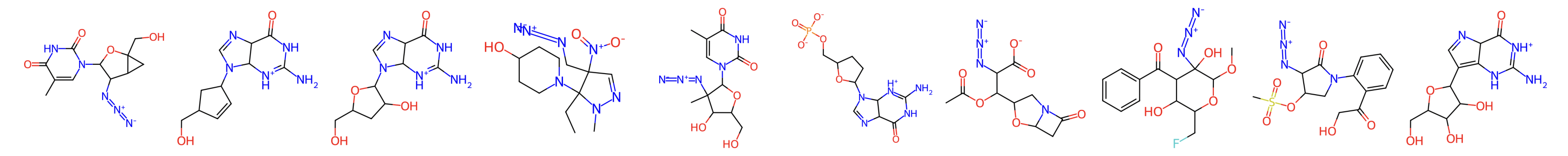}\\
    \captionsetup{width=8cm}
    \caption{Zidovudine OEChem (1-10)}
    \label{fig:mols_azt_oechem_1-10}
\end{subfigure}\\
\begin{subfigure}{\textwidth}
    \centering
    \includegraphics[width=\textwidth]{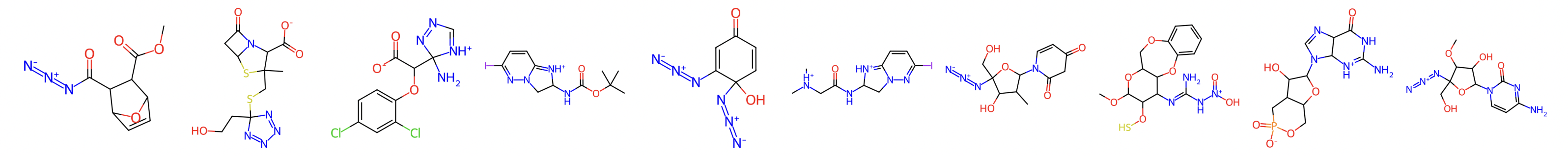}\\
    \captionsetup{width=8cm}
    \caption{Zidovudine OEChem (11-20)}
    \label{fig:mols_azt_oechem_11-20}
\end{subfigure}\\
\begin{subfigure}{\textwidth}
    \centering
    \includegraphics[width=\textwidth]{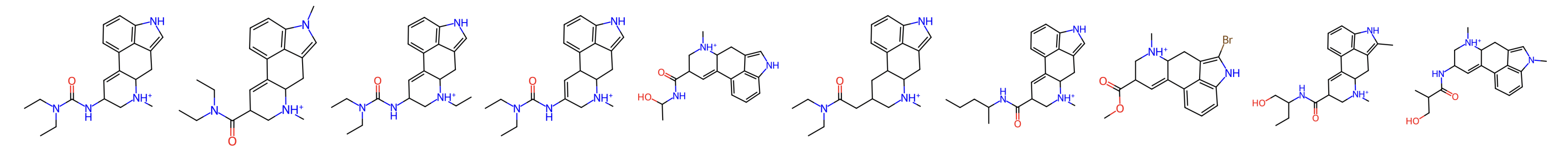}\\
    \captionsetup{width=8cm}
    \caption{LSD RDKit Atom 0 (1-10)}
    \label{fig:mols_lsd_rdkit_atom_0_1-10}
\end{subfigure}\\
\begin{subfigure}{\textwidth}
    \centering
    \includegraphics[width=\textwidth]{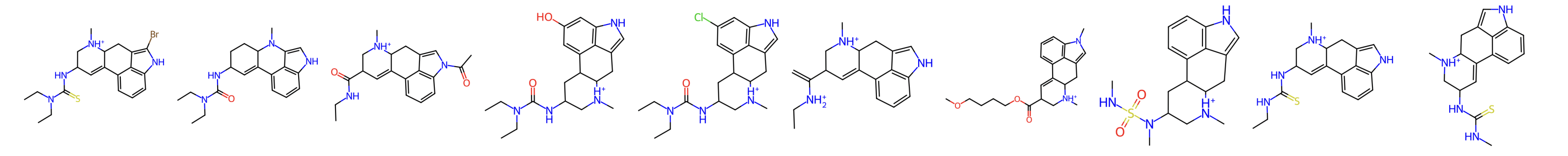}\\
    \captionsetup{width=8cm}
    \caption{LSD RDKit Atom 0 (11-20)}
    \label{fig:mols_lsd_rdkit_atom_0_11-20}
\end{subfigure}\\
\captionsetup{singlelinecheck=false}
\caption{\textbf{Structures of top 20 results for each query.}}
\label{fig:mols_supp_2}
\end{figure}%
\begin{figure}[h]\ContinuedFloat
\centering
\begin{subfigure}{\textwidth}
    \centering
    \includegraphics[width=\textwidth]{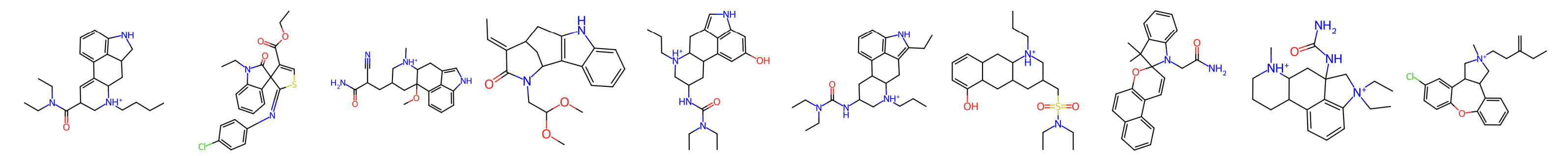}\\
    \captionsetup{width=8cm}
    \caption{LSD RDKit Atom n (1-10)}
    \label{fig:mols_lsd_rdkit_atom_n_1-10}
\end{subfigure}\\
\begin{subfigure}{\textwidth}
    \centering
    \includegraphics[width=\textwidth]{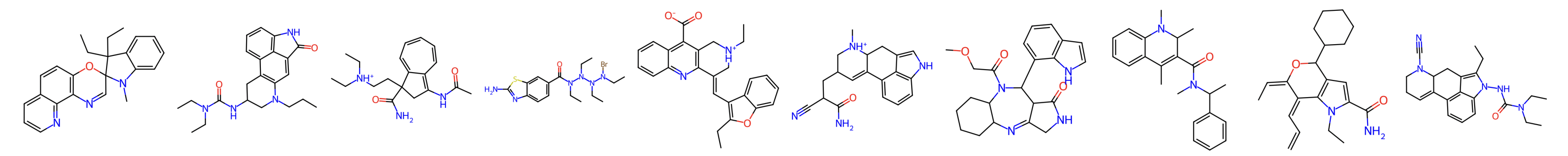}\\
    \captionsetup{width=8cm}
    \caption{LSD RDKit Atom n (11-20)}
    \label{fig:mols_lsd_rdkit_atom_n_11-20}
\end{subfigure}\\
\begin{subfigure}{\textwidth}
    \centering
    \includegraphics[width=\textwidth]{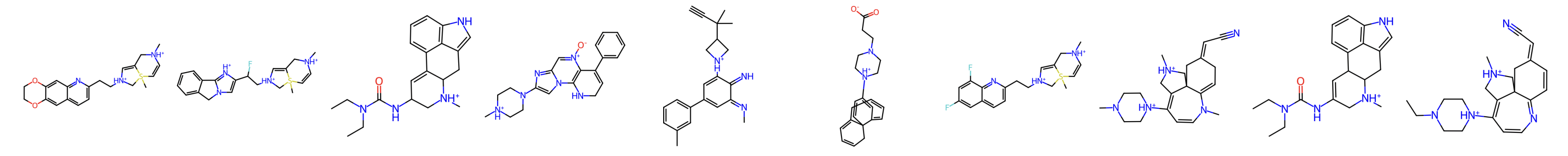}\\
    \captionsetup{width=8cm}
    \caption{LSD OEChem (1-10)}
    \label{fig:mols_lsd_oechem_1-10}
\end{subfigure}\\
\begin{subfigure}{\textwidth}
    \centering
    \includegraphics[width=\textwidth]{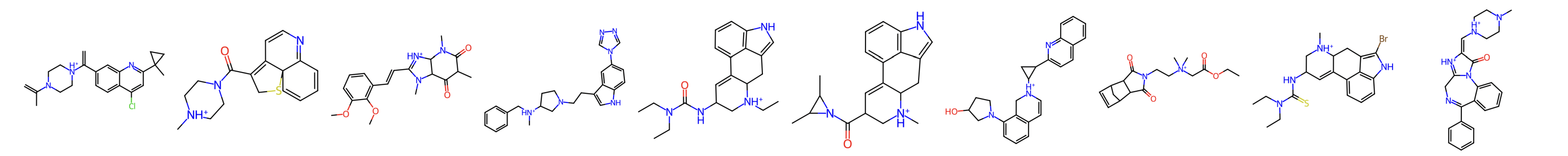}\\
    \captionsetup{width=8cm}
    \caption{LSD OEChem (11-20)}
    \label{fig:mols_lsd_oechem_11-20}
\end{subfigure}\\
\begin{subfigure}{\textwidth}
    \centering
    \includegraphics[width=\textwidth]{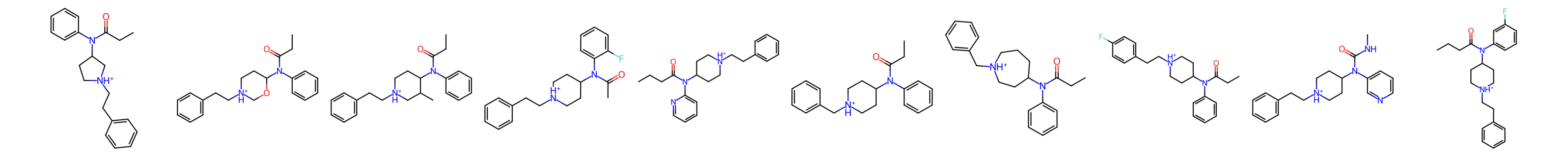}\\
    \captionsetup{width=8cm}
    \caption{Fentanyl RDKit Atom 0 (1-10)}
    \label{fig:mols_fentanyl_rdkit_atom_0_1-10}
\end{subfigure}\\
\begin{subfigure}{\textwidth}
    \centering
    \includegraphics[width=\textwidth]{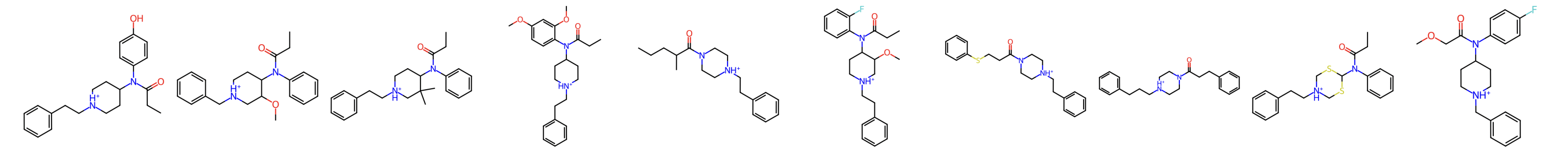}\\
    \captionsetup{width=8cm}
    \caption{Fentanyl RDKit Atom 0 (11-20)}
    \label{fig:mols_fentanyl_rdkit_atom_0_11-20}
\end{subfigure}\\
\begin{subfigure}{\textwidth}
    \centering
    \includegraphics[width=\textwidth]{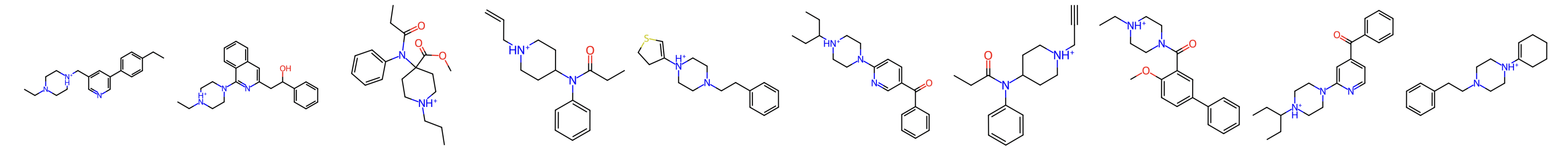}\\
    \captionsetup{width=8cm}
    \caption{Fentanyl RDKit Atom n (1-10)}
    \label{fig:mols_fentanyl_rdkit_atom_n_1-10}
\end{subfigure}\\
\begin{subfigure}{\textwidth}
    \centering
    \includegraphics[width=\textwidth]{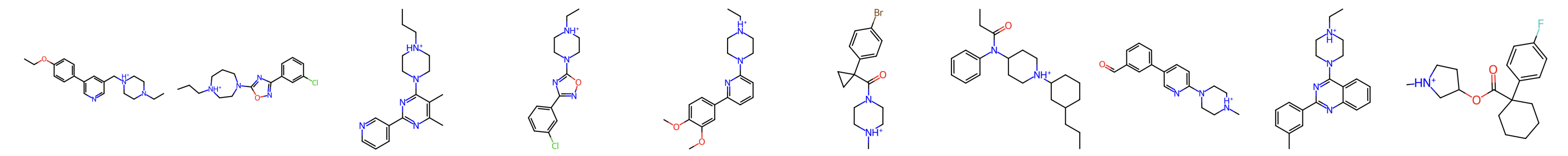}\\
    \captionsetup{width=8cm}
    \caption{Fentanyl RDKit Atom n (11-20)}
    \label{fig:mols_fentanyl_rdkit_atom_n_11-20}
\end{subfigure}\\
\begin{subfigure}{\textwidth}
    \centering
    \includegraphics[width=\textwidth]{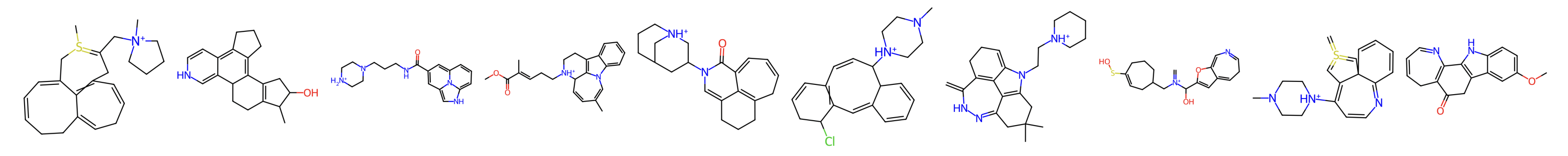}\\
    \captionsetup{width=8cm}
    \caption{Fentanyl OEChem (1-10)}
    \label{fig:mols_fentanyl_oechem_1-10}
\end{subfigure}\\
\begin{subfigure}{\textwidth}
    \centering
    \includegraphics[width=\textwidth]{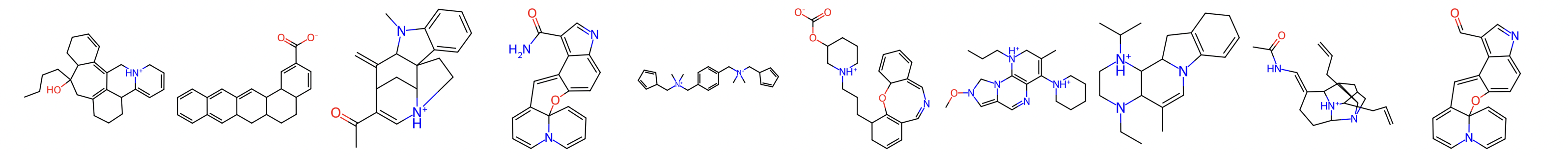}\\
    \captionsetup{width=8cm}
    \caption{Fentanyl OEChem (11-20)}
    \label{fig:mols_fentanyl_oechem_11-20}
\end{subfigure}\\
\captionsetup{singlelinecheck=false}
\caption{\textbf{Structures of top 20 results for each query.}}
\label{fig:mols_supp_3}
\end{figure}%
\begin{figure}[h]\ContinuedFloat
\centering
\begin{subfigure}{\textwidth}
    \centering
    \includegraphics[width=\textwidth]{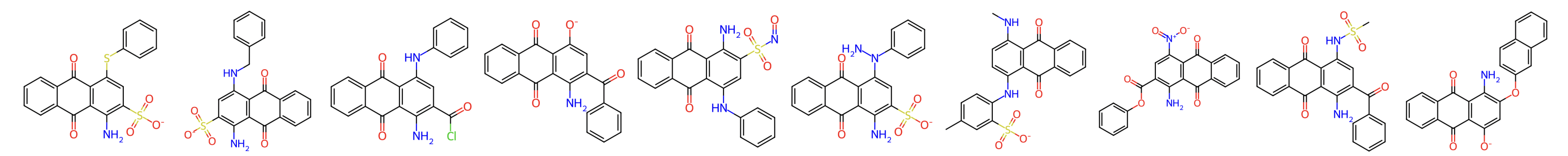}\\
    \captionsetup{width=8cm}
    \caption{Acid Blue 25 FA RDKit Atom 0 (1-10)}
    \label{fig:mols_acid_blue_rdkit_atom_0_1-10}
\end{subfigure}\\
\begin{subfigure}{\textwidth}
    \centering
    \includegraphics[width=\textwidth]{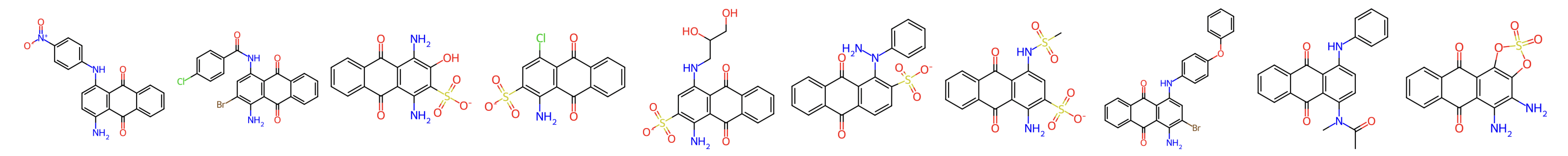}\\
    \captionsetup{width=8cm}
    \caption{Acid Blue 25 FA RDKit Atom 0 (11-20)}
    \label{fig:mols_acid_blue_rdkit_atom_0_11-20}
\end{subfigure}\\
\begin{subfigure}{\textwidth}
    \centering
    \includegraphics[width=\textwidth]{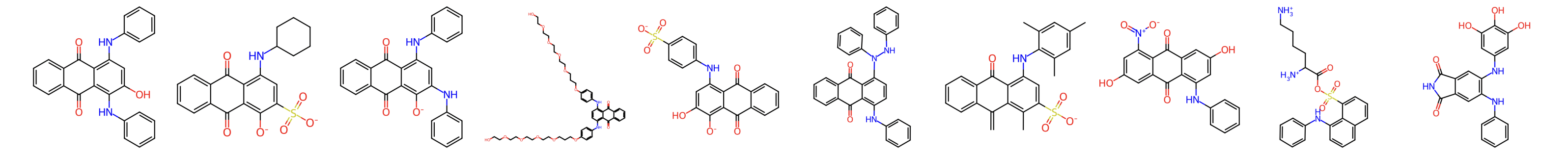}\\
    \captionsetup{width=8cm}
    \caption{Acid Blue 25 FA RDKit Atom n (1-10)}
    \label{fig:mols_acid_blue_rdkit_atom_n_1-10}
\end{subfigure}\\
\begin{subfigure}{\textwidth}
    \centering
    \includegraphics[width=\textwidth]{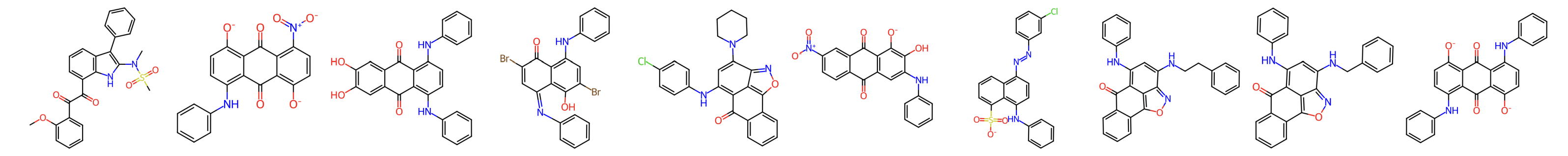}\\
    \captionsetup{width=8cm}
    \caption{Acid Blue 25 FA RDKit Atom n (11-20)}
    \label{fig:mols_acid_blue_rdkit_atom_n_11-20}
\end{subfigure}\\
\begin{subfigure}{\textwidth}
    \centering
    \includegraphics[width=\textwidth]{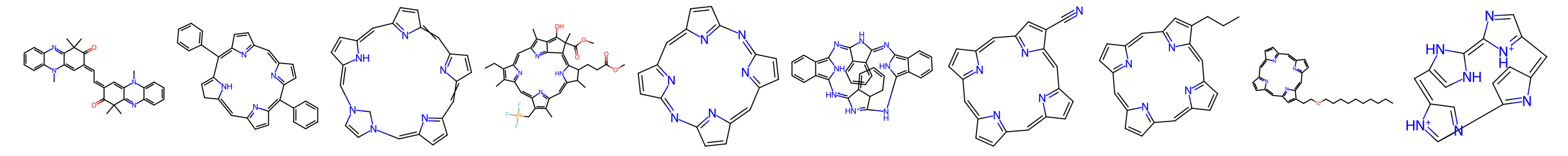}\\
    \captionsetup{width=8cm}
    \caption{Acid Blue 25 FA OEChem (1-10)}
    \label{fig:mols_acid_blue_oechem_1-10}
\end{subfigure}\\
\begin{subfigure}{\textwidth}
    \centering
    \includegraphics[width=\textwidth]{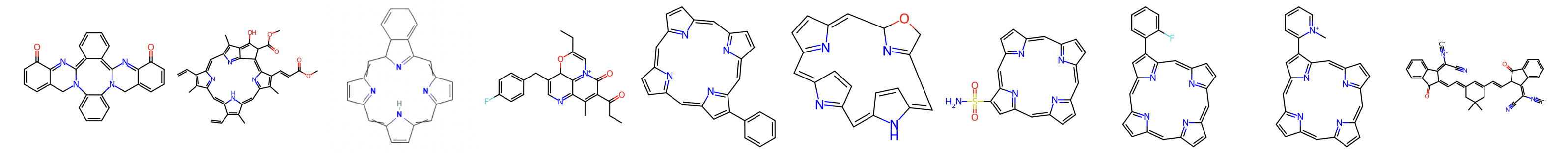}\\
    \captionsetup{width=8cm}
    \caption{Acid Blue 25 FA OEChem (11-20)}
    \label{fig:mols_acid_blue_oechem_11-20}
\end{subfigure}\\
\begin{subfigure}{\textwidth}
    \centering
    \includegraphics[width=\textwidth]{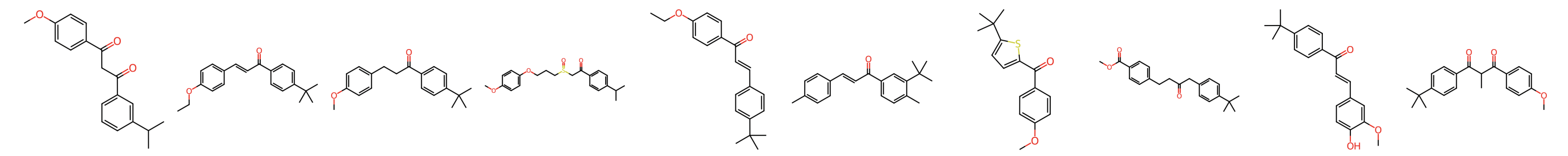}\\
    \captionsetup{width=8cm}
    \caption{Avobenzone RDKit Atom 0 (1-10)}
    \label{fig:mols_avobenzone_rdkit_atom_0_1-10}
\end{subfigure}\\
\begin{subfigure}{\textwidth}
    \centering
    \includegraphics[width=\textwidth]{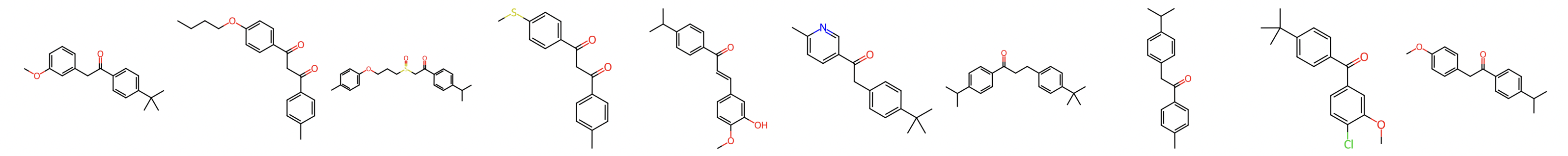}\\
    \captionsetup{width=8cm}
    \caption{Avobenzone RDKit Atom 0 (11-20)}
    \label{fig:mols_avobenzone_rdkit_atom_0_11-20}
\end{subfigure}\\
\begin{subfigure}{\textwidth}
    \centering
    \includegraphics[width=\textwidth]{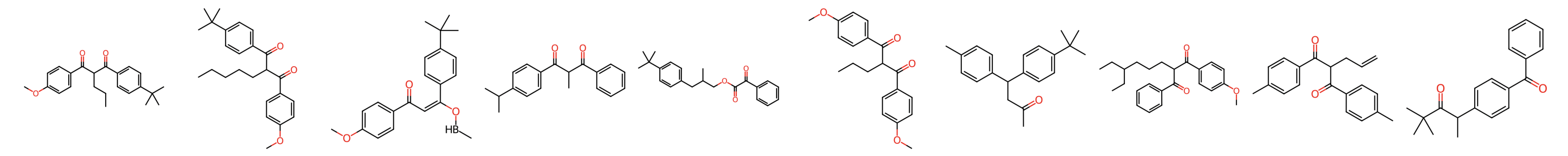}\\
    \captionsetup{width=8cm}
    \caption{Avobenzone RDKit Atom n (1-10)}
    \label{fig:mols_avobenzone_rdkit_atom_n_1-10}
\end{subfigure}\\
\begin{subfigure}{\textwidth}
    \centering
    \includegraphics[width=\textwidth]{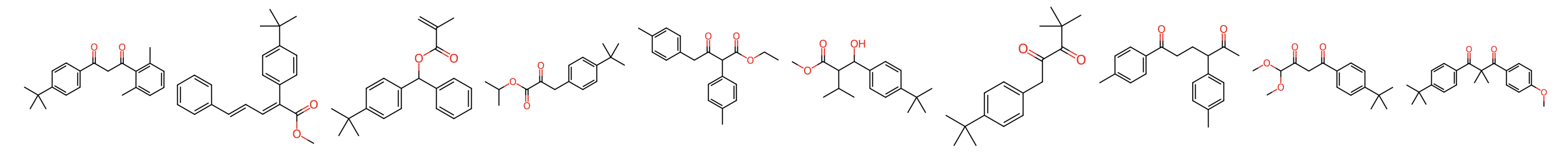}\\
    \captionsetup{width=8cm}
    \caption{Avobenzone RDKit Atom n (11-20)}
    \label{fig:mols_avobenzone_rdkit_atom_n_11-20}
\end{subfigure}\\
\captionsetup{singlelinecheck=false}
\caption{\textbf{Structures of top 20 results for each query.}}
\label{fig:mols_supp_4}
\end{figure}%
\begin{figure}[h]\ContinuedFloat
\centering
\begin{subfigure}{\textwidth}
    \centering
    \includegraphics[width=\textwidth]{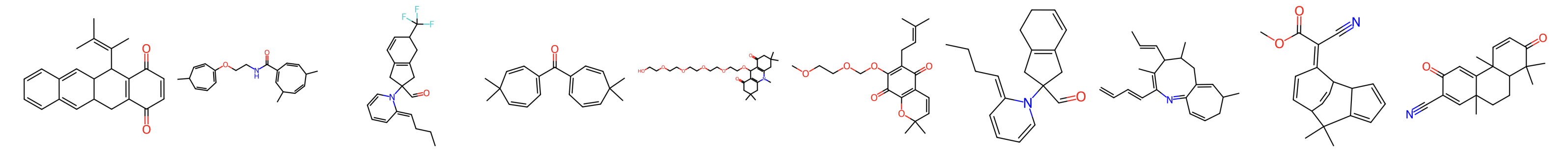}\\
    \captionsetup{width=8cm}
    \caption{Avobenzone OEChem (1-10)}
    \label{fig:mols_avobenzone_oechem_1-10}
\end{subfigure}\\
\begin{subfigure}{\textwidth}
    \centering
    \includegraphics[width=\textwidth]{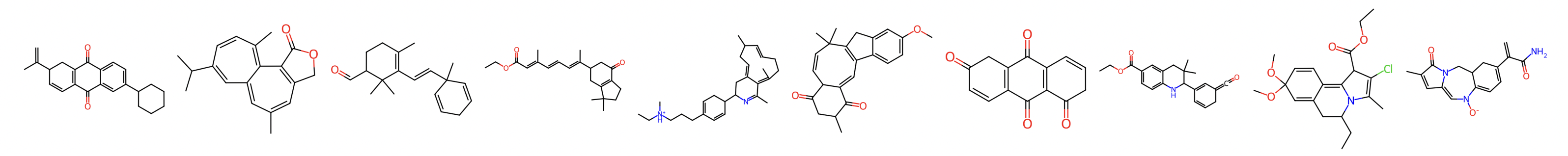}\\
    \captionsetup{width=8cm}
    \caption{Avobenzone OEChem (11-20)}
    \label{fig:mols_avobenzone_oechem_11-20}
\end{subfigure}\\
\begin{subfigure}{\textwidth}
    \centering
    \includegraphics[width=\textwidth]{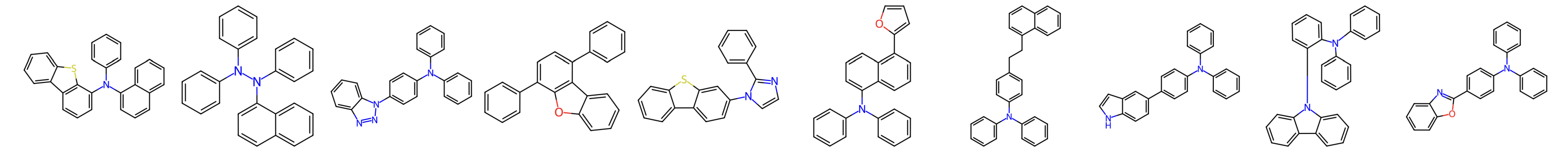}\\
    \captionsetup{width=8cm}
    \caption{2-dPAC RDKit Atom 0 (1-10)}
    \label{fig:mols_2dpac_rdkit_atom_0_1-10}
\end{subfigure}\\
\begin{subfigure}{\textwidth}
    \centering
    \includegraphics[width=\textwidth]{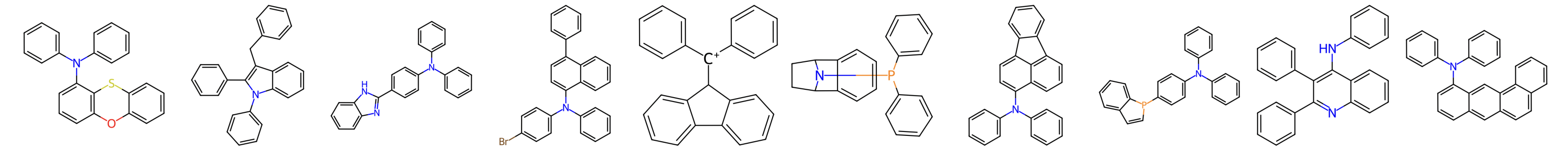}\\
    \captionsetup{width=8cm}
    \caption{2-dPAC RDKit Atom 0 (11-20)}
    \label{fig:mols_2dpac_rdkit_atom_0_11-20}
\end{subfigure}\\
\begin{subfigure}{\textwidth}
    \centering
    \includegraphics[width=\textwidth]{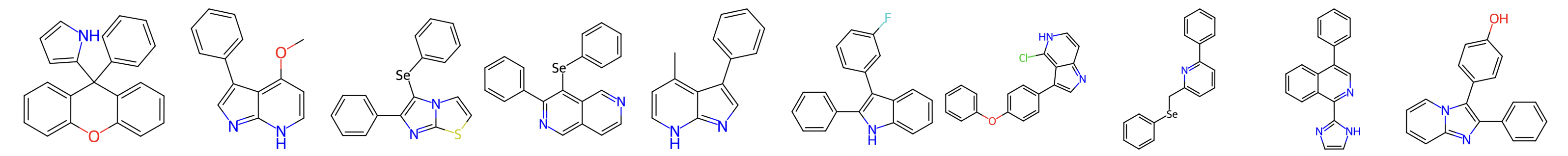}\\
    \captionsetup{width=8cm}
    \caption{2-dPAC RDKit Atom n (1-10)}
    \label{fig:mols_2dpac_rdkit_atom_n_1-10}
\end{subfigure}\\
\begin{subfigure}{\textwidth}
    \centering
    \includegraphics[width=\textwidth]{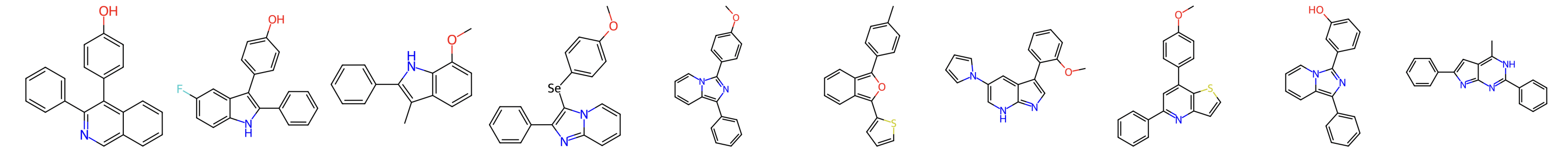}\\
    \captionsetup{width=8cm}
    \caption{2-dPAC RDKit Atom n (11-20)}
    \label{fig:mols_2dpac_rdkit_atom_n_11-20}
\end{subfigure}\\
\begin{subfigure}{\textwidth}
    \centering
    \includegraphics[width=\textwidth]{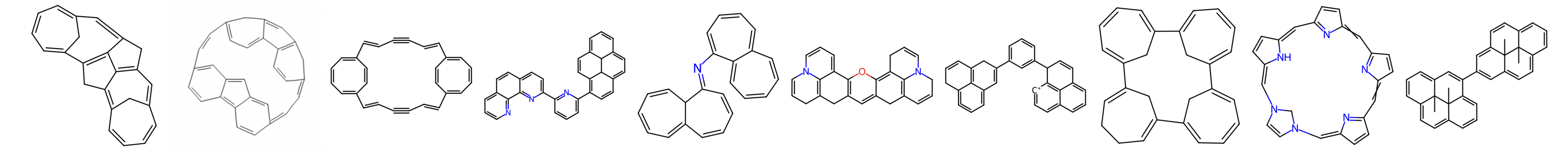}\\
    \captionsetup{width=8cm}
    \caption{2-dPAC OEChem (1-10)}
    \label{fig:mols_2dpac_oechem_1-10}
\end{subfigure}\\
\begin{subfigure}{\textwidth}
    \centering
    \includegraphics[width=\textwidth]{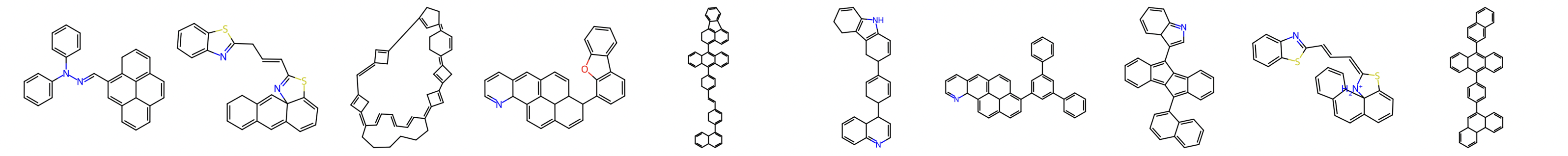}\\
    \captionsetup{width=8cm}
    \caption{2-dPAC OEChem (11-20)}
    \label{fig:mols_2dpac_oechem_11-20}
\end{subfigure}\\
\captionsetup{singlelinecheck=false}
\caption{\textbf{Structures of top 20 results for each query.}}
\label{fig:mols_supp_5}
\end{figure}%
\FloatBarrier

\begin{landscape}
\begin{longtable}{llllllllllll}
    \captionsetup{justification=raggedright}
    
    \caption{\textbf{CheSS Top Results Information.} Includes query, canonicalization, search rank, PubChem CID, Patent ID/DOI, functional descriptor, categorized drug/dye-likeness based on functionality, same functionality categorization, fingerprint Tanimoto coefficient between query \& result, categorized Structurally Distinct Functional Analogue (SDFA), categorized Non-Derivative Functional Analogue (NDFA).} \\
    \label{tab:patents_list}
    \centering
        \textbf{Query} & \textbf{Canon.} & \textbf{Rank} & \textbf{CID} & \textbf{Patent ID / DOI} & \textbf{Functional Descriptor} & \textbf{Drug} & \textbf{Dye} & \textbf{Sim. Fun.} & \textbf{Tc} & \textbf{SDFA} & \textbf{NDFA}\\
        \hline
        \endhead
    \hline
        Pen. & RDKit 0 & 1 & 18643908 & US-4996313-A & Beta-Lactam Antibiotic & 1 & 0 & 1 & 0.78 & 0 & 0 \\ 
        Pen. & RDKit 0 & 2 & 70630416 & US-4171304-A & Beta-Lactam Antibiotic & 1 & 0 & 1 & 0.92 & 0 & 0 \\ 
        Pen. & RDKit 0 & 3 & 23616098 & n/a & n/a & 0 & 0 & 0 & 0.72 & 0 & 0 \\ 
        Pen. & RDKit 0 & 4 & 70367018 & n/a & n/a & 0 & 0 & 0 & 0.92 & 0 & 0 \\ 
        Pen. & RDKit 0 & 5 & 21500059 & US-4010156-A & Beta-Lactam Antibiotic & 1 & 0 & 1 & 0.75 & 0 & 0 \\ 
        Pen. & RDKit 0 & 6 & 57624296 & US-2009081766-A1 & Beta-Lactam Antibiotic & 1 & 0 & 1 & 0.49 & 1 & 0 \\ 
        Pen. & RDKit 0 & 7 & 58112358 & US20190315770A1 & Beta-Lactam Antibiotic & 1 & 0 & 1 & 0.7 & 0 & 0 \\ 
        Pen. & RDKit 0 & 8 & 20389584 & US-4171304-A & Beta-Lactam Antibiotic & 1 & 0 & 1 & 0.83 & 0 & 0 \\ 
        Pen. & RDKit 0 & 9 & 56988098 & US-4461726-A & Beta-Lactam Antibiotic & 1 & 0 & 1 & 0.68 & 0 & 0 \\ 
        Pen. & RDKit 0 & 10 & 102119253 & n/a & n/a & 0 & 0 & 0 & 0.71 & 0 & 0 \\ 
        Pen. & RDKit 0 & 11 & 13277548 & US-4942229-A & Beta-Lactam Antibiotic & 1 & 0 & 1 & 0.94 & 0 & 0 \\ 
        Pen. & RDKit 0 & 12 & 152766084 & US-3954732-A & Beta-Lactam Antibiotic & 1 & 0 & 1 & 0.69 & 0 & 0 \\ 
        Pen. & RDKit 0 & 13 & 22874634 & US-5656623-A & Beta-Lactam Antibiotic & 1 & 0 & 1 & 0.83 & 0 & 0 \\ 
        Pen. & RDKit 0 & 14 & 4393031 & n/a & n/a & 0 & 0 & 0 & 0.81 & 0 & 0 \\ 
        Pen. & RDKit 0 & 15 & 56628131 & US-4847247-A & Elastase Inhibitor & 1 & 0 & 0 & 0.9 & 0 & 0 \\ 
        Pen. & RDKit 0 & 16 & 11741115 & n/a & n/a & 0 & 0 & 0 & 0.8 & 0 & 0 \\ 
        Pen. & RDKit 0 & 17 & 54281720 & US4077967A & Beta-Lactam Antibiotic & 1 & 0 & 1 & 0.96 & 0 & 0 \\ 
        Pen. & RDKit 0 & 18 & 70547535 & EP-0005889-B1 & Beta-Lactam Antibiotic & 1 & 0 & 1 & 0.68 & 0 & 0 \\ 
        Pen. & RDKit 0 & 19 & 14829431 & US-4159272-A & Beta-Lactam Antibiotic & 1 & 0 & 1 & 0.9 & 0 & 0 \\ 
        Pen. & RDKit 0 & 20 & 5046855 & n/a & n/a & 0 & 0 & 0 & 0.82 & 0 & 0 \\ 
        Pen. & RDKit n & 1 & 108789963 & n/a & n/a & 0 & 0 & 0 & 0.9 & 0 & 0 \\ 
        Pen. & RDKit n & 2 & 108788832 & n/a & n/a & 0 & 0 & 0 & 0.81 & 0 & 0 \\ 
        Pen. & RDKit n & 3 & 57624293 & US-2009081766-A1 & Beta-Lactam Antibiotic & 1 & 0 & 1 & 0.49 & 1 & 0 \\ 
        Pen. & RDKit n & 4 & 108795879 & n/a & n/a & 0 & 0 & 0 & 0.9 & 0 & 0 \\ 
        Pen. & RDKit n & 5 & 20341961 & US-4200572-A & Beta-Lactam Antibiotic & 1 & 0 & 1 & 0.46 & 1 & 0 \\ 
        Pen. & RDKit n & 6 & 203884 & n/a & n/a & 0 & 0 & 0 & 0.85 & 0 & 0 \\ 
        Pen. & RDKit n & 7 & 44458116 & n/a & n/a & 0 & 0 & 0 & 0.82 & 0 & 0 \\ 
        Pen. & RDKit n & 8 & 123234728 & WO-2014138046-A1 & Formyl Peptide Receptor Agonist & 1 & 0 & 0 & 0.26 & 0 & 0 \\ 
        Pen. & RDKit n & 9 & 59069194 & US-6815440-B2 & Metalloprotease Inhibitor & 1 & 0 & 0 & 0.41 & 0 & 0 \\ 
        Pen. & RDKit n & 10 & 50877043 & n/a & n/a & 0 & 0 & 0 & 0.31 & 0 & 0 \\ 
        Pen. & RDKit n & 11 & 5211680 & n/a & n/a & 0 & 0 & 0 & 0.35 & 0 & 0 \\ 
        Pen. & RDKit n & 12 & 3288488 & n/a & n/a & 0 & 0 & 0 & 0.33 & 0 & 0 \\ 
        Pen. & RDKit n & 13 & 11047455 & n/a & n/a & 0 & 0 & 0 & 0.41 & 0 & 0 \\ 
        Pen. & RDKit n & 14 & 11860261 & n/a & n/a & 0 & 0 & 0 & 0.26 & 0 & 0 \\ 
        Pen. & RDKit n & 15 & 121879937 & n/a & n/a & 0 & 0 & 0 & 0.4 & 0 & 0 \\ 
        Pen. & RDKit n & 16 & 72811315 & US-2016039825-A1 & S1P / ATX Receptor Agonist & 1 & 0 & 0 & 0.4 & 0 & 0 \\ 
        Pen. & RDKit n & 17 & 53715169 & CA-2297988-A1 & Metalloprotease Inhibitor & 1 & 0 & 0 & 0.43 & 0 & 0 \\ 
        Pen. & RDKit n & 18 & 56796420 & n/a & n/a & 0 & 0 & 0 & 0.26 & 0 & 0 \\ 
        Pen. & RDKit n & 19 & 59847560 & n/a & n/a & 0 & 0 & 0 & 0.24 & 0 & 0 \\ 
        Pen. & RDKit n & 20 & 11859780 & n/a & n/a & 0 & 0 & 0 & 0.26 & 0 & 0 \\ 
        Pen. & OEChem & 1 & 135259185 & WO-2018134148-A1 & MKNK Inhibitor & 1 & 0 & 0 & 0.45 & 0 & 0 \\ 
        Pen. & OEChem & 2 & 20559382 & US-4304779-A & Beta-Lactam Antibiotic & 1 & 0 & 1 & 0.54 & 1 & 0 \\ 
        Pen. & OEChem & 3 & 53944004 & US-4077967-A & Beta-Lactam Antibiotic & 1 & 0 & 1 & 0.74 & 0 & 0 \\ 
        Pen. & OEChem & 4 & 88832135 & US-3954731-A & Beta-Lactam Antibiotic & 1 & 0 & 1 & 0.8 & 0 & 0 \\ 
        Pen. & OEChem & 5 & 20536617 & US4333944A & Sedative & 1 & 0 & 0 & 0.43 & 0 & 0 \\ 
        Pen. & OEChem & 6 & 88762822 & n/a & n/a & 0 & 0 & 0 & 0.46 & 0 & 0 \\ 
        Pen. & OEChem & 7 & 88151512 & EP1091959B1 & Sedative & 1 & 0 & 0 & 0.44 & 0 & 0 \\ 
        Pen. & OEChem & 8 & 140604038 & US6342340 & Dye & 0 & 0 & 0 & 0.34 & 0 & 0 \\ 
        Pen. & OEChem & 9 & 101392657 & n/a & n/a & 0 & 0 & 0 & 0.4 & 0 & 0 \\ 
        Pen. & OEChem & 10 & 17176746 & n/a & n/a & 0 & 0 & 0 & 0.48 & 0 & 0 \\ 
        Pen. & OEChem & 11 & 10001910 & n/a & n/a & 0 & 0 & 0 & 0.43 & 0 & 0 \\ 
        Pen. & OEChem & 12 & 72736793 & n/a & n/a & 0 & 0 & 0 & 0.46 & 0 & 0 \\ 
        Pen. & OEChem & 13 & 67836365 & EP-0275888-A2$\backslash$ & Cardiovascular & 1 & 0 & 0 & 0.42 & 0 & 0 \\ 
        Pen. & OEChem & 14 & 59898268 & US7173041B2/en & Topoisomerase Inhibitor & 1 & 0 & 0 & 0.44 & 0 & 0 \\ 
        Pen. & OEChem & 15 & 53877724 & US-4077967-A & Beta-Lactam Antibiotic & 1 & 0 & 1 & 0.8 & 0 & 0 \\ 
        Pen. & OEChem & 16 & 88640462 & WO-8807534-A1 & Beta-Lactamase Inhibitor & 1 & 0 & 1 & 0.57 & 1 & 0 \\ 
        Pen. & OEChem & 17 & 24811111 & n/a & n/a & 0 & 0 & 0 & 0.48 & 0 & 0 \\ 
        Pen. & OEChem & 18 & 86733573 & US-3965111-A & Sedative & 1 & 0 & 0 & 0.43 & 0 & 0 \\ 
        Pen. & OEChem & 19 & 58637593 & n/a & n/a & 0 & 0 & 0 & 0.28 & 0 & 0 \\ 
        Pen. & OEChem & 20 & 20559367 & US-4304779-A & Beta-Lactam Antibiotic & 1 & 0 & 1 & 0.55 & 1 & 0 \\ 
        Nir. & RDKit 0 & 1 & 89072452 & US-7192957-B2 & HCV Protease Inhibitor & 1 & 0 & 1 & 0.75 & 0 & 0 \\ 
        Nir. & RDKit 0 & 2 & 513166 & 10.1021/ol035826v & HCV Protease Inhibitor & 1 & 0 & 1 & 0.6 & 0 & 0 \\ 
        Nir. & RDKit 0 & 3 & 89032648 & US-7579320-B2 & IAP BIR Ligand & 1 & 0 & 0 & 0.55 & 0 & 0 \\ 
        Nir. & RDKit 0 & 4 & 88251520 & WO-2008141227-A1 & HCV Protease Inhibitor & 1 & 0 & 1 & 0.62 & 0 & 0 \\ 
        Nir. & RDKit 0 & 5 & 89128765 & n/a & n/a & 0 & 0 & 0 & 0.54 & 0 & 0 \\ 
        Nir. & RDKit 0 & 6 & 89386569 & n/a & n/a & 0 & 0 & 0 & 0.67 & 0 & 0 \\ 
        Nir. & RDKit 0 & 7 & 163831729 & WO-2006113942-A2 & Peptidase Inhibitor & 1 & 0 & 1 & 0.73 & 0 & 0 \\ 
        Nir. & RDKit 0 & 8 & 58846078 & US-2006276406-A1 & HCV Protease Inhibitor & 1 & 0 & 1 & 0.65 & 0 & 0 \\ 
        Nir. & RDKit 0 & 9 & 10146053 & n/a & n/a & 0 & 0 & 0 & 0.65 & 0 & 0 \\ 
        Nir. & RDKit 0 & 10 & 70672489 & US-8067379-B2 & HCV Protease Inhibitor & 1 & 0 & 1 & 0.56 & 1 & 0 \\ 
        Nir. & RDKit 0 & 11 & 49777413 & 10.1021/jm9016027 & HCV Protease Inhibitor & 1 & 0 & 1 & 0.56 & 1 & 0 \\ 
        Nir. & RDKit 0 & 12 & 70332592 & US-7192957-B2 & HCV Protease Inhibitor & 1 & 0 & 1 & 0.64 & 0 & 0 \\ 
        Nir. & RDKit 0 & 13 & 58799854 & US-2006276406-A1 & HCV Protease Inhibitor & 1 & 0 & 1 & 0.63 & 0 & 0 \\ 
        Nir. & RDKit 0 & 14 & 11758272 & US-2006276405-A1 & HCV Protease Inhibitor & 1 & 0 & 1 & 0.65 & 0 & 0 \\ 
        Nir. & RDKit 0 & 15 & 117757300 & EP-1730110-B9 & HCV Protease Inhibitor & 1 & 0 & 1 & 0.61 & 0 & 0 \\ 
        Nir. & RDKit 0 & 16 & 70672229 & US-8067379-B2 & HCV Protease Inhibitor & 1 & 0 & 1 & 0.6 & 0 & 0 \\ 
        Nir. & RDKit 0 & 17 & 69712881 & US-2004209897-A1 & Kinase Inhibitor & 1 & 0 & 0 & 0.34 & 0 & 0 \\ 
        Nir. & RDKit 0 & 18 & 58815070 & US-2007042968-A1 & HCV Protease Inhibitor & 1 & 0 & 1 & 0.59 & 1 & 0 \\ 
        Nir. & RDKit 0 & 19 & 91211330 & US-2006281689-A1 & HCV Protease Inhibitor & 1 & 0 & 1 & 0.56 & 1 & 0 \\ 
        Nir. & RDKit 0 & 20 & 57842534 & US-7244721-B2 & HCV Protease Inhibitor & 1 & 0 & 1 & 0.59 & 1 & 0 \\ 
        Nir. & RDKit n & 1 & 58799854 & US-2006276406-A1 & HCV Protease Inhibitor & 1 & 0 & 1 & 0.63 & 0 & 0 \\ 
        Nir. & RDKit n & 2 & 89386569 & n/a & n/a & 0 & 0 & 0 & 0.67 & 0 & 0 \\ 
        Nir. & RDKit n & 3 & 58799868 & US-2006276406-A1 & HCV Protease Inhibitor & 1 & 0 & 1 & 0.58 & 1 & 0 \\ 
        Nir. & RDKit n & 4 & 58799834 & US-2006276406-A1 & HCV Protease Inhibitor & 1 & 0 & 1 & 0.6 & 0 & 0 \\ 
        Nir. & RDKit n & 5 & 58908615 & US-7244721-B2 & HCV Protease Inhibitor & 1 & 0 & 1 & 0.66 & 0 & 0 \\ 
        Nir. & RDKit n & 6 & 58815070 & US-2007042968-A1 & HCV Protease Inhibitor & 1 & 0 & 1 & 0.59 & 1 & 0 \\ 
        Nir. & RDKit n & 7 & 58845760 & US-2006276405-A1 & HCV Protease Inhibitor & 1 & 0 & 1 & 0.68 & 0 & 0 \\ 
        Nir. & RDKit n & 8 & 91211330 & US-2006281689-A1 & HCV Protease Inhibitor & 1 & 0 & 1 & 0.56 & 1 & 0 \\ 
        Nir. & RDKit n & 9 & 57842534 & US-7244721-B2 & HCV Protease Inhibitor & 1 & 0 & 1 & 0.59 & 1 & 0 \\ 
        Nir. & RDKit n & 10 & 89128765 & n/a & n/a & 0 & 0 & 0 & 0.54 & 0 & 0 \\ 
        Nir. & RDKit n & 11 & 70332562 & US-7759499-B2 & HCV Protease Inhibitor & 1 & 0 & 1 & 0.61 & 0 & 0 \\ 
        Nir. & RDKit n & 12 & 58751308 & US-2006252698-A1 & Peptidase Inhibitor & 1 & 0 & 1 & 0.65 & 0 & 0 \\ 
        Nir. & RDKit n & 13 & 90824016 & US-2006276406-A1 & HCV Protease Inhibitor & 1 & 0 & 1 & 0.55 & 1 & 0 \\ 
        Nir. & RDKit n & 14 & 89072393 & n/a & n/a & 0 & 0 & 0 & 0.58 & 0 & 0 \\ 
        Nir. & RDKit n & 15 & 58845752 & US-2006276406-A1 & HCV Protease Inhibitor & 1 & 0 & 1 & 0.62 & 0 & 0 \\ 
        Nir. & RDKit n & 16 & 10146053 & n/a & n/a & 0 & 0 & 0 & 0.65 & 0 & 0 \\ 
        Nir. & RDKit n & 17 & 6483344 & US-2006276406-A1 & HCV Protease Inhibitor & 1 & 0 & 1 & 0.62 & 0 & 0 \\ 
        Nir. & RDKit n & 18 & 143314765 & WO2006113942 & Peptidase Inhibitor & 1 & 0 & 1 & 0.61 & 0 & 0 \\ 
        Nir. & RDKit n & 19 & 58821357 & US-7192957-B2 & HCV Protease Inhibitor & 1 & 0 & 1 & 0.59 & 1 & 0 \\ 
        Nir. & RDKit n & 20 & 58821916 & US-7192957-B2 & HCV Protease Inhibitor & 1 & 0 & 1 & 0.55 & 1 & 0 \\ 
        Nir. & OEChem & 1 & 123307518 & WO-2013107405-A1 & Dehydrogenase Inhibitor & 1 & 0 & 0 & 0.4 & 0 & 0 \\ 
        Nir. & OEChem & 2 & 87441437 & CA-2757866-A1 & Dehydrogenase Inhibitor & 1 & 0 & 0 & 0.38 & 0 & 0 \\ 
        Nir. & OEChem & 3 & 123215879 & WO-2014089379-A1 & ATR Kinase Inhibitor & 1 & 0 & 0 & 0.44 & 0 & 0 \\ 
        Nir. & OEChem & 4 & 5065329 & n/a & n/a & 0 & 0 & 0 & 0.37 & 0 & 0 \\ 
        Nir. & OEChem & 5 & 71268558 & US-2013053410-A1 & MDM2 Inhibitor & 1 & 0 & 0 & 0.4 & 0 & 0 \\ 
        Nir. & OEChem & 6 & 53344667 & n/a & n/a & 0 & 0 & 0 & 0.36 & 0 & 0 \\ 
        Nir. & OEChem & 7 & 58262781 & US-2011117057-A1 & HCV Protease Inhibitor & 1 & 0 & 1 & 0.56 & 1 & 1 \\ 
        Nir. & OEChem & 8 & 137371163 & US-2019002436-A1 & RSV Inhibitor & 1 & 0 & 1 & 0.35 & 1 & 1 \\ 
        Nir. & OEChem & 9 & 89114369 & US-2012270838-A1 & GPCR Agonist & 1 & 0 & 0 & 0.37 & 0 & 0 \\ 
        Nir. & OEChem & 10 & 11997330 & n/a & n/a & 0 & 0 & 0 & 0.37 & 0 & 0 \\ 
        Nir. & OEChem & 11 & 123849611 & WO-2011143521-A2 & CD26 Inhibitor & 1 & 0 & 0 & 0.41 & 0 & 0 \\ 
        Nir. & OEChem & 12 & 130253486 & US-9714243-B2 & HIV Integrase Inhibitor & 1 & 0 & 0 & 0.36 & 0 & 0 \\ 
        Nir. & OEChem & 13 & 126592125 & WO-2017025989-A1 & Dehydrogenase Inhibitor & 1 & 0 & 0 & 0.34 & 0 & 0 \\ 
        Nir. & OEChem & 14 & 67152605 & US-2011118283-A1 & MDM2 Inhibitor & 1 & 0 & 0 & 0.39 & 0 & 0 \\ 
        Nir. & OEChem & 15 & 134321551 & EP-2941432-B1 & ATR Protein Kinase Inhibitor & 1 & 0 & 0 & 0.46 & 0 & 0 \\ 
        Nir. & OEChem & 16 & 155276168 & EP-2970283-B1 & Peptidase Inhibitor & 1 & 0 & 1 & 0.44 & 1 & 1 \\ 
        Nir. & OEChem & 17 & 70239211 & WO-2010021680-A2 & Aspartyl Protease Inhibitor & 1 & 0 & 1 & 0.36 & 1 & 1 \\ 
        Nir. & OEChem & 18 & 88630878 & US-4889857-A & Antibiotic & 1 & 0 & 0 & 0.37 & 0 & 0 \\ 
        Nir. & OEChem & 19 & 91802477 & WO-2010043893-A1 & DNA Gyrase Inhibitor & 1 & 0 & 0 & 0.38 & 0 & 0 \\ 
        Nir. & OEChem & 20 & 49966021 & n/a & n/a & 0 & 0 & 0 & 0.35 & 0 & 0 \\ 
        Zid. & RDKit 0 & 1 & 44326539 & 10.1021/jm00113a016 & HSV Inhibitor & 1 & 0 & 1 & 0.54 & 1 & 0 \\ 
        Zid. & RDKit 0 & 2 & 57303117 & US-5064946-A & HIV Antiviral & 1 & 0 & 1 & 0.77 & 0 & 0 \\ 
        Zid. & RDKit 0 & 3 & 157607985 & US-2013072458-A1 & HIV Antiviral & 1 & 0 & 1 & 0.89 & 0 & 0 \\ 
        Zid. & RDKit 0 & 4 & 71335580 & n/a & n/a & 0 & 0 & 0 & 0.73 & 0 & 0 \\ 
        Zid. & RDKit 0 & 5 & 21145774 & 10.1016/S0960-894X(01)80809-X & n/a & 0 & 0 & 0 & 0.37 & 0 & 0 \\ 
        Zid. & RDKit 0 & 6 & 9816797 & n/a & n/a & 0 & 0 & 0 & 0.72 & 0 & 0 \\ 
        Zid. & RDKit 0 & 7 & 53721880 & US-6040297-A & HIV Antiviral & 1 & 0 & 1 & 0.86 & 0 & 0 \\ 
        Zid. & RDKit 0 & 8 & 464363 & n/a & n/a & 0 & 0 & 0 & 0.74 & 0 & 0 \\ 
        Zid. & RDKit 0 & 9 & 57085874 & US-4681933-A & Antiviral & 1 & 0 & 1 & 0.88 & 0 & 0 \\ 
        Zid. & RDKit 0 & 10 & 91933148 & n/a & n/a & 0 & 0 & 0 & 0.76 & 0 & 0 \\ 
        Zid. & RDKit 0 & 11 & 44286560 & 10.1021/jm9600095 & HIV Antiviral & 1 & 0 & 1 & 0.88 & 0 & 0 \\ 
        Zid. & RDKit 0 & 12 & 101438451 & n/a & n/a & 0 & 0 & 0 & 0.53 & 0 & 0 \\ 
        Zid. & RDKit 0 & 13 & 13989283 & US-2005026902-A1 & HIV Antiviral & 1 & 0 & 1 & 0.77 & 0 & 0 \\ 
        Zid. & RDKit 0 & 14 & 129675745 & 10.1007/978-1-4615-2824-1$\backslash$\_5 & HIV Antiviral & 1 & 0 & 1 & 0.69 & 0 & 0 \\ 
        Zid. & RDKit 0 & 15 & 10469965 & n/a & n/a & 0 & 0 & 0 & 0.44 & 0 & 0 \\ 
        Zid. & RDKit 0 & 16 & 451826 & US-4681933-A & HIV Antiviral & 1 & 0 & 1 & 0.79 & 0 & 0 \\ 
        Zid. & RDKit 0 & 17 & 12922948 & n/a & n/a & 0 & 0 & 0 & 0.49 & 0 & 0 \\ 
        Zid. & RDKit 0 & 18 & 20158804 & US-2020399304-A1 & HIV Antiviral & 1 & 0 & 1 & 0.91 & 0 & 0 \\ 
        Zid. & RDKit 0 & 19 & 49769999 & 10.1021/jm070824s & HIV Antiviral & 1 & 0 & 1 & 0.5 & 1 & 0 \\ 
        Zid. & RDKit 0 & 20 & 10332953 & n/a & n/a & 0 & 0 & 0 & 0.57 & 0 & 0 \\ 
        Zid. & RDKit n & 1 & 10494229 & n/a & n/a & 0 & 0 & 0 & 0.73 & 0 & 0 \\ 
        Zid. & RDKit n & 2 & 476466 & 10.1016/0968-0896(95)00030-k & HIV Antiviral & 1 & 0 & 1 & 0.56 & 1 & 0 \\ 
        Zid. & RDKit n & 3 & 102306428 & n/a & n/a & 0 & 0 & 0 & 0.29 & 0 & 0 \\ 
        Zid. & RDKit n & 4 & 89763182 & EP-2631239-A1 & HIV/HCV Antiviral & 1 & 0 & 1 & 0.55 & 1 & 0 \\ 
        Zid. & RDKit n & 5 & 10501259 & n/a & n/a & 0 & 0 & 0 & 0.53 & 0 & 0 \\ 
        Zid. & RDKit n & 6 & 89763189 & EP-2631239-A1 & HIV/HCV Antiviral & 1 & 0 & 1 & 0.55 & 1 & 0 \\ 
        Zid. & RDKit n & 7 & 162488274 & US20210371447 & n/a & 0 & 0 & 0 & 0.6 & 0 & 0 \\ 
        Zid. & RDKit n & 8 & 90085164 & US-2014088117-A1 & Kinase Inhibitor & 1 & 0 & 0 & 0.36 & 0 & 0 \\ 
        Zid. & RDKit n & 9 & 89670654 & US-2013172402-A1 & DNA-Binding Oligo & 1 & 0 & 0 & 0.7 & 0 & 0 \\ 
        Zid. & RDKit n & 10 & 25185465 & US-2012070415-A1 & Antiviral & 1 & 0 & 1 & 0.54 & 1 & 0 \\ 
        Zid. & RDKit n & 11 & 10618341 & n/a & n/a & 0 & 0 & 0 & 0.54 & 0 & 0 \\ 
        Zid. & RDKit n & 12 & 89433120 & US-2013064794-A1 & HCV Antiviral & 1 & 0 & 1 & 0.37 & 1 & 0 \\ 
        Zid. & RDKit n & 13 & 135016033 & 10.1055/s-0029-1217566 & n/a & 0 & 0 & 0 & 0.35 & 0 & 0 \\ 
        Zid. & RDKit n & 14 & 124122254 & AU-2015220560-B2 & Kinase Inhibitor & 1 & 0 & 0 & 0.31 & 0 & 0 \\ 
        Zid. & RDKit n & 15 & 71834937 & n/a & n/a & 0 & 0 & 0 & 0.78 & 0 & 0 \\ 
        Zid. & RDKit n & 16 & 23384164 & US-2002028933-A1 & Inflammation Treatment & 1 & 0 & 0 & 0.25 & 0 & 0 \\ 
        Zid. & RDKit n & 17 & 122418602 & WO-2016113335-A1 & Galectin Binding Inhibitor & 1 & 0 & 0 & 0.32 & 0 & 0 \\ 
        Zid. & RDKit n & 18 & 57155433 & WO-9961583-A2 & n/a & 0 & 0 & 0 & 0.39 & 0 & 0 \\ 
        Zid. & RDKit n & 19 & 117680541 & US-2012070415-A1 & Antiviral & 1 & 0 & 1 & 0.38 & 1 & 0 \\ 
        Zid. & RDKit n & 20 & 71477389 & n/a & n/a & 0 & 0 & 0 & 0.37 & 0 & 0 \\ 
        Zid. & OEChem & 1 & 101424608 & n/a & n/a & 0 & 0 & 0 & 0.69 & 0 & 0 \\ 
        Zid. & OEChem & 2 & 76847725 & n/a & n/a & 0 & 0 & 0 & 0.3 & 0 & 0 \\ 
        Zid. & OEChem & 3 & 56679233 & ES-2458358-T3 & HCV Antiviral & 1 & 0 & 1 & 0.33 & 1 & 0 \\ 
        Zid. & OEChem & 4 & 89479439 & EP-2760857-A1 & PIM Kinase Inhibitor & 1 & 0 & 0 & 0.35 & 0 & 0 \\ 
        Zid. & OEChem & 5 & 90445142 & US-2014271547-A1 & HCV Antiviral & 1 & 0 & 1 & 0.68 & 0 & 0 \\ 
        Zid. & OEChem & 6 & 130375412 & US-10449144-B2 & n/a* & 1 & 0 & 0 & 0.33 & 0 & 0 \\ 
        Zid. & OEChem & 7 & 57284777 & US-5585373-A & Anticancer Agent & 1 & 0 & 0 & 0.36 & 0 & 0 \\ 
        Zid. & OEChem & 8 & 57155433 & WO1999061583A2 & Carbohydrate Scaffold & 0 & 0 & 0 & 0.39 & 0 & 0 \\ 
        Zid. & OEChem & 9 & 10594262 & n/a & n/a & 0 & 0 & 0 & 0.34 & 0 & 0 \\ 
        Zid. & OEChem & 10 & 136288884 & n/a & n/a & 0 & 0 & 0 & 0.36 & 0 & 0 \\ 
        Zid. & OEChem & 11 & 102576629 & n/a & n/a & 0 & 0 & 0 & 0.29 & 0 & 0 \\ 
        Zid. & OEChem & 12 & 88490101 & EP-0236074-A1 & Beta-Lactam Inhibitor & 1 & 0 & 0 & 0.35 & 0 & 0 \\ 
        Zid. & OEChem & 13 & 57103302 & EP-0351021-A2 & Herbicide & 1 & 0 & 0 & 0.25 & 0 & 0 \\ 
        Zid. & OEChem & 14 & 68220216 & US-8034812-B2 & Kinase Inhibitor & 1 & 0 & 0 & 0.25 & 0 & 0 \\ 
        Zid. & OEChem & 15 & 19967798 & EP-0729070-A3 & Industrial & 0 & 1 & 0 & 0.24 & 0 & 0 \\ 
        Zid. & OEChem & 16 & 68220235 & US-8034812-B2 & Kinase Inhibitor & 1 & 0 & 0 & 0.24 & 0 & 0 \\ 
        Zid. & OEChem & 17 & 117680541 & US-2012070415-A1 & RSV/Influenaza Antiviral & 1 & 0 & 1 & 0.38 & 1 & 0 \\ 
        Zid. & OEChem & 18 & 6333312 & n/a & n/a & 0 & 0 & 0 & 0.36 & 0 & 0 \\ 
        Zid. & OEChem & 19 & 130290413 & US-2017216275-A1 & Phosphodiesterase Inhibitor & 1 & 0 & 0 & 0.38 & 0 & 0 \\ 
        Zid. & OEChem & 20 & 23519654 & US-2003236216-A1 & HCV Antiviral & 1 & 0 & 1 & 0.57 & 1 & 0 \\ 
        LSD & RDKit 0 & 1 & 3938 & WO-2020068832-A1 & Dopaminergic Agonist & 1 & 0 & 1 & 0.79 & 0 & 0 \\ 
        LSD & RDKit 0 & 2 & 165200 & US-2023026731-A1 & LSD Derivative & 1 & 0 & 1 & 0.92 & 0 & 0 \\ 
        LSD & RDKit 0 & 3 & 56988771 & DE-2924102-A1 & Dopaminergic Agonist & 1 & 0 & 1 & 0.79 & 0 & 0 \\ 
        LSD & RDKit 0 & 4 & 9884289 & n/a & n/a & 0 & 0 & 0 & 0.6 & 0 & 0 \\ 
        LSD & RDKit 0 & 5 & 134553 & US20220096504 & LSD Derivative & 1 & 0 & 1 & 0.97 & 0 & 0 \\ 
        LSD & RDKit 0 & 6 & 24837792 & US-3944582-A & LSD Derivative & 1 & 0 & 0 & 0.59 & 0 & 0 \\ 
        LSD & RDKit 0 & 7 & 10427207 & 10.1021/jm00006a015 & LSD Derivative & 1 & 0 & 1 & 0.97 & 0 & 0 \\ 
        LSD & RDKit 0 & 8 & 13059865 & US-4853390-A & Antidopaminergic & 1 & 0 & 1 & 0.8 & 0 & 0 \\ 
        LSD & RDKit 0 & 9 & 76683560 & US-2017174684-A1 & Dopaminergic Agonist & 1 & 0 & 1 & 0.89 & 0 & 0 \\ 
        LSD & RDKit 0 & 10 & 130318308 & US9777016B2 & 5HT / Dopaminergic & 1 & 0 & 1 & 0.77 & 0 & 0 \\ 
        LSD & RDKit 0 & 11 & 13770164 & US-4826852-A & Dopaminergic & 1 & 0 & 1 & 0.73 & 0 & 0 \\ 
        LSD & RDKit 0 & 12 & 45357584 & n/a & n/a & 0 & 0 & 0 & 0.59 & 0 & 0 \\ 
        LSD & RDKit 0 & 13 & 3039342 & Book & see PubChem & 1 & 0 & 1 & 0.85 & 0 & 0 \\ 
        LSD & RDKit 0 & 14 & 13823178 & US-4863929-A & Antidopaminergic & 1 & 0 & 1 & 0.57 & 1 & 0 \\ 
        LSD & RDKit 0 & 15 & 67937682 & US-4863929-A & Antidopaminergic & 1 & 0 & 1 & 0.57 & 1 & 0 \\ 
        LSD & RDKit 0 & 16 & 88985978 & WO2011003988A1 & Somatic cell self renewel & 1 & 0 & 0 & 0.88 & 0 & 0 \\ 
        LSD & RDKit 0 & 17 & 70430592 & n/a & n/a & 0 & 0 & 0 & 0.81 & 0 & 0 \\ 
        LSD & RDKit 0 & 18 & 57175320 & US4348392A & Dopaminergic & 1 & 0 & 1 & 0.56 & 1 & 0 \\ 
        LSD & RDKit 0 & 19 & 13192408 & US-4826852-A & Dopaminergic & 1 & 0 & 1 & 0.78 & 0 & 0 \\ 
        LSD & RDKit 0 & 20 & 13192405 & US-4826852-A & Dopaminergic & 1 & 0 & 1 & 0.78 & 0 & 0 \\ 
        LSD & RDKit n & 1 & 44457782 & US-2021346346-A1 & Serotonergic Agonist & 1 & 0 & 1 & 0.58 & 1 & 0 \\ 
        LSD & RDKit n & 2 & 132584757 & n/a & n/a & 0 & 0 & 0 & 0.52 & 0 & 0 \\ 
        LSD & RDKit n & 3 & 3058122 & n/a & n/a & 0 & 0 & 0 & 0.57 & 0 & 0 \\ 
        LSD & RDKit n & 4 & 10640901 & n/a & n/a & 0 & 0 & 0 & 0.54 & 0 & 0 \\ 
        LSD & RDKit n & 5 & 67938090 & EP-0220129-B1 & Dopaminergic Agonist & 1 & 0 & 1 & 0.57 & 1 & 0 \\ 
        LSD & RDKit n & 6 & 14723139 & US-5411966-A & Dopaminergic Agonist & 1 & 0 & 1 & 0.57 & 1 & 0 \\ 
        LSD & RDKit n & 7 & 58459676 & US-9730923-B2 & Dopaminergic Agonist & 1 & 0 & 1 & 0.3 & 1 & 1 \\ 
        LSD & RDKit n & 8 & 4371364 & n/a & n/a & 0 & 0 & 0 & 0.51 & 0 & 0 \\ 
        LSD & RDKit n & 9 & 67821466 & n/a & n/a & 0 & 0 & 0 & 0.48 & 0 & 0 \\ 
        LSD & RDKit n & 10 & 68168631 & US20120202823A1 & Prodrug Carrier Moiety & 1 & 0 & 0 & 0.35 & 0 & 0 \\ 
        LSD & RDKit n & 11 & 56989523 & n/a & Low-migration Coating & 0 & 0 & 0 & 0.51 & 0 & 0 \\ 
        LSD & RDKit n & 12 & 101648098 & n/a & n/a & 0 & 0 & 0 & 0.44 & 0 & 0 \\ 
        LSD & RDKit n & 13 & 23621001 & n/a & n/a & 0 & 0 & 0 & 0.35 & 0 & 0 \\ 
        LSD & RDKit n & 14 & 88840811 & US3966706A & Dye & 0 & 1 & 0 & 0.38 & 0 & 0 \\ 
        LSD & RDKit n & 15 & 8882005 & n/a & n/a & 0 & 0 & 0 & 0.52 & 0 & 0 \\ 
        LSD & RDKit n & 16 & 56843556 & n/a & n/a & 0 & 0 & 0 & 0.84 & 0 & 0 \\ 
        LSD & RDKit n & 17 & 123178010 & US20120177730A1 & Chemosensory Receptor Ligand & 1 & 0 & 0 & 0.54 & 0 & 0 \\ 
        LSD & RDKit n & 18 & 101872812 & n/a & n/a & 0 & 0 & 0 & 0.45 & 0 & 0 \\ 
        LSD & RDKit n & 19 & 126638676 & WO-2017033019-A1 & Kinase Inhibitor & 1 & 0 & 0 & 0.5 & 0 & 0 \\ 
        LSD & RDKit n & 20 & 57240530 & US-5037832-A & Dopaminergic Agonist & 1 & 0 & 1 & 0.72 & 0 & 0 \\ 
        LSD & OEChem & 1 & 89651741 & EP-2794606-A1 & Neurodegen & 1 & 0 & 0 & 0.41 & 0 & 0 \\ 
        LSD & OEChem & 2 & 89705589 & WO-2013107856-A1 & Neurodegen & 1 & 0 & 0 & 0.43 & 0 & 0 \\ 
        LSD & OEChem & 3 & 3938 & WO-2020068832-A1 & LSD Derivative & 1 & 0 & 1 & 0.79 & 0 & 0 \\ 
        LSD & OEChem & 4 & 135838366 & US-4696928-A & Anti-cancer & 1 & 0 & 0 & 0.42 & 0 & 0 \\ 
        LSD & OEChem & 5 & 129198569 & EP-2888252-B1 & Anti-Amyloid & 1 & 0 & 0 & 0.38 & 0 & 0 \\ 
        LSD & OEChem & 6 & 88368790 & JPH0920755A/en & Alpha-1 receptor agonist & 1 & 0 & 0 & 0.28 & 0 & 0 \\ 
        LSD & OEChem & 7 & 89651520 & EP-2794606-A1 & Neurodegen & 1 & 0 & 0 & 0.38 & 0 & 0 \\ 
        LSD & OEChem & 8 & 18634859 & US-4866057-A & Dopaminergic Agonist & 1 & 0 & 1 & 0.39 & 1 & 1 \\ 
        LSD & OEChem & 9 & 9884289 & n/a & LSD Derivative & 0 & 0 & 0 & 0.6 & 0 & 0 \\ 
        LSD & OEChem & 10 & 19850770 & US-4866057-A & Dopaminergic Agonist & 1 & 0 & 1 & 0.39 & 1 & 1 \\ 
        LSD & OEChem & 11 & 129029504 & WO-2017061957-A1 & Methyltransferase Inhibitor & 1 & 0 & 0 & 0.38 & 0 & 0 \\ 
        LSD & OEChem & 12 & 69959271 & US5126448A & Psychotropic & 1 & 0 & 0 & 0.43 & 0 & 0 \\ 
        LSD & OEChem & 13 & 23815464 & n/a & n/a & 0 & 0 & 0 & 0.44 & 0 & 0 \\ 
        LSD & OEChem & 14 & 10621020 & 10.1021/jm9805687 & 5-HT1 Receptor Agonist & 1 & 0 & 1 & 0.45 & 1 & 1 \\ 
        LSD & OEChem & 15 & 56988771 & DE-2924102-A1 & LSD Derivative & 1 & 0 & 1 & 0.79 & 0 & 0 \\ 
        LSD & OEChem & 16 & 42609829 & 10.1038/nchembio.188 & LSD Derivative & 1 & 0 & 1 & 0.97 & 0 & 0 \\ 
        LSD & OEChem & 17 & 123495938 & WO-2013068470-A1$\backslash$ & Neurological Treatment & 1 & 0 & 0 & 0.46 & 0 & 0 \\ 
        LSD & OEChem & 18 & 45124493 & n/a & n/a & 0 & 0 & 0 & 0.26 & 0 & 0 \\ 
        LSD & OEChem & 19 & 13770164 & US-4826852-A & LSD Derivative & 1 & 0 & 1 & 0.73 & 0 & 0 \\ 
        LSD & OEChem & 20 & 67563282 & US-2010004226-A1$\backslash$ & Anxiolytic & 1 & 0 & 0 & 0.43 & 0 & 0 \\ 
        Fen. & RDKit 0 & 1 & 210867 & US-RE29828-E & Analgesic & 1 & 0 & 1 & 0.74 & 0 & 0 \\ 
        Fen. & RDKit 0 & 2 & 101682792 & n/a & n/a & 0 & 0 & 0 & 0.51 & 0 & 0 \\ 
        Fen. & RDKit 0 & 3 & 61996 & WO-2021216450-A1 & Mefentanyl & 1 & 0 & 1 & 0.87 & 0 & 0 \\ 
        Fen. & RDKit 0 & 4 & 619324 & n/a & n/a & 0 & 0 & 0 & 0.79 & 0 & 0 \\ 
        Fen. & RDKit 0 & 5 & 21496449 & US-3993762-A & Analgesic & 1 & 0 & 1 & 0.63 & 0 & 0 \\ 
        Fen. & RDKit 0 & 6 & 7365707 & n/a & n/a & 0 & 0 & 0 & 0.81 & 0 & 0 \\ 
        Fen. & RDKit 0 & 7 & 12666095 & 10.1021/jm00182a016 & Analgesic & 1 & 0 & 1 & 0.75 & 0 & 0 \\ 
        Fen. & RDKit 0 & 8 & 451052 & n/a & n/a & 0 & 0 & 0 & 0.93 & 0 & 0 \\ 
        Fen. & RDKit 0 & 9 & 85863080 & n/a & n/a & 0 & 0 & 0 & 0.55 & 0 & 0 \\ 
        Fen. & RDKit 0 & 10 & 129522083 & n/a & n/a & 0 & 0 & 0 & 0.86 & 0 & 0 \\ 
        Fen. & RDKit 0 & 11 & 21951518 & US-2014336214-A1 & Analgesic & 1 & 0 & 1 & 0.93 & 0 & 0 \\ 
        Fen. & RDKit 0 & 12 & 20068771 & US-5100903-A & Analgesic & 1 & 0 & 1 & 0.58 & 1 & 0 \\ 
        Fen. & RDKit 0 & 13 & 10499027 & n/a & n/a & 0 & 0 & 0 & 0.82 & 0 & 0 \\ 
        Fen. & RDKit 0 & 14 & 12741855 & 10.1021/jm00139a003 & Analgesic & 1 & 0 & 1 & 0.71 & 0 & 0 \\ 
        Fen. & RDKit 0 & 15 & 133809799 & n/a & n/a & 0 & 0 & 0 & 0.36 & 0 & 0 \\ 
        Fen. & RDKit 0 & 16 & 19902260 & US-5100903-A & Analgesic & 1 & 0 & 1 & 0.6 & 0 & 0 \\ 
        Fen. & RDKit 0 & 17 & 55456487 & n/a & n/a & 0 & 0 & 0 & 0.31 & 0 & 0 \\ 
        Fen. & RDKit 0 & 18 & 113074725 & n/a & n/a & 0 & 0 & 0 & 0.26 & 0 & 0 \\ 
        Fen. & RDKit 0 & 19 & 101682787 & n/a & n/a & 0 & 0 & 0 & 0.43 & 0 & 0 \\ 
        Fen. & RDKit 0 & 20 & 46048033 & n/a & n/a & 0 & 0 & 0 & 0.64 & 0 & 0 \\ 
        Fen. & RDKit n & 1 & 46044596 & n/a & n/a & 0 & 0 & 0 & 0.22 & 0 & 0 \\ 
        Fen. & RDKit n & 2 & 22503272 & US-2004204421-A1 & Serotonergic Agonist & 1 & 0 & 0 & 0.25 & 0 & 0 \\ 
        Fen. & RDKit n & 3 & 12298017 & US-4179569-A & Analgesic & 1 & 0 & 1 & 0.51 & 1 & 0 \\ 
        Fen. & RDKit n & 4 & 15075634 & n/a & n/a & 0 & 0 & 0 & 0.81 & 0 & 0 \\ 
        Fen. & RDKit n & 5 & 71258753 & WO2013024291A2 & Various Diseases & 1 & 0 & 0 & 0.27 & 0 & 0 \\ 
        Fen. & RDKit n & 6 & 57919355 & US-2003236259-A1 & Histamine H3 Agonist & 1 & 0 & 0 & 0.28 & 0 & 0 \\ 
        Fen. & RDKit n & 7 & 91753695 & n/a & n/a & 0 & 0 & 0 & 0.81 & 0 & 0 \\ 
        Fen. & RDKit n & 8 & 110408477 & n/a & n/a & 0 & 0 & 0 & 0.23 & 0 & 0 \\ 
        Fen. & RDKit n & 9 & 57919390 & US-2003236259-A1 & Histamine H3 Agonist & 1 & 0 & 0 & 0.27 & 0 & 0 \\ 
        Fen. & RDKit n & 10 & 120540 & 10.1016/j.bmcl.2012.04.098 & 5-HTR1A Agonist & 1 & 0 & 0 & 0.31 & 0 & 0 \\ 
        Fen. & RDKit n & 11 & 46044468 & n/a & n/a & 0 & 0 & 0 & 0.19 & 0 & 0 \\ 
        Fen. & RDKit n & 12 & 133342710 & n/a & n/a & 0 & 0 & 0 & 0.23 & 0 & 0 \\ 
        Fen. & RDKit n & 13 & 133316530 & n/a & n/a & 0 & 0 & 0 & 0.24 & 0 & 0 \\ 
        Fen. & RDKit n & 14 & 133282119 & n/a & n/a & 0 & 0 & 0 & 0.19 & 0 & 0 \\ 
        Fen. & RDKit n & 15 & 68820800 & US-2006122187-A1 & Histamine H3 Agonist & 1 & 0 & 0 & 0.22 & 0 & 0 \\ 
        Fen. & RDKit n & 16 & 60716433 & US-8420657-B2 & Kinase Inhibitor & 1 & 0 & 0 & 0.3 & 0 & 0 \\ 
        Fen. & RDKit n & 17 & 90268870 & WO-2014102590-A1 & Analgesic & 1 & 0 & 1 & 0.74 & 0 & 0 \\ 
        Fen. & RDKit n & 18 & 46208762 & WO-2010057833-A1 & Muscular Dystrophy Treatment & 1 & 0 & 0 & 0.23 & 0 & 0 \\ 
        Fen. & RDKit n & 19 & 23961496 & n/a & n/a & 0 & 0 & 0 & 0.24 & 0 & 0 \\ 
        Fen. & RDKit n & 20 & 46213640 & KR-100990872-B1 & Muscarinic Receptor Agonist & 1 & 0 & 1 & 0.2 & 1 & 1 \\ 
        Fen. & OEChem & 1 & 123511870 & WO-2012113860-A2 & Photosensitiziing Antibiotic & 1 & 1 & 0 & 0.21 & 0 & 0 \\ 
        Fen. & OEChem & 2 & 87843970 & US-7074779-B2 & Estrogen Receptor Modulator & 1 & 0 & 0 & 0.16 & 0 & 0 \\ 
        Fen. & OEChem & 3 & 57150779 & WO-9740051-A1 & Antihyperlipidemic Agent & 1 & 0 & 0 & 0.24 & 0 & 0 \\ 
        Fen. & OEChem & 4 & 57343443 & n/a & n/a & 0 & 0 & 0 & 0.26 & 0 & 0 \\ 
        Fen. & OEChem & 5 & 19769681 & US-5202333-A & 5-HT3 Agonist & 1 & 0 & 0 & 0.25 & 0 & 0 \\ 
        Fen. & OEChem & 6 & 53890204 & US4616023A & Calcium Channel Inhibitor & 1 & 0 & 0 & 0.21 & 0 & 0 \\ 
        Fen. & OEChem & 7 & 130393870 & EP-2797921-B1 & PARP Inhibitor & 1 & 0 & 0 & 0.25 & 0 & 0 \\ 
        Fen. & OEChem & 8 & 123921833 & WO2013127266A1 & NAMPT Inhibitor & 1 & 0 & 0 & 0.22 & 0 & 0 \\ 
        Fen. & OEChem & 9 & 88122657 & US4885278A/en & Psychoactive (nonspecific) & 1 & 0 & 0 & 0.21 & 0 & 0 \\ 
        Fen. & OEChem & 10 & 70137168 & WO-9965910-A1 & Kinase Inhibitor & 1 & 0 & 0 & 0.24 & 0 & 0 \\ 
        Fen. & OEChem & 11 & 70633255 & US4054569A & CNS Depressant & 1 & 0 & 0 & 0.2 & 0 & 0 \\ 
        Fen. & OEChem & 12 & 123142262 & WO2012006321A2 & Antibacterial & 1 & 0 & 0 & 0.18 & 0 & 0 \\ 
        Fen. & OEChem & 13 & 74349214 & n/a & n/a & 0 & 0 & 0 & 0.32 & 0 & 0 \\ 
        Fen. & OEChem & 14 & 88208824 & WO2000042045A2 & Chemokine Receptor Mod. & 1 & 0 & 0 & 0.24 & 0 & 0 \\ 
        Fen. & OEChem & 15 & 100937756 & n/a & n/a & 0 & 0 & 0 & 0.19 & 0 & 0 \\ 
        Fen. & OEChem & 16 & 70204314 & EP0820450B1 & C-Fiber Analgesic & 1 & 0 & 0 & 0.21 & 0 & 0 \\ 
        Fen. & OEChem & 17 & 68869271 & US-2009143392-A1 & PDE10a Inhibitor & 1 & 0 & 0 & 0.24 & 0 & 0 \\ 
        Fen. & OEChem & 18 & 18597963 & US-4515949-A & Hypertension & 1 & 0 & 0 & 0.25 & 0 & 0 \\ 
        Fen. & OEChem & 19 & 163122578 & n/a & n/a & 0 & 0 & 0 & 0.26 & 0 & 0 \\ 
        Fen. & OEChem & 20 & 88209026 & WO2000042045A2 & Chemokine Receptor Mod. & 1 & 0 & 0 & 0.24 & 0 & 0 \\ 
        AB25 & RDKit 0 & 1 & 53758252 & US-3963763-A & Dye & 0 & 1 & 1 & 0.69 & 0 & 0 \\ 
        AB25 & RDKit 0 & 2 & 20658940 & US-6241789-B1 & Dye & 0 & 1 & 1 & 0.78 & 0 & 0 \\ 
        AB25 & RDKit 0 & 3 & 20314677 & US-RE29724-E & Dye & 0 & 1 & 1 & 0.62 & 0 & 0 \\ 
        AB25 & RDKit 0 & 4 & 28468134 & US3211755A & Dye & 0 & 1 & 1 & 0.52 & 1 & 0 \\ 
        AB25 & RDKit 0 & 5 & 123191991 & WO-2012035122-A1 & Dye & 0 & 1 & 1 & 0.81 & 0 & 0 \\ 
        AB25 & RDKit 0 & 6 & 88847454 & US-3763159-A & Dye & 0 & 1 & 1 & 0.89 & 0 & 0 \\ 
        AB25 & RDKit 0 & 7 & 80836 & US5385842A & Bacterial Sulfide Inhibitor & 1 & 0 & 0 & 0.76 & 0 & 0 \\ 
        AB25 & RDKit 0 & 8 & 23316284 & US-4596666-A & Dye & 0 & 1 & 1 & 0.51 & 1 & 0 \\ 
        AB25 & RDKit 0 & 9 & 117662018 & US-9228063-B2 & Dye & 0 & 1 & 1 & 0.51 & 1 & 0 \\ 
        AB25 & RDKit 0 & 10 & 118582126 & US-2021060995-A1 & Dye & 0 & 1 & 1 & 0.5 & 1 & 0 \\ 
        AB25 & RDKit 0 & 11 & 3100898 & US-4191566-A & Dye & 0 & 1 & 1 & 0.64 & 0 & 0 \\ 
        AB25 & RDKit 0 & 12 & 23454406 & US-4276213-A & Dye & 0 & 1 & 1 & 0.51 & 1 & 0 \\ 
        AB25 & RDKit 0 & 13 & 20136182 & US-4749521-A & Dye & 0 & 1 & 1 & 0.77 & 0 & 0 \\ 
        AB25 & RDKit 0 & 14 & 12260350 & CN-103965648-A & Dye & 0 & 1 & 1 & 0.73 & 0 & 0 \\ 
        AB25 & RDKit 0 & 15 & 57573599 & US-2009166583-A1 & Dye & 0 & 1 & 1 & 0.77 & 0 & 0 \\ 
        AB25 & RDKit 0 & 16 & 19083394 & US-5759211-A & Dye & 0 & 1 & 1 & 0.79 & 0 & 0 \\ 
        AB25 & RDKit 0 & 17 & 154153477 & US-3324150-A & Dye & 0 & 1 & 1 & 0.76 & 0 & 0 \\ 
        AB25 & RDKit 0 & 18 & 21195065 & US-5973038-A & Dye & 0 & 1 & 1 & 0.6 & 0 & 0 \\ 
        AB25 & RDKit 0 & 19 & 5027999 & n/a & n/a & 0 & 0 & 0 & 0.6 & 0 & 0 \\ 
        AB25 & RDKit 0 & 20 & 129772523 & 10.1134/S0026261713040140 & Dye & 0 & 1 & 1 & 0.51 & 1 & 0 \\ 
        AB25 & RDKit n & 1 & 16205090 & DE-1644589-B2 & Dye & 0 & 1 & 1 & 0.61 & 0 & 0 \\ 
        AB25 & RDKit n & 2 & 59123478 & US-2003231237-A1 & Dye & 0 & 1 & 1 & 0.65 & 0 & 0 \\ 
        AB25 & RDKit n & 3 & 66487 & US-6140517-A & Dye & 0 & 1 & 1 & 0.59 & 1 & 0 \\ 
        AB25 & RDKit n & 4 & 22975881 & US-2003110581-A1 & Dye & 0 & 1 & 1 & 0.62 & 0 & 0 \\ 
        AB25 & RDKit n & 5 & 71451728 & n/a & n/a & 0 & 0 & 0 & 0.64 & 0 & 0 \\ 
        AB25 & RDKit n & 6 & 70619779 & US-4128396-A & Dye & 0 & 1 & 1 & 0.59 & 1 & 0 \\ 
        AB25 & RDKit n & 7 & 89357654 & WO2007053227A1 & Dye & 0 & 1 & 1 & 0.71 & 0 & 0 \\ 
        AB25 & RDKit n & 8 & 101530693 & n/a & n/a & 0 & 0 & 0 & 0.57 & 0 & 0 \\ 
        AB25 & RDKit n & 9 & 91433520 & 10.1016/j.jphotobiol.2010.08.010 & Fluorescence Probe & 0 & 1 & 1 & 0.37 & 1 & 1 \\ 
        AB25 & RDKit n & 10 & 22918414 & US-5663336-A & Kinase Inhibitor & 1 & 0 & 0 & 0.33 & 0 & 0 \\ 
        AB25 & RDKit n & 11 & 134848698 & n/a & n/a & 0 & 0 & 0 & 0.4 & 0 & 0 \\ 
        AB25 & RDKit n & 12 & 5483103 & US-2015225584-A1 & Dye & 0 & 1 & 1 & 0.57 & 1 & 0 \\ 
        AB25 & RDKit n & 13 & 86048149 & n/a & n/a & 0 & 0 & 0 & 0.62 & 0 & 0 \\ 
        AB25 & RDKit n & 14 & 137212270 & n/a & n/a & 0 & 0 & 0 & 0.45 & 0 & 0 \\ 
        AB25 & RDKit n & 15 & 1727626 & n/a & n/a & 0 & 0 & 0 & 0.47 & 0 & 0 \\ 
        AB25 & RDKit n & 16 & 101530913 & n/a & n/a & 0 & 0 & 0 & 0.47 & 0 & 0 \\ 
        AB25 & RDKit n & 17 & 105919 & n/a & n/a & 0 & 0 & 0 & 0.39 & 0 & 0 \\ 
        AB25 & RDKit n & 18 & 1992311 & n/a & n/a & 0 & 0 & 0 & 0.46 & 0 & 0 \\ 
        AB25 & RDKit n & 19 & 1914542 & n/a & n/a & 0 & 0 & 0 & 0.46 & 0 & 0 \\ 
        AB25 & RDKit n & 20 & 117567 & JP-2009072919-A & Dye & 0 & 1 & 1 & 0.6 & 0 & 0 \\ 
        AB25 & OEChem & 1 & 102017117 & n/a & n/a & 0 & 0 & 0 & 0.39 & 0 & 0 \\ 
        AB25 & OEChem & 2 & 10895852 & CN-108715591-A & Dye & 0 & 1 & 1 & 0.36 & 1 & 1 \\ 
        AB25 & OEChem & 3 & 10178481 & US6919448B2 & Dye & 0 & 1 & 1 & 0.32 & 1 & 1 \\ 
        AB25 & OEChem & 4 & 136132324 & US-2005020559-A1 & Photosensitizing Agent & 1 & 1 & 1 & 0.44 & 1 & 1 \\ 
        AB25 & OEChem & 5 & 90309725 & EP-2159227-B1 & Dye & 0 & 1 & 1 & 0.2 & 1 & 1 \\ 
        AB25 & OEChem & 6 & 4426440 & US20100055044A1 & Anticancer (ETC Inhibitor) & 1 & 0 & 0 & 0.41 & 0 & 0 \\ 
        AB25 & OEChem & 7 & 67741728 & US-5382662-A & Catalyst & 0 & 0 & 0 & 0.19 & 0 & 0 \\ 
        AB25 & OEChem & 8 & 68519583 & US20100009927A1 & Anti-inflammatory & 1 & 0 & 0 & 0.18 & 0 & 0 \\ 
        AB25 & OEChem & 9 & 69299605 & n/a & n/a & 0 & 0 & 0 & 0.2 & 0 & 0 \\ 
        AB25 & OEChem & 10 & 136003417 & n/a & n/a & 0 & 0 & 0 & 0.29 & 0 & 0 \\ 
        AB25 & OEChem & 11 & 101659428 & n/a & n/a & 0 & 0 & 0 & 0.39 & 0 & 0 \\ 
        AB25 & OEChem & 12 & 136851558 & n/a & n/a & 0 & 0 & 0 & 0.44 & 0 & 0 \\ 
        AB25 & OEChem & 13 & 54159364 & US-2015209428-A1 & Photosensitizing Agent & 1 & 1 & 1 & 0.33 & 1 & 1 \\ 
        AB25 & OEChem & 14 & 91212111 & WO2010011959A1 & Antiviral & 1 & 0 & 0 & 0.43 & 0 & 0 \\ 
        AB25 & OEChem & 15 & 66696851 & WO-2009100800-A1 & Fluroescent & 0 & 1 & 1 & 0.26 & 1 & 1 \\ 
        AB25 & OEChem & 16 & 136079295 & WO-2009099673-A1 & Electron Transfer & 0 & 1 & 0 & 0.3 & 0 & 0 \\ 
        AB25 & OEChem & 17 & 69636218 & WO2006015714A1 & Optical Data Carrier & 0 & 1 & 1 & 0.25 & 1 & 1 \\ 
        AB25 & OEChem & 18 & 66835451 & WO-2006007184-A1 & Phosphorescent & 0 & 1 & 1 & 0.29 & 1 & 1 \\ 
        AB25 & OEChem & 19 & 69328133 & US-6916799-B2 & Ionophore & 1 & 0 & 0 & 0.33 & 0 & 0 \\ 
        AB25 & OEChem & 20 & 91003879 & US-6423469-B1 & Dye & 0 & 1 & 1 & 0.4 & 1 & 1 \\ 
        Avo & RDKit 0 & 1 & 69397178 & EP-1181000-B1 & Sunscreen & 0 & 1 & 1 & 0.74 & 0 & 0 \\ 
        Avo & RDKit 0 & 2 & 39444601 & n/a & n/a & 0 & 0 & 0 & 0.52 & 0 & 0 \\ 
        Avo & RDKit 0 & 3 & 56836196 & EP2600854B1/de & Sunscreen & 0 & 1 & 1 & 0.93 & 0 & 0 \\ 
        Avo & RDKit 0 & 4 & n/a & n/a & n/a & 0 & 0 & 0 & 0.41 & 0 & 0 \\ 
        Avo & RDKit 0 & 5 & 131632684 & n/a & n/a & 0 & 0 & 0 & 0.54 & 0 & 0 \\ 
        Avo & RDKit 0 & 6 & n/a & n/a & n/a & 0 & 0 & 0 & 0.35 & 0 & 0 \\ 
        Avo & RDKit 0 & 7 & 81633789 & n/a & n/a & 0 & 0 & 0 & 0.29 & 0 & 0 \\ 
        Avo & RDKit 0 & 8 & 102107511 & n/a & n/a & 0 & 0 & 0 & 0.43 & 0 & 0 \\ 
        Avo & RDKit 0 & 9 & 8862174 & n/a & n/a & 0 & 0 & 0 & 0.45 & 0 & 0 \\ 
        Avo & RDKit 0 & 10 & 67831086 & WO-9116034-A1 & Sunscreen & 0 & 1 & 1 & 0.9 & 0 & 0 \\ 
        Avo & RDKit 0 & 11 & 43337768 & n/a & n/a & 0 & 0 & 0 & 0.49 & 0 & 0 \\ 
        Avo & RDKit 0 & 12 & 18428769 & US-6448304-B1 & UV Absorbtion & 0 & 1 & 1 & 0.81 & 0 & 0 \\ 
        Avo & RDKit 0 & 13 & n/a & n/a & n/a & 0 & 0 & 0 & 0.45 & 0 & 0 \\ 
        Avo & RDKit 0 & 14 & 43321130 & n/a & n/a & 0 & 0 & 0 & 0.55 & 0 & 0 \\ 
        Avo & RDKit 0 & 15 & 72034670 & n/a & n/a & 0 & 0 & 0 & 0.44 & 0 & 0 \\ 
        Avo & RDKit 0 & 16 & 105089397 & n/a & n/a & 0 & 0 & 0 & 0.3 & 0 & 0 \\ 
        Avo & RDKit 0 & 17 & 56836133 & WO-2012016619-A1 & Sunscreen & 0 & 1 & 1 & 0.73 & 0 & 0 \\ 
        Avo & RDKit 0 & 18 & 63411706 & n/a & n/a & 0 & 0 & 0 & 0.38 & 0 & 0 \\ 
        Avo & RDKit 0 & 19 & 115781741 & n/a & n/a & 0 & 0 & 0 & 0.38 & 0 & 0 \\ 
        Avo & RDKit 0 & 20 & 43625029 & US-2018085348-A1 & RORI Inhibitor Anticancer & 1 & 0 & 0 & 0.55 & 0 & 0 \\ 
        Avo & RDKit n & 1 & 102420223 & US20230113128 & Skin Protectant & 0 & 1 & 1 & 0.75 & 0 & 0 \\ 
        Avo & RDKit n & 2 & 123816547 & US-2018155288-A1 & UV Absorbing & 0 & 1 & 1 & 0.73 & 0 & 0 \\ 
        Avo & RDKit n & 3 & 89752401 & n/a & n/a & 0 & 0 & 0 & 0.41 & 0 & 0 \\ 
        Avo & RDKit n & 4 & 69262951 & WO-9423693-A1 & UV Absorbing & 0 & 1 & 1 & 0.67 & 0 & 0 \\ 
        Avo & RDKit n & 5 & 89601532 & US20140323376A1 & Microcapsules & 0 & 1 & 0 & 0.37 & 0 & 0 \\ 
        Avo & RDKit n & 6 & 10592228 & n/a & n/a & 0 & 0 & 0 & 0.65 & 0 & 0 \\ 
        Avo & RDKit n & 7 & 102454724 & n/a & n/a & 0 & 0 & 0 & 0.37 & 0 & 0 \\ 
        Avo & RDKit n & 8 & 66756943 & US-8119107-B2 & UV Absorbing & 0 & 1 & 1 & 0.61 & 0 & 0 \\ 
        Avo & RDKit n & 9 & 15262337 & n/a & n/a & 0 & 0 & 0 & 0.52 & 0 & 0 \\ 
        Avo & RDKit n & 10 & 121423935 & US9452220B2 & n/a & 0 & 0 & 0 & 0.33 & 0 & 0 \\ 
        Avo & RDKit n & 11 & 9857795 & n/a & n/a & 0 & 0 & 0 & 0.58 & 0 & 0 \\ 
        Avo & RDKit n & 12 & 134950302 & n/a & n/a & 0 & 0 & 0 & 0.25 & 0 & 0 \\ 
        Avo & RDKit n & 13 & 89393392 & US-8362177-B1 & Refractive Copolymer & 0 & 1 & 1 & 0.27 & 1 & 1 \\ 
        Avo & RDKit n & 14 & 18370144 & US-6348617-B1 & Pyruvate Purification & 0 & 0 & 0 & 0.35 & 0 & 0 \\ 
        Avo & RDKit n & 15 & 237566 & n/a & n/a & 0 & 0 & 0 & 0.28 & 0 & 0 \\ 
        Avo & RDKit n & 16 & 62393856 & n/a & n/a & 0 & 0 & 0 & 0.32 & 0 & 0 \\ 
        Avo & RDKit n & 17 & 117535120 & n/a & n/a & 0 & 0 & 0 & 0.36 & 0 & 0 \\ 
        Avo & RDKit n & 18 & 69247116 & n/a & Synthesis-Like Patents & 0 & 0 & 0 & 0.41 & 0 & 0 \\ 
        Avo & RDKit n & 19 & 62745021 & n/a & n/a & 0 & 0 & 0 & 0.51 & 0 & 0 \\ 
        Avo & RDKit n & 20 & 67376073 & WO-2004105712-A1 & Sunscreen & 0 & 1 & 1 & 0.84 & 0 & 0 \\ 
        Avo & OEChem & 1 & 90148382 & US-9059408-B2 & Semiconductor & 0 & 1 & 0 & 0.12 & 0 & 0 \\ 
        Avo & OEChem & 2 & 89110687 & WO-2010003533-A2 & Tubercular Antibacterial & 1 & 0 & 0 & 0.13 & 0 & 0 \\ 
        Avo & OEChem & 3 & 88264669 & EP-2388002-A2 & Antidepressant & 1 & 0 & 0 & 0.14 & 0 & 0 \\ 
        Avo & OEChem & 4 & 123262979 & WO-2012084941-A1 & Photovoltaic & 0 & 1 & 1 & 0.08 & 1 & 1 \\ 
        Avo & OEChem & 5 & 102300581 & n/a & n/a & 0 & 0 & 0 & 0.12 & 0 & 0 \\ 
        Avo & OEChem & 6 & 101259804 & n/a & n/a & 0 & 0 & 0 & 0.14 & 0 & 0 \\ 
        Avo & OEChem & 7 & 88267104 & n/a & n/a & 0 & 0 & 0 & 0.14 & 0 & 0 \\ 
        Avo & OEChem & 8 & 91595142 & WO-2010050781-A1 & Electroluminescent & 0 & 1 & 1 & 0.12 & 1 & 1 \\ 
        Avo & OEChem & 9 & 10266401 & n/a & n/a & 0 & 0 & 0 & 0.13 & 0 & 0 \\ 
        Avo & OEChem & 10 & 58802136 & US7176237B2 & Many drug-like & 1 & 0 & 0 & 0.1 & 0 & 0 \\ 
        Avo & OEChem & 11 & 89652998 & EP-1775335-B9 & Electroluminescent & 0 & 1 & 1 & 0.24 & 1 & 1 \\ 
        Avo & OEChem & 12 & 101683095 & n/a & n/a & 0 & 0 & 0 & 0.12 & 0 & 0 \\ 
        Avo & OEChem & 13 & 10945525 & n/a & n/a & 0 & 0 & 0 & 0.09 & 0 & 0 \\ 
        Avo & OEChem & 14 & 57236739 & US-4193931-A & Anti-Tumor & 1 & 0 & 0 & 0.11 & 0 & 0 \\ 
        Avo & OEChem & 15 & 123440338 & WO2014086032A1 & Kinase Inhibitor & 1 & 0 & 0 & 0.12 & 0 & 0 \\ 
        Avo & OEChem & 16 & 122224185 & n/a & n/a & 0 & 0 & 0 & 0.21 & 0 & 0 \\ 
        Avo & OEChem & 17 & 23208426 & EP0654488B1 & Copolymer & 0 & 0 & 0 & 0.11 & 0 & 0 \\ 
        Avo & OEChem & 18 & 87402810 & US20130023518A1 & AMP Kinase Activator & 1 & 0 & 0 & 0.2 & 0 & 0 \\ 
        Avo & OEChem & 19 & 68630174 & EP-1641792-B1 & PDE10A Inhibitor & 1 & 0 & 0 & 0.15 & 0 & 0 \\ 
        Avo & OEChem & 20 & 88746062 & US-4011140-A & Anti-Tumor & 1 & 0 & 0 & 0.13 & 0 & 0 \\ 
        2-dPAC & RDKit 0 & 1 & 118405815 & US-2015280136-A1 & Optoelectronic & 0 & 1 & 1 & 0.28 & 1 & 0 \\ 
        2-dPAC & RDKit 0 & 2 & 57120863 & US-2021135127-A1 & Electroluminescent & 0 & 1 & 1 & 0.28 & 1 & 1 \\ 
        2-dPAC & RDKit 0 & 3 & 117633923 & EP-1802706-B1 & Electroluminescent & 0 & 1 & 1 & 0.26 & 1 & 0 \\ 
        2-dPAC & RDKit 0 & 4 & 57702177 & US-2009131673-A1 & Electroluminescent & 0 & 1 & 1 & 0.18 & 1 & 0 \\ 
        2-dPAC & RDKit 0 & 5 & 58401023 & n/a & n/a & 0 & 0 & 0 & 0.18 & 0 & 0 \\ 
        2-dPAC & RDKit 0 & 6 & 135173910 & US11289654B2/en & Optoelectronic & 0 & 1 & 1 & 0.28 & 1 & 0 \\ 
        2-dPAC & RDKit 0 & 7 & 54511588 & US-4931350-A & Electrophotographic & 0 & 1 & 1 & 0.33 & 1 & 0 \\ 
        2-dPAC & RDKit 0 & 8 & 53473286 & n/a & n/a & 0 & 0 & 0 & 0.51 & 0 & 0 \\ 
        2-dPAC & RDKit 0 & 9 & 66650204 & CN-114163447-A & Electroluminescent & 0 & 1 & 1 & 0.43 & 1 & 0 \\ 
        2-dPAC & RDKit 0 & 10 & 85666158 & US-2018040829-A1 & Electroluminescent & 0 & 1 & 1 & 0.27 & 1 & 0 \\ 
        2-dPAC & RDKit 0 & 11 & 118134114 & US-2015162542-A1 & Optoelectronic & 0 & 1 & 1 & 0.22 & 1 & 0 \\ 
        2-dPAC & RDKit 0 & 12 & 15641729 & n/a & n/a & 0 & 0 & 0 & 0.21 & 0 & 0 \\ 
        2-dPAC & RDKit 0 & 13 & 72736479 & JP2014034654 & Electrochromic & 0 & 1 & 1 & 0.32 & 1 & 0 \\ 
        2-dPAC & RDKit 0 & 14 & 131980099 & US-2017288148-A1 & Electroluminescent & 0 & 1 & 1 & 0.29 & 1 & 0 \\ 
        2-dPAC & RDKit 0 & 15 & 101115479 & n/a & n/a & 0 & 0 & 0 & 0.13 & 0 & 0 \\ 
        2-dPAC & RDKit 0 & 16 & 102333726 & n/a & n/a & 0 & 0 & 0 & 0.16 & 0 & 0 \\ 
        2-dPAC & RDKit 0 & 17 & 59133370 & US-2018090687-A1 & Electroluminescent & 0 & 1 & 1 & 0.29 & 1 & 0 \\ 
        2-dPAC & RDKit 0 & 18 & 132502906 & n/a & n/a & 0 & 0 & 0 & 0.26 & 0 & 0 \\ 
        2-dPAC & RDKit 0 & 19 & 12505268 & n/a & n/a & 0 & 0 & 0 & 0.32 & 0 & 0 \\ 
        2-dPAC & RDKit 0 & 20 & 88364885 & US-2012126180-A1 & Electroluminescent & 0 & 1 & 1 & 0.39 & 1 & 0 \\ 
        2-dPAC & RDKit n & 1 & 10892657 & n/a & n/a & 0 & 0 & 0 & 0.15 & 0 & 0 \\ 
        2-dPAC & RDKit n & 2 & 123135311 & n/a & n/a & 0 & 0 & 0 & 0.17 & 0 & 0 \\ 
        2-dPAC & RDKit n & 3 & 122210717 & n/a & n/a & 0 & 0 & 0 & 0.15 & 0 & 0 \\ 
        2-dPAC & RDKit n & 4 & 11810541 & n/a & n/a & 0 & 0 & 0 & 0.19 & 0 & 0 \\ 
        2-dPAC & RDKit n & 5 & 121228673 & WO-2021012717-A1 & Antitumor & 1 & 0 & 0 & 0.16 & 0 & 0 \\ 
        2-dPAC & RDKit n & 6 & 12065985 & n/a & n/a & 0 & 0 & 0 & 0.25 & 0 & 0 \\ 
        2-dPAC & RDKit n & 7 & 58966992 & US-2007213327-A1 & Kinase Inhibitor & 1 & 0 & 0 & 0.15 & 0 & 0 \\ 
        2-dPAC & RDKit n & 8 & 102163265 & n/a & n/a & 0 & 0 & 0 & 0.14 & 0 & 0 \\ 
        2-dPAC & RDKit n & 9 & 68969564 & US-2009326006-A1 & Beta-Secretase Inhibitor & 1 & 0 & 0 & 0.21 & 0 & 0 \\ 
        2-dPAC & RDKit n & 10 & 71456541 & n/a & n/a & 0 & 0 & 0 & 0.14 & 0 & 0 \\ 
        2-dPAC & RDKit n & 11 & 101082530 & n/a & n/a & 0 & 0 & 0 & 0.2 & 0 & 0 \\ 
        2-dPAC & RDKit n & 12 & 9836049 & n/a & n/a & 0 & 0 & 0 & 0.23 & 0 & 0 \\ 
        2-dPAC & RDKit n & 13 & 10585958 & n/a & n/a & 0 & 0 & 0 & 0.23 & 0 & 0 \\ 
        2-dPAC & RDKit n & 14 & 134852402 & n/a & n/a & 0 & 0 & 0 & 0.16 & 0 & 0 \\ 
        2-dPAC & RDKit n & 15 & 699711 & n/a & n/a & 0 & 0 & 0 & 0.17 & 0 & 0 \\ 
        2-dPAC & RDKit n & 16 & 101767888 & n/a & n/a & 0 & 0 & 0 & 0.16 & 0 & 0 \\ 
        2-dPAC & RDKit n & 17 & 59699922 & WO-2007106236-A2 & Kinase Modulator & 1 & 0 & 0 & 0.17 & 0 & 0 \\ 
        2-dPAC & RDKit n & 18 & 102515925 & n/a & n/a & 0 & 0 & 0 & 0.18 & 0 & 0 \\ 
        2-dPAC & RDKit n & 19 & 729459 & n/a & n/a & 0 & 0 & 0 & 0.18 & 0 & 0 \\ 
        2-dPAC & RDKit n & 20 & 5378237 & n/a & n/a & 0 & 0 & 0 & 0.16 & 0 & 0 \\ 
        2-dPAC & OEChem & 1 & 101378249 & n/a & n/a & 0 & 0 & 0 & 0.09 & 0 & 0 \\ 
        2-dPAC & OEChem & 2 & 11374507 & n/a & n/a & 0 & 0 & 0 & 0.16 & 0 & 0 \\ 
        2-dPAC & OEChem & 3 & 134982929 & n/a & n/a & 0 & 0 & 0 & 0.06 & 0 & 0 \\ 
        2-dPAC & OEChem & 4 & 121277727 & n/a & n/a & 0 & 0 & 0 & 0.21 & 0 & 0 \\ 
        2-dPAC & OEChem & 5 & 91515383 & US-2004151944-A1 & Electroluminescent & 0 & 1 & 1 & 0.13 & 1 & 1 \\ 
        2-dPAC & OEChem & 6 & 122204293 & n/a & n/a & 0 & 0 & 0 & 0.16 & 0 & 0 \\ 
        2-dPAC & OEChem & 7 & 136880633 & n/a & n/a & 0 & 0 & 0 & 0.18 & 0 & 0 \\ 
        2-dPAC & OEChem & 8 & 632403 & n/a & n/a & 0 & 0 & 0 & 0.07 & 0 & 0 \\ 
        2-dPAC & OEChem & 9 & 10178481 & US6919448B2 & Dye & 0 & 1 & 1 & 0.16 & 1 & 1 \\ 
        2-dPAC & OEChem & 10 & 102090697 & n/a & n/a & 0 & 0 & 0 & 0.09 & 0 & 0 \\ 
        2-dPAC & OEChem & 11 & 57599395 & US-2006216621-A1 & Electrophotographic & 0 & 1 & 1 & 0.19 & 1 & 1 \\ 
        2-dPAC & OEChem & 12 & 54275018 & US-5041366-A & Electrophotographic & 0 & 1 & 1 & 0.18 & 1 & 1 \\ 
        2-dPAC & OEChem & 13 & 137052079 & WO-2013105026-A1 & Nanoparticle Carrier & 0 & 0 & 0 & 0.08 & 0 & 0 \\ 
        2-dPAC & OEChem & 14 & 89304021 & EP-2535331-A1 & Electroluminescent & 0 & 1 & 1 & 0.18 & 1 & 1 \\ 
        2-dPAC & OEChem & 15 & 124008654 & WO-2013178041-A1 & Electroluminescent & 0 & 1 & 1 & 0.17 & 1 & 1 \\ 
        2-dPAC & OEChem & 16 & 118122961 & EP-2878599-A1 & Electroluminescent & 0 & 1 & 1 & 0.17 & 1 & 0 \\ 
        2-dPAC & OEChem & 17 & 89296775 & EP-2535331-A1 & Electroluminescent & 0 & 1 & 1 & 0.17 & 1 & 1 \\ 
        2-dPAC & OEChem & 18 & 90180369 & US-9899600-B2 & Electroluminescent & 0 & 1 & 1 & 0.19 & 1 & 1 \\ 
        2-dPAC & OEChem & 19 & 136436068 & n/a & n/a & 0 & 0 & 0 & 0.19 & 0 & 0 \\ 
        2-dPAC & OEChem & 20 & 60136245 & US-2010314615-A1 & Electroluminescent & 0 & 1 & 1 & 0.19 & 1 & 1 \\ 
\end{longtable}
\end{landscape}

\end{document}